\documentclass[pre,10pt,a4paper,nofootinbib]{revtex4}

\usepackage{graphicx}   
\usepackage{amsmath}
\usepackage{amssymb}

\newcommand {\be}  {\begin{equation}}
\newcommand {\bea} {\begin{eqnarray} \nonumber }
\newcommand {\ee}  {\end{equation}}
\newcommand {\eea} {\end{eqnarray}}
\newcommand {\al}  {\alpha}
\newcommand {\de}  {\delta}
\newcommand {\eps} {\epsilon}

\newcommand {\si}  {\sigma}

\renewcommand{\d}    {\mbox{ d}}

\newcommand {\sign}  { \mbox{sign} }

\newcommand {\ba}      {\overline}
\newcommand {\lan}     {\langle}
\newcommand {\ran}     {\rangle}

\def \parziale#1#2  {{\partial {#1} \over \partial {#2}}}

\begin{document}

\title{Replica Symmetry Breaking in Short-Range Spin Glasses:\\
  Theoretical Foundations and Numerical Evidences}

\author{
  Enzo Marinari$^{(1)}$, 
  Giorgio Parisi$^{(2)}$,
  Federico Ricci-Tersenghi$^{(3)}$,\\ 
  Juan J. Ruiz-Lorenzo$^{(4)}$, 
  Francesco Zuliani$^{(1)}$}

\affiliation{
1) Dipartimento di Fisica and INFN, Universit\`a di Cagliari,\\
Cittadella Universitaria, S.P. Monserrato-Sestu km 0.7,\\
09042 Monserrato (CA), Italy\\
2) Dipartimento di Fisica and INFN Universit\`a di Roma {\em La Sapienza},\\
P. A. Moro 2, 00185 Roma, Italy\\
3) International Center for Theoretical Physics (ICTP),\\
P.O. Box 586, 34100 Trieste, Italy\\
4) Departamento de F\'{\i}sica Te\'orica I Universidad Complutense de Madrid,\\
Ciudad Universitaria, 28040 Madrid, Spain}

\date{November 1999}

\begin{abstract}
We discuss replica symmetry breaking (RSB) in spin glasses. We update
work in this area, from both the analytical and numerical points of
view. We give particular attention to the difficulties stressed by
Newman and Stein concerning the problem of constructing pure states
in spin glass systems. We mainly discuss what happens in
finite-dimensional, realistic spin glasses. Together with a detailed
review of some of the most important features, facts, data, and
phenomena, we present some new theoretical ideas and numerical
results. We discuss among others the basic idea of the RSB theory,
correlation functions, interfaces, overlaps, pure states, random
field, and the dynamical approach. We present new numerical results
for the behaviors of coupled replicas and about the numerical
verification of sum rules, and we review some of the available
numerical results that we consider of larger importance (for example,
the determination of the phase transition point, the correlation
functions, the window overlaps, and the dynamical behavior of the
system).
\end{abstract} 

\maketitle

\section{Introduction} 

The concept of replica symmetry breaking (RSB) as a crucial tool in
the study of spin glass systems has been introduced close to twenty
years ago, and a successful Ansatz for the breaking pattern has been
proposed \cite{RSBONE,BREAKING,PARDUE}.  Nowadays there are no doubts
that it describes correctly what happens in infinite range models
\cite{MEPAVI,GUERRA,AIZCON}.  On the contrary its correctness in the
case of short range models has remained controversial.  For short
range models we cannot exhibit a solution in closed form, and things
are made more difficult from the fact that the picture that emerges
from the RSB is substantially different from the usual scenario valid
for normal ferromagnets, with which we are familiar.

Moreover we have to deal with rather complex theoretical predictions,
that can be interpreted in a simpler way by recasting them as the
claim of the existence of an extremely large number of pure states (or
phases) for very large systems.  However this notion of many
equilibrium states (not related by a trivial symmetry) is rather
counterintuitive in the case of spin glasses (since it is difficult to
visualize the situation in a fruitful way) while it would be quite
natural in other contexts.  For example the picture is clear in the
case of protein folding, where a different equilibrium state
corresponds to a different folding (tertiary structure) of the
protein.  Such intrinsic novelty and unusual character of the RSB
formalism (and maybe some lack of precision in the mathematical
definitions of some of the relevant literature \cite{MEPAVI}) has led
to some degree of controversy on the precise definition and meaning of
RSB and on its range of validity \cite{NS0,NSnew,NS1,NSREP,NS3,NS4} (for the
possible application of the Migdal-Kadanoff approximation, MKA, to
spin glass systems, see \cite{MIGKAD,MCMILL,BRAMOO,FISHUS}: it is
interesting to notice that Gardner \cite{GARDNER} has shown that it
fails already at the mean field level).

In this paper we try to bring a positive contribution to the 
discussion by stating in the most clear and complete way which are the 
predictions implied by a RSB scenario, and summarizing the quite ample 
numerical evidence for its validity in finite dimensional systems.  
This paper integrates and supplements the review \cite{BOOKYO} (for 
relevant books and reviews see \cite{MEPAVI,BINYOU,FISHER,PARISI}).  
In order not to mix different conceptual issues we will first state 
the RSB predictions without using the concept of finite volume pure 
state.  However this concept will be used later, when discussing the 
physical interpretation of the Ansatz.  We will also discuss the 
relation among the finite volume pure states and the infinite volume 
ones.

We will analyze the two cases of the infinite range model and of short 
range one.  We will discuss the problem of the number of {\sl phases} 
for a finite system in the infinite volume limit and for an actual 
infinite system (these are two related but {\sl different} problems).  
In doing that one should distinguish clearly the predictions coming 
from RSB from their consequences on the equilibrium states of the 
system.  We will also discuss in detail the available numerical 
evidence.  We will present new data which should help in clarifying 
the situation further.

\section{The Replica Method}

\subsection{The Basic Results}

The first steps are quite simple.  We consider a system of $N$ spins 
whose Hamiltonian depends on some quenched random couplings $J$.  In 
order to avoid all the complications present in the definition of the 
Boltzmann-Gibbs distribution for an actual infinite volume system (see 
section (\ref{INFINITE})) we follow the traditional approach of 
defining a probability distribution of the configuration $C$ as $P(C) 
\propto \exp (- \beta H)$ for a finite volume system.  It is evident 
that as it stands this definition cannot be used naively for an actual 
infinite system, since the exponent would be always infinite.  A more 
careful treatment is needed.

For definiteness we shall consider two different models:

\begin{itemize}

\item 
A long range model, i.e. the SK model \cite{SK} with Hamiltonian

\begin{equation}
  {\cal H}_{J}=\sum_{i,k} \si_{i}J_{i,k}\si_{k} -\sum_{i}h_{i}\si_{i} \ ,
\end{equation}
where $\sigma_{i}=\pm 1$, the first sum is over all couples of sites 
of the lattice and the second one over all sites.  The couplings $J$ 
are quenched random variables with zero expectation value and for a 
system with $N$ sites $\overline{J^{2}}=N^{-1}$ (we denote by brackets 
the thermal average and by an upper bar the averages over the quenched 
disorder).

\item A short range model, i.e. the EA model \cite{EA} with 
Hamiltonian:

\begin{equation}
  {\cal H}_{J}= \sum_{i,k}' \si_{i} J_{i,k}\si_{k} 
-\sum_{i}h_{i}\si_{i}\ ,
  \protect\label{ham}
\end{equation}
where now the primed sum runs over couples of first neighboring 
sites, 
and $\overline{J^{2}}=1$.

\end{itemize}

In order to study the average over the couplings $J$ of the free 
energy and of the correlation functions it is useful to consider a 
system where the spin configurations are replicated $n$ times (in the 
same realization of the random interaction), with Hamiltonian

\begin{equation}
  {\cal H}^{(n)}=\sum_{a=1}^n {\cal H}_{J}(\{\si^{(a)}\}) \ .
  \protect\label{REPHAM}
\end{equation}
After integration over the $J$ one arrives to a new Hamiltonian in 
which different replicas are coupled.  In order to study the model it 
is convenient to introduce the quantity

\begin{equation}
  q_{a,b}\equiv \frac{1}{N} \sum_{i}\si_{i}^{(a)}\si_{i}^{(b)}\ ,
  \protect\label{E-OVERLAP}
\end{equation} 
where $a$ and $b$ characterize two of the different replicas that we 
have introduced, and $N$ is the number of spins per configuration.

We follow here a standard procedure (used in Statistical Mechanics and 
Field Theory in the last $75$ years \cite{STAFIE}): we first derive a 
mean field approximation, in which some kind of fluctuations are 
neglected (for a ferromagnet we would get the equation $m=\tanh(\beta 
z J m)$, where $m$ is the magnetization, $z$ is the coordination 
number of the lattice and $J$ does not depend on the site).  At a 
second stage we deduce the correlation from the mean field 
approximation, obtaining the equivalent of the Ornstein-Zernike (OZ) 
expression for the correlation functions.  In a third stage, which 
even for ferromagnets is relatively recent, we compute systematically 
the corrections to the mean field approximation and (when this is 
possible) we use the renormalization group to sum them up.  In the 
case of spin glasses this last stage is still lacking, so that we are 
going to discuss only the mean field predictions with the computation 
of OZ correlations.

In the mean field approximation (i.e.  when we neglect fluctuations) 
we find a free energy $F[Q]$, where $Q_{a,b}\equiv\lan q_{a,b}\ran$.  
It is usually assumed (and it can be proved for $n>1$) that the value 
of $Q$ can be found by solving the equation

\begin{equation}
\parziale{F}{Q_{a,b}} =0 \ ,
\end{equation}
under the condition that all the eigenvalues of the Hessian

\begin{equation}
 (\mbox{Hess})_{ab,cd} \equiv 
 \frac{\partial^{2} F}{\partial Q_{a,b} \partial Q_{c,d}}
\end{equation}
are non negative.  

The real problem arises when the saddle point equations admit more
than one solution, and these solutions are related by a symmetry.
This phenomenon is well known (it was discovered by Archimedes a long
time ago), and in the context of Statistical Mechanics it goes under
the name of {\em symmetry breaking}.  In this particular case the
obvious symmetry\footnote{Exactly at zero magnetic field more
symmetries would appear, but we do not deal with this case and we do
not take them into account.} is the permutation of different replicas.
This symmetry is broken as soon as the non-diagonal elements of
$Q_{a,b}$ depend on $a$ and $b$.  In this case the usual prescription
is to average over all the different solutions of the saddle point
equations in replica space (which being related by a symmetry have
exactly the same free energy).  Therefore if the value of an
observable $A$ depends on $Q$ we find that

\begin{equation}
  \ba{\lan A \ran}= \frac{1}{n!} \sum_{\Pi} A(Q_{\Pi(a),\Pi(b)}) \ ,
\end{equation} 
where the sum runs over the $n!$ permutations ($\Pi$) of $n$ elements.

If we apply this procedure to the correlation functions we find that 
\cite{PARDUE,ITZYKS}

\begin{equation}
  \ba{
    \frac{1}{N^2} \langle \sum_{i,k}\si_{i}\si_{k}\rangle^{2 s}
  }   = 
  \lim_{n\to 0}\frac{1}{n(n-1)} \sum_{a,b;a\ne b} Q_{a,b}^{s} \ .
  \protect\label{E-PQ-A}
\end{equation}

The previous equation can also be written as

\begin{equation}
  \int dq\ P(q)\  q^s =  
  \lim_{n\to 0}\frac{1}{n(n-1)} \sum_{a,b;a\ne b} Q_{a,b}^{s} \ , 
  \protect\label{E-PQ-B}
\end{equation}
where 

\begin{equation}
  P(q)=\ba{P_{J}(q)}\ ,
\end{equation}
and $P_{J}(q)$ is the probability that two configurations selected 
according to the Boltzmann-Gibbs (B-G) distribution are such that 
their overlap (\ref{E-OVERLAP}) is equal to $q$.

The properties of the systems are related to those of the matrix
$Q_{a,b}$. At first view the introduction of this matrix could appear
a strange step. It has been shown however that it represents a compact
way to encode the probability distribution of the overlaps among an
arbitrary number of equilibrium configurations, and to describe how
this probability changes when we change the couplings $J$.

The form of the matrix $Q$ is crucial. In the replica approach we are
forced by stability considerations (see section \ref{S-STOSTA} on
stochastic stability) to consider matrices in which every line is a
permutation of the other lines. The consequences of this form of the
matrix $Q$, or equivalently of {\em stochastic stability}, will be
discussed in the rest of this paper.  There are some extra predictions
which come from more specific assumptions about the form of the matrix
$Q$. For example if we also assume ultrametricity we will refer to
them as the {\em Mean Field predictions}.  Indeed in the mean field
approximation one finds that the matrix $Q$ which makes the mean field
free energy stationary is ultrametric.  When we will meet a prediction
which depends on the specific form of the matrix $Q$ we will note it
explicitly.

If the function $P_{J}(q)$ is trivial, i.e.  if 

\begin{equation}
  P_{J}(q)=\delta\left(q-q_{EA}\right)\ ,
  \label{TRIVIAL}
\end{equation}
independently on the coupling $J$ (as it happens in ferromagnets), the
use of the replica method would be not interesting\footnote{For
simplifying the discussion we assume that a non-zero, albeit small,
magnetic field is turned on, so that the symmetry $\si \to - \si$ is
broken.  If this is not true the function $P(q)$ is be symmetric under
the transformation $q \to -q$.  Usually, even in a ferromagnet, in the
cold phase the function $P(q)$ is not continuous at $h=0$: its value
at $h=0$ is different from the limits $h \to 0^{\pm}$.}. If the
function $P(q)$ is non-trivial we say that {\em replica symmetry is
broken}.

It is convenient to introduce the function $x(q)$ 

\begin{equation}
  x(q)=\int_{0}^{q} dq \ P(q)\ ,
\end{equation}
where we assume that the function $P(q)$ has support for positive
values of $q$. We define a function $q(x)$ as the inverse
of $x(q)$. Of course $q(x)$ is defined only in the interval $0\le x
\le 1$. We thus have

\begin{equation}
  P(q)= \frac{dx}{dq}\ .
  \protect\label{E-XQ}
\end{equation}
It is clear that the correlation functions are not clustering unless
the function $q(x)$ is constant, i.e.  (see (\ref{E-QEA}))
$q(x)=q_{EA}$ (this is needed so that in a translational invariant
system intensive quantities do not fluctuate in the infinite volume
limit and to make all connected correlation functions vanishing at
large distance): for clustering systems the function $P(q)$ must be
equal\footnote{We remind the reader that we are in finite volume and
that the delta functions are slightly smoothed.  It is implicit that
all our statements become valid (obviously in distribution sense)
without corrections only when the volume of the system goes to
infinity.} to $\delta(q-q_{EA})$.

The replica formalism also allows to study the dependence of the
function $P_{J}(q)$ over the overlap $q$. Under general assumptions
one finds \cite{MPSTV} that universal relations like

\begin{equation}
  \ba{P_{J}(q_1)P_{J}(q_2)} = \frac23 P(q_1)  P(q_2) +
  \frac13 P(q_1) \delta(q_1-q_2)\ ,
  \protect\label{GUEREL}
\end{equation}
hold ($q_{1}$ and $q_{2}$ are two given values of $q$).  A set of
relations of this type has been proven under general hypothesis
(i.e. stochastic stability) in \cite{GUERRA} and in \cite{AIZCON}.

In the most popular scheme for replica symmetry breaking
\cite{BREAKING,PARDUE}, which is supposed to be correct for the
infinite range Sherrington Kirkpatrick (SK) model, given any function
function $q(x)$ (where $x$ goes from $0$ to $1$) in the limit $n \to
0$ we can find a matrix $Q_{a,b}$ such that the previous equations are
satisfied.

In the infinite volume limit the typical form of the function $q(x)$ 
is such that (at least in the mean field approximation)

\begin{equation}
  P(q)=a\de(q-q_{m})+b\de(q-q_{M})+r(q) \ ,
  \protect\label{E-DUEDEL}
\end{equation}
where the function $r(q)$ is smooth\footnote{The form of the function
$P(q)$ is discontinuous in the limit of zero magnetic field.}.

Some further relations which are not necessarily valid on general
grounds turn out to hold in the RSB Ansatz, that is verified in the
solution of the mean field theory of spin glasses: the most popular
one is the ultrametric relation \cite{MPSTV}, which states that the
overlap distribution of three overlaps ($q_{1,2},q_{2,3}, q_{3,1}$)
among three replicas ($1$, $2$, and $3$) in the same realization of
the quenched disorder is zero as soon as the ultrametricity inequality

\begin{equation}
  \min\left(q_{1,2},q_{2,3}\right)\ge q_{3,1}
\end{equation}
(or one of the two other permutations of the former
relation) is violated.  It can be shown \cite{INPARU96} that these
inequalities (plus the general conditions which come from stochastic
stability) imply that the average probability distribution of three
overlaps is given by:

\begin{eqnarray}
  \nonumber
  \ba{P_{J}(q_{1,2})P_{J}(q_{2,3})P_{J}(q_{3,1})}
  &=& \frac{1}{2} P(q_{1,2}) x(q_{1,2}) 
  \delta(q_{1,2}-q_{2,3}) \delta(q_{2,3} - q_{3,1})\\
  &+& \frac{1}{2} \left[ P(q_{1,2}) P(q_{2,3})
  \theta(q_{1,2}-q_{2,3}) \delta(q_{2,3}-
  q_{3,1}) +\mbox{two permutations }  \right] \ .
\end{eqnarray}

Moreover all the probability distributions of more that three overlaps
and their variation from system to system are completely determined by
the function $P(q)$, or equivalently by the function $q(x)$. In other
words the usual replica symmetry breaking of mean field theory is the
only possible pattern of replica symmetry breaking that is
ultrametric.

In principle it is possible that ultrametricity is valid in the
infinite range model and it is violated in finite dimensional short
range models, the function $P(q)$ remaining non-trivial and
consequently replica symmetry being still broken. In this case one
would have to consider, in the short range models, more complex forms
of the matrix $Q$. Indications that this does not happen and that
ultrametricity will hold in finite dimensional models will be
discussed later.

Summarizing we are considering three possibilities

\begin{itemize}

\item
The function $P(q)$ is a single delta function at $q=q_{EA}$, and
$q_{EA}$ is non-zero below the critical temperature. This situation,
with no breaking of the replica symmetry, is realized in the Migdal
Kadanoff approximation or equivalently in the droplet model.

\item
The $P(q)$ is a not a single delta function: in this case replica
symmetry is broken. The condition of replica equivalence strongly
constrains the behavior of the system.

\item
The ultrametricity condition is satisfied. This is the mean field
case, where many more detailed predictions can be obtained.

\end{itemize}

\subsection{The Correlation Functions\label{SS-CF}}

We have already commented on the fact that a non trivial form of the 
function $P(q)$ implies that the correlation functions are 
non-clustering.  In this case the study of the correlation functions 
can be involved, since we can select different values of the overlap. 
In all generality we can try to determine correlations of $\Theta$ 
operators at different fixed overlap $q_{\theta}$. Here we will focus 
only on the simplest kind of such correlation functions. 

As before we consider two copies of the system in the same realization 
of the disorder, and we denote their spin variables by $\sigma$ and 
$\tau$.  We define the local overlap on the site $i$ by

\begin{equation}
  q_{i} \equiv \si_{i}\tau_{i} \ ,
\end{equation}
which can take the values $\pm 1$.
We define the average $\lan \ \cdot\  \ran_{\hat{q}}$, 
restricted to the configurations of the system such that

\begin{equation}
  \frac{1}{N} \sum_{i}q_{i}= {\hat{q}} \ .
\end{equation}
We define now a (non-connected) correlation function 

\begin{equation}
G^{N}_{ \hat q}(i)\equiv \ba{\lan q_{i} q_{0} \ran_{\hat{q}}} \ ,
\end{equation}
where we are considering systems with $N$ spins.  These $\hat q$
dependent correlation functions can be computed in the replica
formalism by using the analogue of the OZ formulae.  We consider the
case, relevant for spin glasses, where the function $r$ of
(\ref{E-DUEDEL}) is non-zero and $q_{m}\le {\hat q} \le q_{M}$ (i.e.
$\hat{q}$ is in the support of $P(q)$).  In the usual Ansatz for
replica symmetry one finds that \cite{DKTYOU,DOKO86,DEKOTE,DOKO94}

\begin{equation}
  \lim_{i \to \infty} \lim_{N \to \infty} G^{N}_{ \hat q}(i) 
  \equiv \lim_{i \to \infty} G^{\infty}_{ \hat q}(i)={ \hat q}^{2} \ .
\end{equation}
Therefore for all $q_{m}\le {\hat q} \le q_{M}$ $q-q$ connected
correlation functions at fixed $q$ value satisfy the usual cluster
decomposition.

The total free energy of the constrained system (with fixed $\hat{q}$)
is (apart from corrections of order one) equal to the unconstrained
free energy because $P({\hat{q}}) \ne 0$.

A behavior of this type is unusual, and we will discuss it in better
detail in the next sections.  One can think about a similar
construction in a ferromagnet: in that case when forcing a value of
the magnetization smaller than the spontaneous magnetization one would
create a situation where the correlation function of the magnetization
would not be clustering (there would be at least an interface).

What is happening here is that for each value of the overlap $\hat q$
that we can select we find a different asymptotic limit for the value
of the correlation functions.  The approach of these correlation
functions to their large distance asymptotic value is very
interesting: in mean field using the OZ approximation one finds that
for large $i$

\begin{equation}
\lim_{N \to \infty} G^{N}_{ \hat q}(i) \approx{ \hat q}^{2} + A(\hat
q) i^{-\theta( \hat q)} \ .
\end{equation}

The exponent (which is a non trivial function of ${\hat q}$) is known
only at zero loops in the mean field approximation
\cite{DEKOTE,DKTYOU}. It is an extremely important and open question
to verify if its value is correct beyond tree level. We can
distinguish three cases:

\begin{itemize}

\item $\theta(q_{M}) =D-2$. There is a feeling that this result may 
be exact,  and that it corresponds to some kind of Goldstone theorem
(see \cite{DKTYOU} and references therein, and \cite{FERPAR}).

\item $\theta(q)= D-3$ for $q_M > q \ge q_{m}$ (where the last
equivalence holds only for $q_m$ different from $0$). There is a
feeling that this result may be modified by corrections to the mean
field approximation \cite{DKTYOU}, although there is no final evidence
of this effect.  On the contrary, if the exponent would be exact even
in three dimensions, one should conclude that replica symmetry
breaking disappears in $D=3$, and $3$ should be identified as the
lower critical dimension (but for the existence of a phase transition
in $3D$ see \cite{KAWYOU,BERJAN,BOOKYO,MPRUNP,INMAPARU}).

\item $\theta(q_m)= D-4$ for $q=q_{m}=0$ (i.e. only in zero magnetic
field). This result does   not apply in $D<6$.  The
following formula has been conjectured (see \cite{DKTYOU} and 
references therein):

\begin{equation}
  \theta(q=0) = \frac{(D-2+\eta_{c})}{2}\ ,
\end{equation} 
where $\eta_{c}$ is the exponent $\eta$ for the $q-q$ correlations at
the critical point.

\end{itemize}

A phenomenon which is related to this behavior of the
correlation functions is the non-triviality of the window overlap
distributions. Let us consider the case of $3$ dimensions. We select a
cubic region $B$ of $B^{3}$ spins in the center of our system (of $N$
spins) and we define the {\em window overlap}

\begin{equation}
  q_{B} \equiv \frac{1}{B^{3}} \sum_{i\in B} q_{i} \ .
\end{equation}
We can define the probability distribution

\begin{equation}
  P_{B}(q) \equiv \lim_{N\to \infty}P_{B}^{N}(q) \ ,
\end{equation}
where $P_{B}^{N}(q)$ is the probability distribution of $q_{B}$ in a
system with $N$ spins.  The prediction of the replica approach is then
that $P_{B}(q)$ is non trivial.  One finds that

\begin{equation}
  \lim_{B \to \infty}  P_{B}(q)= P(q) \ .
\end{equation}
If we define the restricted probability distributions $P_{B}(q|{\hat
q})$, i.e.  we consider only those pairs of configurations with a
fixed value of the overlap $\hat q$, we find that

\begin{equation}
  \lim_{B \to \infty} P_{B}(q|{\hat q})=\delta({\hat q}-q) \ ,
\end{equation}
where the rate of the approach to the asymptotic limit is controlled
by the appropriate exponent $\theta({\hat q})$ that we have just
discussed.

In other words if we fix the value of the total overlap we expect that
the average overlap in each large region will become equal to the
global overlap. This mean that without paying any price in free
energy density we can change the average value of the overlap in each
region of the space. This in sharp contrast with what would happen
in a ferromagnet, where locally the system would remain in the phase
of positive or negative magnetization.

\subsection{No Disguised Interfaces\protect\label{SS-NODISG}}

The results shown in the previous sections are the main predictions of
the RSB Ansatz.  Broken Replica Theory predicts that for realistic
spin glasses the function $P(q)$ is non trivial not because of the
presence of interfaces among two different phases, which are located
in random position.  In order to stress this point and to clarify the
implications of these predictions we will discuss now a negative
example: we will examine a very simple physical system with a
non-trivial $P(q)$ which {\bf does not} satisfy these predictions, and
we will show how RSB theory is adamant in differentiating this case.

The system is simple: we consider a three dimensional ferromagnet that 
in the infinite volume limit develops a non-zero spontaneous 
magnetization $m$ (i.e.  with $T<T_c$).  For sake of simplicity we 
also suppose that the temperature is sufficiently small that the 
interface among the positive and negative magnetized phases is not 
rough.  We also consider a finite volume realization of the system in 
a cubic box of $N$ spins with periodic boundary conditions in the $y$ 
and $z$ direction and antiperiodic boundary conditions in the $x$ 
direction.  At a sufficiently low temperature the antiperiodic 
boundary conditions force the formation of a domain wall.  For very 
large volumes $N$ we can classify the equilibrium configurations by 
the magnetization profile

\begin{equation}
  m(i_{x})\equiv \frac{1}{N_y N_z} \sum_{i_y=1,N_y;i_z=1,N_z} 
  \si(i_x,i_y,i_z)\ ,
\end{equation}
that will be of the form

\begin{equation}
  m(i_{x}) = f(i_{x}-I) \ ,
\end{equation}
where $I$ is the position of the interface, and the function $f$
characterizes the profile. If $i_x$ takes integer values from $1$ to
$N_x$, $f$ is or $+m$ for $i_x\to 0$ and $-m$ for $i_x\to N_x$ or it
is $-m$ for $i_x\to 0$ and $+m$ for $i_x\to N_x$ and it satisfies
$f(\pm \infty)=\pm m$.  This will be true for almost all
equilibrium configurations, with exceptional configurations whose
probability goes to zero when $N$ goes to infinity.
If we now  consider two generic equilibrium configurations a trivial
computation shows that in the limit $N\to \infty$ 
$P(q)=\frac{1}{2m^2}$ for
$-m^2\le q\le m^2$.

Apart from the absence of fluctuation of  $P_J(q)$ with $J$ (this
system is not a random system) the results for the correlation are
different from the prediction of the RSB Ansatz.
The fact that in each configuration there is a single domain wall
implies that

\begin{equation}
  \lim_{B\to \infty}P_{B}(q) =
  \frac12 \delta (q-m^{2}) +
  \frac12 \delta (q+m^{2}) \ ,
\end{equation}
because if we look to a local observable we have zero probability of
hitting an interface.  In this example the window overlaps are
insensitive to the boundary conditions.

With the same token we get that here

\begin{equation}
  \lim_{i \to \infty} G^{\infty}_{ \hat q}(i)= m^{4} \ .
\end{equation}
In other words in this case the system may exist in two different
phases separated by an interface.  The arbitrariness in the position
of the interface implies a non trivial form of the function $P(q)$.
However local quantities are insensitive to the presence of an
interface and the replica predictions are not valid.  This
situation is not described by replica symmetry breaking, and a study
of the relevant quantities makes it clear.

If we add a random term to the Hamiltonian, for example a random
dilution, it is possible to create a situation where the interface
will be pinned in one single position. In this case the function
$P(q)$ will become a single trivial delta function. If we allow to the
interface two or more positions we will obtain again a non trivial
function $P(q)$. The local overlaps are insensitive to all these
variations. In the two cases that we have described where the function
$P(q)$ is non-trivial (an infinite number of positions or few
positions allowed to the interface) the relation (\ref{GUEREL}) is not
satisfied (unless some magic coincidence happens in the last case,
where the interface may be located in a few selected positions).

\subsection{The Energy Overlap\protect\label{S-ENEOVE}}

An alternative way to understand how much the predictions of the RSB
Ansatz differ from the situation described in the previous section
(i.e.  where there are two or more different phases related by a
symmetry separated by an interface) is considering different kinds of
overlaps among spin configurations, in particular the energy
overlap.

Generally speaking, given a local quantity\footnote{In the infinite 
volume limit $A_{i}[\si]$ is a continuous function of $\si$ in the 
usual product topology.} $A_{i}[\si]$ which depends only over the 
spins of the configuration $\si$ that are close to the site $i$, we 
can define the $A$-overlap of two configurations as

\begin{equation}
  q_{A}[\si,\tau] \equiv \frac{1}{N}\sum_{i} A_{i}[\si] A_{i}[\tau] \ . 
  \protect\label{E-OVEGEN}
\end{equation}
A very interesting case is taking as local operator the energy
$E$, i.e. writing

\begin{equation}
  {\cal H}[\si]=\sum_{i} E_{i}[\si]\ ,
\end{equation}
and using $E_i$ to compute what we call the {\em energy overlap}
($q_{E}$).

The crucial point is that in the infinite volume limit the interfaces
have vanishing weight. If the bulk of the system is in two or more
different phases related by a symmetry, away from the interface the
contribution to the energy overlap is constant.  In this way one sees
that in this case in the infinite volume limit the probability
distribution of $q_{E}$ becomes a delta function.

On the contrary when replica symmetry is broken the probability 
distribution of $q_{E}$ remains non trivial also in the infinite 
volume limit.  In the SK model one has that $q_{E}[\si,\tau]$ is a 
linear function of $q[\si,\tau]^{2}$: it is clear that here the two 
probability distributions $P(q)$ and $P_{E}(q_{E})$ are strongly 
related, and they must become simultaneously different from a delta 
function\footnote{By using some imagination we could say that the 
breaking of the replica symmetry corresponds to a situation where we 
have an interface which occupies a finite fraction of the volume.  On 
the other hand in this situation we could not speak anymore of an 
interface in a proper sense.}.  This linear relation may approximately 
hold also in finite dimensional systems, but there are no reasons why 
it should remain exact.  At least in the mean field approximation 
all possible types of overlaps are functions of the basic overlap $q$, 
so that their probability distribution can be computed by evaluating 
the function $P(q)$ and the relation among the generalized overlap and 
the basic overlap $q$.

An other relevant quantity is the {\em link overlap}

\begin{equation}
  q_{\mbox{link}}[\si,\tau] \equiv
  \frac{2}{zN}\sum_{i,k}\si_{i}\si_{k}\tau_{i}\tau_{k}\ ,
\end{equation}
where the sum is taken over all nearest neighbor pairs, and $z$ is the
coordination number of the lattice.  In the case of a model with
$J=\pm 1$ it coincides with an energy overlap where the sum is done
over the sites and not over the links.  The link overlap is clearly
sensitive to differences in the correlation functions of the two
configurations.  It is also equal to the $q-q$ correlation at distance
one.  The fact that the link overlap has a non trivial distribution is
related to the dependence of the correlation function
$G^{\infty}_{ \hat q}(i)$ on $\hat q$ at $i=1$.

\section{Finite Volume States and RSB Results}

\subsection{The Definition of States of a Finite System}

In order to develop a formalism useful to discuss the physical meaning
of the results contained in the previous section it is convenient to
introduce the concept of {\em pure states in a finite volume}.  This
concept is crystal clear from a physical point of view. However it can
be difficult to state it in a rigorous way (i.e to prove existence
theorems) mostly because the notion of finite volume pure states (or
phases) is deeply conditioned by the physical properties of the system
under consideration.  In order to prove theorems on finite volume pure
states one needs a very strong rigorous command of the physical
properties of finite, large statistical system.

Most of the research in mathematical physics has been devoted to the
study of the pure states of an infinite system.  Unfortunately the
concept of pure states of an infinite system is too rigid to capture
all the statistical properties of a finite system and it is not
relevant for the physical interpretation of the replica predictions:
here we need, as we will see later, more sophisticated tools, that are
exactly finite volume pure states.

Readers who are not interested to enter the details of this
interpretation of the physical picture of RSB may skip this section
and base their considerations on the results discussed in the previous
chapter.  Those who decide to read this section should be aware
that the finite volume pure states we introduce here are
mathematically different from the pure states for an infinite
system which are normally used in the literature.

We must admit that there has been some confusion on this point in the literature on RSB. In the first paper where the interpretation of replica symmetry breaking was presented \cite{PARDUE} the words \textit{pure states, clustering decomposition...} were used, strongly suggesting that the states one was speaking about were the infinite volume states which are normally used in the literature. Only later on, after some reflections, it became clear from the cavity approach \cite{MPV_86} and from the presence of a chaotic dependence on the volume \cite{NS_92}, that the there were some problems in identifying the states that were needed in the replica approach with the infinite volume states.

A first attempt in describing directly what happens in finite volume was done in Ref.~\cite{PAR1}. For a while the problem was not investigated further and the words \emph{pure states} continued to be used in the literature: no explicit discussion appeared about why finite volume states should be used, as opposed to the infinite volume pure states. One of the reasons of this lack of precision was that the decomposition into states is a theoretical tool mainly needed to reach a better understanding of the theoretical framework, and that no ambiguity was present for predictions about numerical simulations and about experimental results.

Let us see how approximate pure states or phases in a large but {\em
finite} system can be defined, introducing a definition of state that
is different, but maybe more physical than the usual one.  We will at
first give a rough definition.  Our strategy is to mimic the
definition of pure states of an infinite system and to apply it to the
physical relevant situation (which is the only one accessible by
numerical simulations and by experiments) of a finite (and large)
system.

We consider a system in a box of linear size $L$, containing a total
of $N$ spins.  We partition the configuration space in regions,
labeled by $\alpha$, and we define averages restricted to these
regions \cite{PAR1,PAR2}: {\em these regions will correspond to our
finite volume pure states or phases}.  It is clear that in order to
produce something useful we have to impose sensible constraints on the
form of these partitions.  We require that the restricted averages on
these regions are such that connected correlation functions are
small\footnote{The precise definition of {\em small at large distance}
in a finite volume system can be phrased in many different ways.  For
example we can introduce a function $g(x)$ which goes to zero when $x$
goes to infinity and require that the connected correlation functions
evaluated in a given phase are smaller than $g(x)$.  Of course the
function $g(x)$ should be carefully chosen such not to give trivial
results; one should also prove the independence of the results from
the choice of $g$ in a given class of functions.  The choice of the
class of functions to which $g$ belongs would be the first step
towards a rigorous proof of the existence of this construction.} at
large distance $x$.

In the case of a ferromagnet the two regions are defined by 
considering the sign of the total magnetization.  One includes 
configurations with a positive total magnetization, the second selects 
negative total magnetization.  There are ambiguities for those 
configurations which have exactly zero total magnetization, but the 
probability that such a configuration can occur is exponentially small 
at low temperature.  

Physical intuition tells us that this decomposition exists (at least 
for familiar systems), otherwise it would make no sense to speak about 
the spontaneous magnetization of a ferromagnetic sample, or to declare 
that a finite amount of water (at the melting point) is in the solid 
or liquid state.  Moreover all numerical simulations gather data that 
are based on these kinds of notions, since the systems that we can 
store in a computer are always finite.  The concept of finite volume 
states is preeminent from the physical point of view: infinite volume 
states are mainly an attempt to capture their properties in an 
amenable mathematical setting.  We strongly believe that this 
decomposition of a finite but large system into phases does make 
sense, although its translation in a rigorous mathematical setting has 
not been done (also because it is much simpler, and in most cases 
informative enough, to work directly in the infinite volume setting).

In order to present an interpretation of these results we assume that 
such decomposition exists also for spin glasses (the contrary would be 
very surprising for any system with a short range Hamiltonian).  
Therefore the {\em finite} volume Boltzmann-Gibbs measure can be 
decomposed in a sum of such finite volume pure states.  The states of 
the system are labeled by $\al$: we can write

\begin{equation}
  \lan\  \cdot\  \ran =\sum_{\alpha} w_{\al}\lan\  \cdot\  
  \ran_{\al} \ ,
  \protect\label{E-WSUM}
\end{equation}
with the normalization condition

\begin{equation}
  \sum_{\alpha}w_{\alpha}=1\ .
  \protect\label{E-WNOR}
\end{equation}
The function $P_{J}(q)$ for a particular sample is given by

\begin{equation}
  P_{J}(q)=\sum_{\al,\beta} w_\al w_\beta \delta(q_{\al,\beta}-q)\ ,
\end{equation}
where $q_{\al,\beta}$ is the overlap among two generic configurations 
in the states $\alpha$ and $\beta$.

We want to stress that in our construction (mainly based, at this
point, on the analysis of results from numerical simulations) the
definition of finite volume pure states that we have introduced here
is only used in order to present an interpretation of the results.
The predictions of the mean field theory concern correlation functions
computed in the appropriate ensemble \cite{BREAKING}, and computer
simulations measure directly these correlation functions.  The
decomposition into finite volume states (which is never used by the
actual computer simulations) is an interpretative tool which
translates to a simple and intuitive framework the complex
phenomenology displayed by the correlation functions.  We stress again
that the analysis of the overlap probability distribution $P(q)$ can
be based on the definition given in equations
(\ref{E-PQ-A},\ref{E-PQ-B}) that is more direct but has, in our
opinion, much less appeal and intuitive meaning than the previous one.

\subsection{More on the Finite Volume State Classification}

We will try now to give a few more details about the definition of 
equilibrium pure states for a finite volume system that we have just 
discussed.  As we have already pointed out our statements will be 
relevant for configurations which appear with a non negligible 
probability in the low temperature region.  In the following we will 
consider again a system with $N$ spins $\sigma_{i}$, which can take 
the values $\pm 1$, and we will not need to specify the details of the 
Hamiltonian of the system.  We assume however that there are no 
continuous symmetries that are spontaneously broken, otherwise the 
states would be labeled by a continuous parameter: the discussion of 
this case would need more details, without bringing anything 
substantially new.

Given two spin configurations ($\si$ and $\tau$) from the overlap 
definition (\ref{E-OVERLAP}) one can introduce a natural concept of 
distance by

\begin{equation}
  d^{2}(\si,\tau)\equiv \frac{1}{N} \sum_{i=1}^N(\si_i-\tau_i)^2 \ ,
\end{equation}
that belongs to the interval [0 - 1], and is zero only if the two 
configurations are equal.  In analogy with our discussion of section 
(\ref{S-ENEOVE}) and with the definition (\ref{E-OVEGEN}), we can 
define more general distances based on the local operator 
$A_{i}(\si)$, which is a function of the spins at a finite distance 
from the site $i$.  In this way we have

\begin{equation}
  d_{A}^{2}(\si,\tau) \equiv 
  \frac{1}{N} \sum_{i=1}^{N}\left(A_{i}(\si)-A_{i}(\tau)\right)^2 \ .
\end{equation}
In the thermodynamical limit, i.e.  for $N\to\infty$, the distance of 
two configurations is zero if the number of different spins remains 
finite.  The percentage of different $\si$'s, not the absolute number, 
is relevant in this definition of the distance.  It is also important 
to notice that at a given temperature $\beta^{-1}$, when $N$ goes to 
infinity the number of relevant configurations is extremely large: it 
is proportional to $\exp (N {\cal S}(\beta))$, where ${\cal S}(\beta)$ 
is the entropy density of the system).

Finite volume pure states will enjoy a set of properties that we will
discuss now: these conditions will be satisfied by the finite volume
states we have defined above, as well as by any other reasonable
definitions of finite volume states, and are by themselves a strong
characterization of the state.

\begin{itemize}

	\item When $N\to\infty$ each state belonging to the
        decomposition (taken for a given $N$ value) includes an
        exponentially large number of configurations\footnote{We warn
        the reader that we will {\em never} consider $N\to\infty$
        limit of a given finite volume pure state. In order to address
        such a question one should compare the properties of systems
        with different values of $N$.  There are some difficulties in
        doing this comparison, due the chaotic dependence of the
        statistical expectation values on the size of the system (see
        section (\ref{S-CHAOS})).}.

        \item The distance of two different generic configurations
        $C_{\alpha}$ and $C_{\beta}$ belonging one to state $\alpha$
        and the other to state $\beta$ does not depend on the
        $C_{\alpha}$ and $C_{\beta}$, but only on $\alpha$ and
        $\beta$. In this way we can define a distance
        $d_{\alpha,\beta}$ among the states $\alpha$ and $\beta$, as
        the distance among two generic configurations in these two
        states. The reader should notice that with this definition the
        distance of a state with itself is not zero. We could define a
        distance

        \begin{equation}
          D_{\alpha,\beta} \equiv
          d_{\alpha,\beta} -
          \frac12\left(d_{\alpha,\alpha}+d_{\beta,\beta}\right)\ ,
        \end{equation}
        in such a way that the distance of a state with itself is zero
        ($D^{A}_{\alpha,\alpha}=0$). In the rest of this paper
        we prefer to stick to the previous definition.

        \item The distance between two configurations belonging to the
        same state $\alpha$ is strictly smaller than the distance
        between one configuration belonging to state $\alpha$ and a
        second configuration belonging to a different state
        $\beta$. This last property can be written as

        \begin{equation} 
          d_{\alpha,\alpha} < d_{\alpha,\beta}\ ,
        \end{equation} 

        where $d_{\alpha,\beta}$ is the distance between the states
        $\alpha$ and $\beta$, i.e.  the distance between two generic
        configurations belonging to the states $\alpha$ and $\beta$.
        This property forbids to have different states such that
        $D_{\alpha,\beta}=0$, and it is crucial in avoiding the
        possibility of introducing arbitrary states, doing a too fine
        classification.  For example if in a ferromagnet at high
        temperature we would classify the configurations into two
        states which we denote by $e$ and $o$, depending on if the
        total number of positive spins is even or odd, we would have
        that $d_{e,e}=d_{e,o}=d_{o,o}$.
 
        \item The classification into states is the finest one which 
        satisfies the three former properties. 

\end{itemize}

The first three conditions forbid a classification too fine, while the
last condition forbids a classification too coarse.

For a given system the classification into states depends on the
temperature of the system.  In some case it can be rigorously proven
that the classification into states is possible and unique
\cite{KASTLE,KASROB,RUELLE} (in these cases all the procedures we will
discuss lead to the same result).  We want to note that the definition
of state is reminiscent of the definition of species which is familiar
to biologists.

States can also be discussed from a slightly different point of 
view, that we  analyze now. One starts by considering a
generic quantity $B$, and by studying its fluctuations

\begin{equation}
  \lan B^2 \ran_c \equiv 
  \lan B ^2 \ran - \lan B \ran^2 = 
  \lan (B - \langle B \rangle)^2 \ran\ .
\end{equation}
Intensive quantities are defined as

\begin{equation}
  b =\frac{1}{N} \sum_{i=1}^N B_i (\si_i) ,
\end{equation}
where the site functions $B_i$ depend only on the value of $\si_i$.  
We then ask ourselves if intensive quantities fluctuate in the infinite volume 
limit.  In general we would expect a negative answer, since intensive 
quantities are averages over the whole system.  This is not the 
case at a first order transition point, where different phases 
coexist.

As we have discussed before in the usual case of a ferromagnet (in the 
broken phase) spin configurations are classified according to the sign 
of the majority of the spins: on averages restricted to one type of 
these configurations the density of spin with a given sign does not 
fluctuate.

The argument is simple.  If the Hamiltonian $\cal H$ is symmetric 
under the global transformation where $\sigma_{i} \to -\sigma_{i}$ in 
all sites $i$, than $\lan \si_i \ran=0$.  In this situation intensive 
quantities do fluctuate.  If we call $\Sigma$ the intensive quantity 
corresponding to the spins (i.e.  $\Sigma \equiv N^{-1} \sum_{i=1}^N 
\si_i$), the expectation value of $\Sigma$ is zero ($\lan \Sigma 
\ran=0)$, while the expectation value of its square is non zero ($\lan 
\Sigma^2 \ran=m^2 \ne 0)$.

By classifying the configurations in two sets, we can define
restricted averages $\lan(\cdot\cdot\cdot) \ran_+$ and
$\lan(\cdot\cdot\cdot) \ran_-$, such that

\begin{equation}
  \lan  \Sigma \ran       =
  \frac{1}{2} \left(\lan  \Sigma \ran_+  +  
  \lan \Sigma \ran_-\right)\ ,
  \ \lan \Sigma \ran_+ = \ + m \ ,
  \ \lan \Sigma \ran_- = \ - m \ .  
\end{equation}
In normal ferromagnetic systems it is possible to prove 
\cite{KASTLE,KASROB,RUELLE} that intensive quantities do no fluctuate 
in $\lan \ \ran_+$ and in $\lan \ \ran_-$.  The decomposition of a 
probability distribution under which intensive quantities fluctuate 
into the linear combination of restricted probability distributions 
where the intensive quantities do not fluctuate works in many physical 
situations.  The restricted probability distributions correspond to 
different states, that are identified by the expectation value of 
intensive quantities.  Equations (\ref{E-WSUM}) and (\ref{E-WNOR}) 
give the decomposition of the expectation value.

If we consider the classification into states for a finite system we
must face the fact the probability distribution is not the
Boltzmann-Gibbs one, since some configurations are excluded. This
amounts to say that the DLR relations which tell us that the
distribution probability is locally a Boltzmann-Gibbs distribution
will be violated, but the violation will go to zero in the large
volume limit.

It is interesting to note that in usual situations in Statistical 
Mechanics the classification in phases is not very rich.  For usual 
materials, in the generic case, there is only one phase: such a 
classification is not very interesting.  In slightly more interesting 
cases (usual symmetry breaking) there may be two states.  For example, 
if we consider the configurations of a large number of water molecules 
at zero degrees, we can classify them as water or ice: here there are 
two states.  In slightly more complex cases, if we tune carefully a 
few external parameters like the pressure or the magnetic field, we 
may have coexistence of three or more phases (a tricritical or 
multicritical point).

In all these cases the classification is simple and the number of
states is small.  On the contrary in the mean field theory of spin
glasses the number of states is very large (it goes to infinity with
$N$), and a very interesting nested classification of states is
possible.  We note ``en passant'' that this behavior implies that the
Gibbs rule is not valid for spin glasses.  The Gibbs rule states that
in order to have coexistence of $n$ phases ($n$-critical point), we
must tune $n$ parameters.  Here no parameters are tuned and the number
of coexisting phases is infinite! This point will be further
elaborated in section (\ref{S-STOSTA}).

\subsection{The RSB Predictions and Finite Volume States}

Now we are ready to rephrase the predictions of the RSB Ansatz under 
the hypothesis that the finite volume Boltzmann-Gibbs state may be 
decomposed into finite volume pure states.  For finite $N$ 
the decomposition (\ref{E-WSUM}) holds, together with the 
normalization condition (\ref{E-WNOR}).

Let us stress once again how crucial it is that this decomposition is 
done for finite large volume.  We are {\em not} assuming that 
quantities like $\lan \cdot \ran_{\al}$ have a limit when the volume 
goes to infinity: that would not be wise because of the chaotic 
dependence of the states on the volume (as we discuss better in 
(\ref{S-CHAOS})).  The set of the weights $w_{\al}$ also changes with 
the volume, and we only assume that the probability distribution of 
this decomposition has a limit when the volume goes to infinity.  The 
potential and not appropriate use of equation (\ref{E-WSUM}) to 
describe an infinite system, which is ironically called the {\em 
standard mean field or SK picture} in \cite{NS1,NS3,NS4}, is 
completely foreign to the RSB Ansatz.

We can rephrase the predictions of the RSB Ansatz as follows:

\begin{itemize}

\item State equivalence.  States cannot be distinguished 
by looking to the expectation value of intensive quantities
like the energy, the magnetization, the Edward-Anderson parameter, i.e

\begin{equation}
  q_{EA}\equiv\frac1{N}\sum_{i}(\lan \si_{i}\ran_{\al})^{2}\ ,
  \protect\label{E-QEA}
\end{equation}
correlation functions or any other quantity which depends in a simple 
way on the couplings $J$ (for example quantities like 
$\sum_{i,k,l,m}\sigma_{i}J_{i,k}J_{k,l}J_{l,m}\si_{m}$).

\item If we compare two or more states all expectation values depend
only on the set of overlap among these states.  For example given two
states we can define the three different correlation functions

\begin{eqnarray} \nonumber 
C_2(k)&=&\frac{1}{N}\sum_{i}\lan \si_{i} \si_{i+k} 
\ran_{\al}\lan \si_{i} \si_{i+k} \ran_{\beta} \ , \\
C_3(k)&=&\frac{1}{N}\sum_{i}\lan \si_{i}\ran_{\al} \lan \si_{i+k} 
\ran_{\al}\lan \si_{i} \si_{i+k} \ran_{\beta} \ , \\
C_4(k)&=&\frac{1}{N}\sum_{i}\lan \si_{i} \ran_{\al} \lan \si_{i+k} 
\ran_{\al}\lan \si_{i} \ran_{\beta} \lan \si_{i+k}
\ran_{\beta}  \nonumber \ .
\end{eqnarray}
All these correlation functions depend on the overlap $q_{\al,\beta}$
and do not fluctuate in the infinite volume limit at fixed
$q_{\al,\beta}$. More precisely for $N$ fixed we can consider the
values of the correlations in the right hand side of the previous
equations when we change $\alpha$ and $\beta$ at fixed overlap
$q_{\alpha,\beta}$: the variance of these values goes to zero when
the volume goes to infinity.

\item As we have remarked for each realization of the couplings $J$ 
one finds a different set of weights $w$ and allowed overlaps $q$.  
The probability distribution of the couplings $J$ induces a 
probability distribution over the set of the $w$ and the $q$.  This 
probability distribution ${\cal P}_{N}[w,q]$ has a limit when $N$ goes 
to infinity.  This statement does not imply that the dependence on $N$ 
of the $w$ and $q$ is smooth.  Only their probability distribution (or 
equivalently the average over the $J$) has a smooth dependence (see 
the discussion of the quenched state in \cite{AIZCON}).  The detailed 
form of ${\cal P}[w,q]$ depends on the exact pattern of the replica 
symmetry breaking.

\item If we order the weights $w$ in a decreasing sequence (from the 
largest weight to the smallest one) we find that \cite{MPSTV},

\begin{equation}
  \sum_{\al=1}^{K} \lan w_{\al}\ran_{\cal P} \to 1 ,
\end{equation}
when $K\to \infty$ ($K$ still being much smaller than $N$).  Here the
expectation value $ \lan \ \cdot\ \ran_{\cal P}$ is taken over ${\cal
P}$.  This last property implies that practically all the weight is
carried by a finite number of finite volume pure states.

\end{itemize}

There are some extra predictions which are characteristic of the RSB
Ansatz of \cite{BREAKING} which describes the mean field
approximation.

\begin{itemize}

\item
If we order the weights as before

\begin{equation}
  \sum_{\al=1}^{K} \lan w_{\al}\ran_{\cal P} = 1- O(K^{-\chi}) \ ,
\end{equation}
where the expectation value $ \lan \ \cdot\ \ran_{\cal P}$ is taken
over ${\cal P}$, and the exponent $\chi$ is given by 

\begin{equation}
  \chi=\frac{1}{x(q_{EA})}\ .  
\end{equation}
We expect a power law decay to hold in general. The exponent may
depend on the form of the RSB Ansatz.  Numerical simulations
\cite{MPRUNP} confirm this behavior.

\item
The different definitions of the distance, depending on the choice of
the local operator $A$, are equivalent. Neglecting terms which go to
zero when $N$ goes to infinity one must have

\begin{equation}
  d^{A}_{\alpha,\beta}=f^{A}\left(q_{\alpha,\beta}\right)\ .  
\end{equation}
In other words any type of distance can be computed in terms of the
overlap.  For given $N$ we can consider the values of the correlations
in the r.h.s. of the previous equations when we change $\alpha$ and
$\beta$ at fixed overlap $q_{\alpha,\beta}$: the variance of these
numbers goes to zero when the volume goes to infinity.

\item   
The overlaps satisfy the {\em ultrametricity condition}

\begin{equation}
  q_{\alpha,\beta} 
  \ge 
  \mbox{min}(q_{\alpha,\gamma},q_{\beta,\gamma}) 
  \,\,\,\,\,\,\,\, 
  \forall \alpha, \beta \mbox{ and } \gamma \ .
\end{equation}
The presence of an ultrametric structure strongly simplifies the 
theoretical analysis: for each realization of the couplings $J$ we can 
construct the tree of the states, where the states are the leaves 
of the tree and the probability distribution ${\cal P}[w,q]$ reduces 
to a probability distribution over the hierarchical tree (which has 
been studied in detail by Ruelle \cite{RUETRE}).

Of course we can restate the ultrametricity property without using the
concept of finite volume states.  It amounts to say that three generic
equilibrium configurations $\si^{1}$, $\si^{3}$ and $\si^{3}$ satisfy
the equivalent inequality

\begin{equation}
  d({\si^1,\si^2}) \le \mbox{max}(d({\si^1,\si^3}),d(\si^2,\si^3)) \ ,
  \protect\label{ULTRA2}
\end{equation}
with probability one when the volume goes to infinity.  In other words 
it says that the joint probability distribution of the overlaps among 
three different equilibrium configurations for finite $N$, averaged 
over the couplings $J$, goes to a limit which is zero in the region of 
overlaps where the previous inequality (\ref{ULTRA2}) is not 
satisfied.

Ultrametricity is quite likely correct in the SK model
and what happens in short range models will be discussed later.

\end{itemize}

This picture based on finite volume states has a clear and direct
meaning.  It has the advantage to be potentially applicable also in
different contests, as for example when discussing protein foldings.
Here the concept of finite volume state is clear: a given tertiary
structure is the equivalent of a state in our spin model.  In this
case theoretical instruments to discuss the behavior of finite systems
are badly needed, also since the limit of a protein of infinite length
is not very interesting from the biological point of view.

\section{The Cavity Approach and Chaos\protect\label{S-CHAOS}}

The original derivation of the properties of the infinite range model 
of spin glasses \cite{BREAKING} was based on the replica method, which 
involves an analytic continuation from integer to non integer values 
of $n$ (the number of replicas).  An alternative approach is the so 
called cavity approach \cite{MEPAVI}.  Here one starts by assuming 
that in a finite volume the decomposition in pure states is possible 
and has the properties that we have described in the previous section.  
Then one compares a system containing $N$ spins to a system with $N+1$ 
spins: in the limit of $N$ large the probability ${\cal 
P}_{N+1}[w,q]$ can be written in explicit form as function of ${\cal 
P}_{N}[w,q]$.  Symbolically we get

\begin{equation} 
  {\cal P}_{N+1}={\cal R}[{\cal P_{N}}] \ .
\end{equation}
The probability ${\cal P}_{\infty}$, which is at the heart of the
replica approach, can be obtained by solving the fixed point equation

\begin{equation}
  {\cal P}_{\infty}={\cal R}[{\cal P_{\infty}}] \ .
\end{equation}
The probability distribution (embedded with an ultrametric structure) 
which was found by using replica theory turns out to be a solution of 
this fixed point equation.  Alas, it is not known if it is the only 
solution.

Chaoticity is a very important property of spin glasses.  
It amounts to say that if we consider  a finite system and we add
to the total Hamiltonian a perturbation $\delta {\cal H}$ such that

\begin{equation}
  1  \ll \delta {\cal H}\ll N \ ,
  \label{E-CHAOS}
\end{equation}
the unperturbed and the perturbed system are as different as possible.
As usual chaoticity may be formulated in terms of equilibrium
expectation values of observables. In the typical chaotic case

\begin{equation}
  q(\delta H) \equiv \frac1{N}
  \sum_{i=1,N}\lan\si_{i}\ran_{H}
              \lan\si_{i}\ran_{H+\delta H}
\end{equation}
is equal to the minimum allowed overlap $q_{m}$ as soon as $\delta H$
satisfies the previous bound (\ref{E-CHAOS}).  Chaoticity can also be
formulated by saying that the states of the perturbed systems have
minimal overlap (i.e. $q_{m}$) with the states of the system in
absence of the perturbation.

Here is a list of some examples:

\begin{itemize}
\item Chaoticity with respect to a random energy perturbation.

This is a trivial effect.  The weights $w$ are of order $1$.  Because 
of this fact a random perturbation in the energy function, which acts 
differently on different states, completely changes the weights $w$ as 
soon as it is much larger than $1$ (or much larger than $\frac{1}{N}$ 
if we look at the energy density).

\item Chaoticity in magnetic field.

This has been the first example of chaos in spin glasses that has been 
discovered \cite{PARDUE}.  If we have two systems with the same 
couplings $J$ but different magnetic fields $h_{1}$ and $h_{2}$, their 
equilibrium overlap becomes the minimum one allowed (that we call 
$q_{m}$) as soon as $(h_{1}-h_{2})^{2}N \gg 1$.  The difference in 
magnetic field acts more or less as a random energy perturbation. 

\item Chaoticity in temperature.

This effect \cite{BMTCH,KONTCH,RITTCH,FNNTCH}
is maybe present in some models and absent in others (surely $p$
spin infinite range spherical models \cite{KUPAVI}). It is related to
the fact that states with the same total free energy may have
different energy and entropy, and therefore a different dependence of
the free energy on the temperature.

\item Chaoticity when changing the number of spin. 

Chaoticity when changing the number of spin \cite{MEPAVI,NS0,NSnew,Kulske} is at the heart of the cavity approach and it is responsible of the differences among the predictions for the free energy of the replica exact and the replica broken theory \cite{MEPAVI}. When we go from $N$ to $N+1$ spin the weights $w$ change of a factor of order 1. Because of that we must write evolution equations for the whole population of states and not for a single state. This crucial and well known effect has been also discussed for the short range model by refs.~\cite{NS0} and \cite{NSnew}: in this context it takes the name of ``chaotic volume dependence'' (see also ref.~\cite{Kulske} for related work).

This effect is at the heart of the cavity approach and it is 
responsible of the differences among the predictions for the free 
energy of the replica exact and the replica broken theory 
\cite{MEPAVI}.  When we go from $N$ to $N+1$ spin the weights $w$ 
change of a factor of order $1$.  Because of that we must write 
evolution equations for the whole population of states and not for a 
single state.  This crucial and well known effect has been 
rediscovered for the short range model by \cite{NS0}: in this context 
it takes the name of ``chaotic volume dependence''. 

This form of chaoticity tell us that we cannot speak of the properties
of a given state in the infinite volume limit, but the decomposition
into states must be done for each value in $N$. Moreover one must pay
a particular attention in taking the infinite volume limit. Local
quantities like the expectation value of a the spin at a given point
do not have a limit, at least in a naive sense, when $N$ goes to
infinity.

A simple case of chaotic dependence on the volume is the one of an
Ising ferromagnet in presence of a random symmetric distribution of
magnetic field at low temperature.  The total magnetization is well
approximated by $\sign\sum_{i=1,N}h_{i}$ (where the $h_{i}$ are the
random fields) and this quantity changes in a random way when $N$ goes
to infinity.

\end{itemize}

\section{The Order Parameter in the RSB Ansatz}

At this point of our discussion the reader could be worried about the
thermodynamical stability of RSB construction.  We are dealing here
with a new and peculiar phenomenon: even in the infinite volume limit
the leading free energy differences are of order $1$, as opposed to a
difference of order $N$ in the usual thermodynamical systems.  It is
important to reach a better understanding of such an unusual behavior,
in order to eliminate possible doubts about the relevance of
RSB. Indeed one may be worried by the fact that by choosing a suitable
perturbation one can select only one of the ground states and arrive
to a situation in which the function $P(q)$ is trivial. This is
correct, but in order to reach this goal the perturbation must be very
carefully tuned. If the perturbation is random the effect will be
different, as discussed in the section \ref{S-STOSTA}.  This section
will be devoted to discussing various methods which can be used to
derive an expression for the function $q(x)$ (or equivalently for
$P(q)$) only considering the expectation values of intensive
quantities, which can be computed in the infinite volume limit.

We will start by introducing a construction that puts the overlap
order parameter in a more familiar perspective (\ref{COUREP}) (like
the magnetic field for ordinary statistical systems).  We will give a
specific example of physical relevance of such order parameter
(\ref{SS-PHYSREV}), and show a computation that is very clarifying
from a theoretical point of view (\ref{SS-RANFIE}) (since it allows to
base the RSB results on the analysis of intensive quantities in the
infinite volume limit).  At last we will show that, under some
hypothesis, the probability distribution of the order parameter can be
reconstructed by means of a dynamical approach (\ref{SS-DYNAPP}).

\subsection{Two Coupled Replicas\protect\label{COUREP}}

Following reference \cite{PARVIR} we will show here that we can 
discuss the two replica overlap as the order parameter of the theory 
in a very natural way.  We consider two copies of the system with the 
same realization of the couplings $J$, and we add to the Hamiltonian 
an external field which couples the two real replicas $\sigma$ and 
$\tau$:

\begin{equation}
  {\cal H}\equiv-\sum_{i,j} J_{ij} \sigma_i \sigma_j 
  -\sum_{i,j} J_{ij} \tau_i \tau_j 
  - h \sum_i  \left(\sigma_i + \tau_i \right)
  -\epsilon \sum_{i} \tau_i \sigma_{i} \ .
  \label{E-HAMRERE}
\end{equation}
Considering the two cases of positive and negative $\eps$ leads to a
discontinuity in the expectation value of the overlap in the limit
where $\epsilon\to 0$ (in complete analogy with the usual symmetry
breaking one has in a ferromagnet when $h\to 0^{+}$ and $h\to 0^{-}$).
In this way we give a definition of the two bounds of the overlap
support, $q_{m}$ and $q_{M}$, that we had already introduced earlier,
by using a purely thermodynamical approach.

In the RSB Ansatz of \cite{BREAKING} one finds that \cite{PARVIR} in 
the mean field theory the expectation value of the overlap $q$ between 
the replicas $\tau$ and $\sigma$, that we denote as $q(\epsilon)$, is 
a discontinuous function of $\epsilon$ as $\epsilon\to 0$.  One has 
that

\begin{equation}
  \lim_{\epsilon \to 0^+} q(\epsilon) = q_{M} 
\end{equation}
(where $q_{M}$ coincides with $q_{EA}$), but

\begin{equation}
  \lim_{\epsilon \to 0^-} q(\epsilon) = q_{m} \ .
\end{equation}
This discontinuity at $\epsilon=0$ is a striking consequence of
replica symmetry breaking: the mean field predictions prediction for
$\epsilon>0$ is $q(\epsilon)= q_{\rm EA} + A \epsilon^{b}$, where
$A$ is a constant and $b=\frac{1}{2}$. It is possible that even
under the same Replica Symmetry Breaking Ansatz of the mean field
theory, the value of $b$ is changed in finite dimensions (a very
naive guess could be $b=\frac{1}{2}$ for $D\ge 6$, $b={D-2 \over
D+2}$ for $D<6$).
  
In a theory without replica symmetry breaking one finds a very 
different result:

\begin{equation}
  \lim_{\epsilon \to 0^+} q(\epsilon)=\lim_{\epsilon \to 0^-}
  q(\epsilon)= q_{\rm EA}\ .
\end{equation}
Continuing our discussion of section (\ref{S-ENEOVE}) about more 
general definitions of overlaps and distances in configuration space 
(we have discussed for example the energy overlap and the link 
overlap) we can base our modified action on the energy overlap, i.e.

\begin{equation}
  {\cal H} \equiv -\sum_{i,j} J_{ij} \sigma_i \sigma_j 
  -\sum_{i,j} J_{ij} \tau_i \tau_j 
  -\epsilon \sum_{i,j} \tau_i \tau_j\sigma_{i}\sigma_j  \ .
  \label{E-HAMENRERE}
\end{equation}
Again the expectation value of the energy overlap $q_E(\epsilon)$ in a 
theory with replica symmetry breaking is a discontinuous function at 
$\epsilon=0$, while $q_{E}(\epsilon)$ is continuous at $\epsilon=0$ in 
systems without replica symmetry breaking.

\subsection{Physical Relevance of the Order Parameter\protect\label{SS-PHYSREV}}

In the previous section we have made clear the physical relevance of 
$q_{M}$, i.e.  of the $q_{EA}$ defined in (\ref{E-QEA}) (the overlap 
of two replica's in the same state). We have to be more careful when 
considering quantities like, for example

\begin{equation}
   q_{J} \equiv \int dq\ P_{J}(q)\  q
   \protect\label{E-QJ}
\end{equation}
(where we have chosen the simplest interesting quantity, and, as 
usual, we assume we are in non-zero magnetic field), and taking their 
disorder average by

\begin{equation}
  q_{a}\equiv \ba{q_{J}}\ .
\end{equation}
$q_{J}$ is not a self-averaging quantity: even for arbitrary large 
volume it takes different values for different realizations of the 
couplings.  Its value is determined by differences in the free 
energies which are of order $1$.

We will show here that in a specific (but very interesting) case 
$q_{a}$ \cite{MAPAZU,PARIRU-4DH} has a direct physical meaning.
We consider  \cite{PARIRU-4DH} the model defined by the Hamiltonian

\begin{equation}
  {\cal H}\equiv-\sum_{ij}\sigma_i J_{ij}\sigma_j -\sum_i\sigma_i 
r_i\ ,
  \protect\label{E-HAMRF}
\end{equation}
where the $r_{i}$ are quenched random Gaussian magnetic fields with 
zero mean and variance $r_{0}^{2}$, and the couplings $J_{ij}$ are the 
usual spin glass quenched random couplings.  We define the infinite 
volume staggered magnetization

\begin{equation}
  m_s\equiv\lim_{N\to \infty}\frac{1}{N} \sum_{i} 
  \overbrace{r_{i}\lan \si_{i}\ran} \ ,
\end{equation}
where $\overbrace{(\cdot\cdot\cdot)}$ denotes the double average over 
the couplings $J_{ij}$ and over the random magnetic field ($r_i$).
A simple integration by parts tells us that 

\begin{equation} 
  m_s = r_{0}^{2} \beta (1-q_{a})\ , 
\end{equation} 
i.e.  establishes a relation among the staggered magnetization and the
average quenched overlap $q_{a}$.  Such a simple identity (in the
style of the relations of \cite{GUERRA,AIZCON,PARSIX}) reveals that
the order parameter can be directly related to a well defined
thermodynamical parameter (note that $m_{s}$ is a self-averaging
quantity).

\subsection{Random Field and Coupled Replicas\protect\label{SS-RANFIE}}

Now, following \cite{PARVIR}, we are able to establish a relation
among the results of the two former sections, and to establish some
further interesting analytic evidence. We consider a spin glass in a
random magnetic field, with the Hamiltonian (\ref{E-HAMRF}).  We call
${\cal Z}^{N}[r]$ the partition function of a system with $N$ spins,
and with a random field $r$ distributed according to $d\mu(r)$.  We
define a generalized free energy density as

\begin{equation}
  F(s,r_{0}^{2}) \equiv
  \lim_{N \to \infty}-\frac{1}{\beta s N}\log \left[\int
  d\mu(r)\left(\frac{{\cal Z}^{N}[r]}{{\cal 
Z}^{N}[0]}\right)^{s}\right] \ .
\end{equation}
$F(s,r_{0}^{2})$ is a well defined
thermodynamic quantity.  For integer values of $s$ the
integrals can be solved and we find a system of $s$ coupled
replicas:

\begin{equation}
  {\cal H}=\sum_{a=1}^s {\cal H}_{0}[\si^{a}]- \frac{r_{0}^{2}}{2} \sum_{i=1}^N 
  \sum_{a,b=1}^s \si^{a}_{i}\si^{b}_{i}\ ,
  \protect\label{E-SCOUPLED}
\end{equation}
where ${\cal H}_{0}$ is the zero field Hamiltonian, and $r_{0}^{2}$ is 
the $\epsilon$ of the Hamiltonian (\ref{E-HAMRERE}).  We are 
interested in the behavior of the free energy $F$ for small positive 
values of $r_{0}^{2}$.  We define

\begin{equation}
  \aleph(s)= \left. 
  {\partial F(s,r_{0}^{2}) \over \partial r_{0}^{2}}
  \right|_{r_{0}^{2}=0^{+}} \ .
\end{equation}
Analogously to what happens in section (\ref{COUREP}) we get that 
\cite{PARVIR}

\begin{equation}
  \aleph(s)=1 +(s-1)q_{\rm EA} \ .
\end{equation}
Let us see why. The term $\aleph(s)$ is proportional to the 
expectation value of the interaction term in equation 
(\ref{E-SCOUPLED}). In this term there are $s$ diagonal entries that 
contribute with one ($a=b$), and $s(s-1)$ non-diagonal entries that 
contribute with $q_{EA}$.

Next we discuss what happens for non-integer values of $s$, $0<s<1$.  
In this case using the mean field RSB Ansatz one can explicitly compute that

\begin{equation} 
  \aleph(s) = 1 - \int _{s}^{1} du \ q(u) \ .  
\end{equation}
It would be interesting to check if this formula is valid in general or
if it is only valid in the mean field ultrametric approach.

The probabilistic meaning of this construction has been discussed in
detail in \cite{PARVIR}.  For example in the simple case where the
function $q(x)$ has a discontinuity at $x=m$ one finds that for $s<m$
the integral over $r$ is dominated by the generic configurations of
the random magnetic field $r$, while for $s>m$ the integral is
dominated by those rare configurations of the magnetic field which
point in the same direction of one of the possible magnetizations of
the system

The non-linearity of the function $\aleph(s)$ for
sufficiently small values of $s$ is a signal of replica symmetry
breaking. One can show in general that

\begin{equation}
  \aleph(0)=\int_{0}^{1} dq\ P(q)\  (1-q)\ ,
\end{equation}
and ${d \aleph(s)\over ds}|_{s=1}=q_{EA}$.

This result is interesting from a theoretical point of view, 
since it leads to the construction of a non trivial order parameter 
which signals the presence of replica symmetry breaking in a purely 
thermodynamic way, i.e.  by only computing intensive quantities.  
Unfortunately the quantity $\aleph(s)$ is extremely difficult to 
compute numerically but for very small systems.

\subsection{The Dynamical Approach\protect\label{SS-DYNAPP}}

In this section we will give a definition of the function $P(q)$ based
on the behavior of the systems when it is slightly off-equilibrium.
This will imply that it is possible to define $P(q)$ as a function of
``well-defined'' physical observable quantities, like the
magnetization of the system.

In principle we can considered two different off-equilibrium situations

\begin{itemize}
    
\item The system is driven off-equilibrium by an external force, or by
a time dependent Hamiltonian. For example we can assume that the
couplings $J$ are not time independent, but that they change at random
on a time scale $\tau$, which is assumed to be very large but
finite. In this case we reach a stationary, off-equilibrium situation,
in which the correlation functions depend only on the time difference
of the observables.
    
\item The system is drifting toward equilibrium because it was not at
equilibrium at time $t_0=0$.  In this case, also if equilibrium is
eventually reached, the system is slightly out equilibrium at finite
time. The correlation functions will be no more translationally
invariant.

\end{itemize}

In this note we will consider only the second situation, but similar
considerations apply to the first case.

We consider the quantity $A(t)$ that depends on the local variables of
the unperturbed Hamiltonian $\cal H$.  We define the associated
autocorrelation function

\begin{equation}
  C(t,t^\prime) \equiv \lan A(t)  A(t^\prime)  \ran \ ,
\label{auto}
\end{equation}
where the brackets $\lan \ldots \ran $ imply a double average, over 
the dynamical process and over the disorder (as it will be in all 
this section). We also define the response function

\begin{equation}
R(t,t^\prime) \equiv \left. \frac{\delta \lan A(t) \ran}{\delta
\epsilon(t^\prime)}\right|_{\epsilon=0} \ ,
\label{res}
\end{equation}
where we have perturbed the original Hamiltonian 
by a small contribution

\begin{equation}
  {\cal H}^\prime= {\cal H} + \epsilon(t) A(t) \ .
\end{equation}
In the following we will specifically select $A(t)=\frac1{N}
\sum_i\sigma_i(t)$.  When looking at the dynamics of the problem and
assuming time translational invariance it is possible to derive the
{\em fluctuation-dissipation} theorem (thereafter FDT, see for example
\cite{STAFIE}), that reads as

\begin{equation}
  R(t,t^\prime)=\beta \theta(t-t^\prime) \frac{\partial
  C(t,t^\prime)}{\partial t^\prime} \ .
  \protect\label{FDT}
\end{equation} 
The fluctuation-dissipation theorem holds in the equilibrium regime, i.e.
when

\begin{equation}
  |t-t^{\prime}|\ll t\ . 
  \label{EQUILIBRIUM}
\end{equation}
We expect a breakdown of its validity in the region where equation
(\ref{EQUILIBRIUM}) does not hold.  Under general assumptions one finds
that \cite{CUKU} the FDT theorem gets modified as

\begin{equation}
  R(t,t^\prime)=\beta X(t,t^\prime) \theta(t-t^\prime) \frac{\partial
  C(t,t^\prime)}{\partial t^\prime} \ .
\end{equation}
It has also been suggested \cite{CUKU,BCKP} that the function
$X(t,t^\prime)$ turns out to be a function of the autocorrelation 
function:

\begin{equation}
  X(t,t^\prime)=X(C(t,t^\prime))\ .
\end{equation}  
Under this hypothesis one can generalize FDT. The off-equilibrium
fluctuation-dissipation relation, which should hold in the aging
regime of the dynamics, reads

\begin{equation}
  R(t,t^\prime)=\beta X(C(t,t^\prime)) \theta(t-t^\prime) 
\frac{\partial
  C(t,t^\prime)}{\partial t^\prime}\ .
  \protect\label{OFDR}
\end{equation}
The previous equation can be  related to observable
quantities like the magnetization. The magnetization in the dynamics
is a function of the time, and a functional of the magnetic field 
(that is itself a function of the time: $h=h(t)$). We denote it by
$m[h](t)$. Using a functional Taylor expansion and the
definition of the response function we can write

\begin{equation}
  m[h](t)=\int_{0}^t dt^\prime\  R(t,t^\prime) 
  h(t^\prime) +{\rm O}(h^2)\ ,
\end{equation}
that is nothing but the linear-response theorem where the terms
proportional to $h^2$ have been neglected.  By applying (\ref{OFDR})
we derive the dependence of the magnetization over time in a generic
time-dependent magnetic field (with a small strength)\footnote{Here
the symbol $\simeq$ means that the equation is valid in the region of
small fields. One must be careful because the strength of the
non-linear effects often increases as a power of $t$
\cite{JOORHA}.}, $h(t)$

\begin{equation}
  m[h](t)\simeq \beta \int_{-\infty}^t dt^\prime\  
X[C(t,t^\prime)]  
  \frac{\partial C(t,t^\prime)}{\partial t^\prime} 
  h(t^\prime) \ .
\end{equation}
Now we can perform the following experiment. We let the system to
evolve in absence of magnetic field from $t=0$ to $t=t_w$, and then we
turn on a constant magnetic field, $h_0$. The time dependent magnetic 
field is\footnote{In our notation we ignore the fact 
that $m[h](t)$ depends on $t_w$.}
$h(t)=h_0\ \theta(t-t_w)$, and

\begin{equation}
m[h](t)\simeq h_0 \beta \int_{t_w}^t dt^\prime \ X[C(t,t^\prime)]  
\frac{\partial C(t,t^\prime)}{\partial t^\prime}
=h_0 \beta \int_{C(t,t_w)}^{1} du \  X[u] \ ,
\protect\label{mag_1}
\end{equation}
where we have performed the change of variables $u=C(t,t^\prime)$ and 
we have used the fact that $C(t,t) \equiv 1$.  In the equilibrium 
regime (where FDT holds, and $X=1$) we must find that

\begin{equation}
  m[h](t)\simeq h_0 \beta \left(1-C(t,t_w)\right)\ ,
  \protect\label{mag_fdt}
\end{equation}
i.e.  $\frac{m[h](t) T}{h_0}$ is a linear function of $C(t,t_w)$ with 
slope $-1$.  A precise relation connecting the function $X$ to the 
equilibrium behavior of the system has been conjectured 
\cite{CUKU,BCKP}.  In the limit where $t$ and $t_w$ go to $\infty$ and 
$C(t,t_w) = q$ one expects that $X(C)$ converges to the $x(q)$ of 
equation (\ref{E-XQ}).  Obviously $x(q)$ is equal to $1$ for all $q > 
q_{\rm EA}$, and in this region we recover FDT. It follows that, if 
the previous conjecture is true, it is possible to define $P(q)$ in 
terms of the magnetization of the system, i.e.

\begin{equation}
  P(q)=-\frac{1}{h_0 \beta} \left. \frac{d^2 m[h](t)}{d 
  C^2}\right|_{C=q}\ .
\end{equation}
In the last section we shall see how this approach gives results for
the function $P(q)$ which are very similar to those one can obtain by
its direct definition.

\section{Stochastic Stability\protect\label{S-STOSTA}}

Stochastic stability is a property which is valid in the mean field
approximation: it is reasonable to conjecture that is valid in general
also for short range models. It has been introduced quite recently
\cite{GUERRA,AIZCON,PARSIX,FMPP1,MARINA} and strong progresses have
been done on the study of its consequences.

In order to decide if a system with Hamiltonian $H$ is stochastically
stable, we have to consider the free energy of an auxiliary system
with the following Hamiltonian:

\begin{equation}
  H+\eps H_{R}\ .
\end{equation}
If the average (with respect to $H_{R}$) free energy is a
differentiable function of $\eps$ (and the limit where the volume goes
to infinity commutes with the derivative with respect to $\eps$), for
a generic choice of the random perturbation $H_{R}$ inside a given
class and $\eps$ close to zero, the system is said to be {\em
stochastically stable}.  In the nutshell stochastic stability tells us
that the Hamiltonian $H$ does not has any special features and that
its properties are analogous to those of similar random systems ($H$
can contain quenched random disorder).

It is simpler to compute the properties of the stochastically
perturbed system than those of the original system, since any possible
accidental symmetry is washed out by the random perturbation. For
example spin glasses at exactly zero field are not stochastically
stable because of the symmetry $\si \to -\si$ which is destroyed by a
random perturbation. Stochastic stability may hold only for spin
glasses in presence of a non-zero magnetic field.
 
The definition of stochastic stability may depend on the class of
random perturbations we consider.  Quite often it is convenient to
chose as a random perturbation an infinite range Hamiltonian, e.g.

\begin{equation}
   H_{R}^{3} =
  \sum_{i,k,l}J_{i,k,l} \si_{i} \si_{k} \si_{l} \ ,
  \label{SSS}
\end{equation}
where the sum runs over all the $N$ sites of the system and the $J$'s
are random uncorrelated variables with variance $\frac{1}{N}$.  In the
same way we can define a random perturbation $H_{R}^{p}$ where the
interaction involves $p$ spins.

A simple integration by part tells us that in finite volume

\begin{equation}
  R^{p}(\eps) = {\lan H_{R} \ran_{\eps} \over \eps} 
  = \int dq\ P_{\eps}(q)\ (1-q^{p})\ ,
\end{equation}
where $ P_{\eps}(q)$ is the average over the random perturbation of
the overlap distribution probability.  Adding terms with different
values of $p$ we can reconstruct the whole function $P(q)$.

The quantity $R^{p}(\eps)$ is a well defined thermodynamic quantity
like the internal energy, whose expectation value may be ambiguous
only at exceptional points of first order phase transitions. It can
also be obtained as the derivative of the $\eps$ dependent free
energy.

Stochastic stability is a very strong property. Many properties can be
derived from stochastic stability, in particular the overlap sum rules
(\ref{GUEREL}) and their generalizations.  Let us discuss a few
examples which should help to understand some implications of
stochastic stability.

\begin{itemize}

    \item Let us consider two different systems with two different
    non-trivial functions $P_{1}(q_{1})$ and $P_{2}(q_{2})$, and let
    us suppose that each of the systems is stochastically stable.  The
    union of the two systems, if they are non interacting, will not be
    stochastically stable anymore, as can be easily seen: the relation
    (\ref{GUEREL}) is no more valid for the union system.  Stochastic
    stability describes a situation in which the whole system remains
    strongly correlated, as opposed to the one in which the function
    $P(q)$ is non trivial because of the formation of interfaces among
    two possible phases.

    \item When the overlap may only take two values, or
    equivalently the function $P(q)$ is simply given by

    \begin{equation} 
       P(q) =A \delta(q-q_{m}) + (1-A) \delta(q-q_{M})\ , 
    \end{equation} 
    stochastic stability implies that the usual Mean Field Ansatz 
    gives the correct result.

    \item The identity among the dynamic $X(q)$ function and the
    static $x(q)$ function can be proved to be a consequence of
    stochastic stability \cite{FMPP1,FMPP2}.

\end{itemize} 

We have seen that stochastic stability strongly constrains the
properties of the system, and that many of the qualitative results of
the replica approach can be derived as mere consequences of stochastic
stability.  Stochastic stability apparently does not imply
ultrametricity, which seems to be an independent property
\cite{PARSIX}.  This independence problem is still open as far as the
only probabilities distribution of the free energies of the states
that have been constructed in an explicit form are ultrametric.

\section{The Infinite Volume Pure States\protect\label{INFINITE}}

The free energy is a very basic quantity to analyze when studying the 
infinite volume limit of a physical system.  In many cases it is easy 
to prove (at least for the lattice model) the existence of the free 
energy in the infinite volume limit.  At first sight that could seem 
to be enough for deriving the thermodynamics and for computing 
intensive quantities, by derivating the free energy with respect to 
the external field.

However when we consider the free energy as a function of some 
parameter we can find out that there are points where it is not 
differentiable, more precisely where the left and the right derivatives do 
differ.  The typical example is the case of the spontaneous 
magnetization where

\begin{equation}
  F(h) = F(h=0) + m|h| + O(h^{2})\ .
\end{equation}
In this case we can obtain different values of the magnetization by
considering the two limits $h\to 0 ^{\pm}$.  The same Hamiltonian
(with zero magnetic field) could give different result for the
magnetization in the limit $N\to\infty$ and $h \to 0$.  In this way we
say that we end up in one of two different allowed states of the
infinite volume system.  A crucial reason for the direct introduction
of pure states for an infinite volume system is to make life simpler,
giving an intrinsic definition of pure states, avoiding the necessity
of considering the response of the system all possible forms of the
external fields.

\subsection{Generalities}

We will try to clarify here the language we need in order to proceed
further, and we will give a few needed mathematical definitions.

The word {\em state} is often used in mathematics, with different
underlying meanings.  A possible definition is the following.  Let
$\cal A$ be the $B^{*}$ algebra of observables, which in classical
statistical mechanics form an Abelian algebra with identity.  We say
that a linear functional $\rho$ is a state if

\begin{eqnarray} \nonumber
  A\ge 0    & \Rightarrow & \rho(A)\ge 0,\\
  ||\rho||  & = &           1\ ,
\end{eqnarray}
where this second condition (the norm of $\rho$ is one) implies 
that $\rho(A=1)=1$.

In a finite volume a state is a normalized probability distribution:
we can associate to it an expectation value.  The notion of a state
can be extended to an infinite system.  If a compact set of states is
convex, we can consider the extremal points of this set and call them
{\em pure states}.  Any state of the set can be written in an unique
way as a linear combination of these pure states.

The introduction of the notion of pure states in statistical physics
has the main goal, as we said before, to allow a clear definition of
symmetry breaking.  A crucial notion is the one of {\em clustering}: a
state is called clustering if connected correlation functions computed
in such state go to zero at large distance.  There exist systems for
which the finite volume equilibrium state goes to an infinite volume
one with the unwanted property that its correlation functions do not
satisfy the clustering properties (consider for example a
ferromagnetic system at zero external magnetic field in the low
temperature region, where there is a spontaneous magnetization: in a
typical situation volume $N$ will have a magnetization plus with
probability $0.5$ or minus with probability $0.5$, volume $N+1$ again
a magnetization plus or minus and so on.  The infinite volume limit of
these pure states will be a statistical mixture of the plus and the
minus states). Physical consistency requires that if a system is in a
given phase intensive quantities do not fluctuate, that is possible
only if correlation functions are clustering.

For a class of models (i.e.  models without quenched disorder with 
translational invariant Hamiltonians) it has been established that, if 
translational invariance is not spontaneously broken, the equilibrium 
state obtained as the infinite volume limit of finite volume states 
can be decomposed in an unique way as the convex combination of 
translational invariant states, in which the clustering property is 
satisfied.  This is a remarkable result.  This procedure has been 
generalized by introducing local equilibrium states (DLR states 
\cite{RUELLE}) for an infinite system with short range Hamiltonian.  A 
DLR state satisfies all the identities on conditional probabilities 
that would be valid for a Boltzmann-Gibbs state and involve only a 
finite number of variable.  The DLR states form a convex set and their 
extremal points are clustering pure states.  To find the structure of 
all extremal clustering infinite volume states for a given system is a 
non trivial task, which has been solved only in a few cases (typically 
for a ferromagnetic model).

This formalism in principle would allow to write formulae of the kind

\begin{equation}
  \lan \cdot \ran_{B-G} =\sum_{\al}w_{\al}\lan \cdot \ran_{\al} \ ,
\end{equation}
where $\lan \cdot \ran_{B-G}$ is the infinite volume limit of the 
Boltzmann-Gibbs probability distribution defined for finite volume, 
and $\alpha$ labels the extremal DLR states.  The sum must be replaced 
by an integral if the DLR states are not a numerable set\footnote{This 
happens when a continuous symmetry group is broken.  The most familiar 
example is the Heisenberg ferromagnet, where the DLR states are 
labeled by an unit vector, giving the direction where the spontaneous 
magnetization points.}.  Of course we are assuming the far from 
evident existence of $\lan \cdot \ran_{B-G}$.

The proofs which are needed are very simple\footnote{The only delicate
point is to prove the clustering property for pure states.} if one
uses the appropriate mathematical setting \cite{KASTLE}.  Hard problems start when one has to show that this construction is not empty, i.e., when one has to prove that local equilibrium states do exist directly for the the actual infinite system without obtaining it as a limit of finite volume measures \footnote{We do not want to arise the point that an actual infinite random system quite likely does not make sense from Brower's constructivistic point of view.}.

The simplest way to define something in the infinite volume limit is to consider a finite volume system and to show that the infinite volume limit of the Boltzmann-Gibbs probability exists. In this construction there is the freedom to chose the boundary conditions of the system, that could lead to different local equilibrium states. If the boundary conditions are chosen in an appropriate way (e.g., all spins up in a ferromagnet) one obtains a pure state. Unfortunately such a procedure can be carried out only in a very few cases: in general when the volume goes to infinity the Boltzmann-Gibbs probability does not have a limit, at the opposite of what we could naively think (we give a
simple example of such a system in section~\ref{SS-NAIVE}).

The proof of the existence of the free energy density in the infinite 
volume limit for short range lattice systems is quite simple in the 
case of non random systems and not too complex for disordered systems.  
The crucial point in the proof is that we can modify the total 
Hamiltonian without changing the free energy density by adding a 
contribution which diverges with the volume, but diverges with a 
slower rate than the volume itself.  It is intuitive that in the 
infinite volume limit the free energy density cannot depend by any 
boundary effect (the surface to volume ratio goes to zero).  On the 
contrary when we try to control the expectation values over the 
probability distribution we find that a change in total Hamiltonian of 
a quantity of order $1$ may completely change the result.

It always true that using compactness argument we can prove that if we consider a sequence of systems inside a box of side $L$ we can find a sequence $L(k)$ (with $\lim_{k\to\infty} L(k)=\infty$), such that the local expectation values go to a finite limit when k goes to infinity \cite{NS1}.
However there can be infinitely many of such sequences, so that in this way we may obtain many different infinite volume expectation values. Given the non constructive nature of compactness arguments it is also difficult to discuss the clustering properties of the resulting states.

The nature and the number of the states of the infinite system is a 
crucial issue.  A priori the RSB Ansatz does not give an answer, that 
can turn out to be different for different systems: detailed 
computations and the analysis of theoretical ideas are needed to 
understand what exactly happens.  One has to consider $3$ main 
possibilities: first the number of states can be finite, second it can 
be infinite with states forming a countable set, and as a third 
possibility the states can form an uncountable set.  The RSB Ansatz 
applied to the theory of spin glasses turns out to imply that the 
number of finite volume states of a system becomes infinite in the 
infinite volume limit.  That implies that the second or the third case 
hold.  A detailed computation (see section (\ref{SS-META})) shows that the 
third possibility is the correct one.

In the following we will also need to define a {\em Ces\`aro average} 
(on which the metastate \cite{AIZWEH} construction is based).  We 
consider a (reasonable enough) function $f_{J}^{(N)} \equiv 
f(\rho_{J}^{(N)})$ defined on a lattice of volume $N$ and define its 
Ces\`aro average as

\begin{equation} 
  f_{J}^{C}\equiv\lim_{M\to \infty} \frac{1}{M} \sum_{N=1}^M f^{(N)}_{J} \ .  
  \protect\label{E-CESARO}
\end{equation}

\subsection{The Naive Infinite Volume Limit Does Not Exist\protect\label{SS-NAIVE}}

We will give now an example of a situation where the B-G distribution 
does not have a limit when the volume diverges.  We consider a 
ferromagnetic Ising model with a small random magnetic field and a 
small temperature in three dimensions.

The heuristic analysis goes as follows.  In a finite box there are two 
relevant finite volume states, the one with positive magnetization and 
the one with negative magnetization.  In a first approximation, valid 
at small field and temperature, the difference in free energy of the 
two states is proportional to $2 \sum_{i}h_{i}$, which is a number of 
order $N^{3/2}$.  Therefore (but for rare choices of the random field) 
for a given value of $L$ one state will dominate.  However when 
increasing $L$ one will switch an infinite number of times from the 
situation where the magnetization is positive to a state with negative 
magnetization.  The magnetization itself does not have a limit when 
the volume goes to infinity.

A definite limit may be obtained by taking the limit by subsequences 
(i.e.  choosing only those value of $L$ for which the magnetization is 
positive or negative).  The other possibility is in taking the average 
over $L$ and using the fact that the Ces\`aro limit (\ref{E-CESARO}) 
of $m$ exists.

Once the infinite volume limit has been taken we can decompose the 
state in extremal DLR states.  Here the typical relevant states for 
the infinite volume limit are those with positive or negative 
magnetization.  If we choose the first approach (limit by appropriate 
subsequences) we will find a pure state that is a clustering state.  
In the case of the Ces\`aro limit we will obtain a 50\% mixture of the
two extremal states.

Similar and more complex phenomena are frequent both in random and in 
non random system (e.g.  those having quasi periodic ground states).  
Care is needed in order to obtain the infinite volume limit.  
Statements about chaoticity are basically about the fact that the 
naive infinite volume limit does not converge.

\subsection{Non Self-Averageness and the Infinite Volume Limit}

At this stage it is clear that we need a lot of ingenuity in order to
be able to define the B-G distribution for an infinite system.

The replica approach only deals with very large but finite systems, and it 
does not ask what happens in an actual infinite systems.  It should also be 
clear that the two approaches, the replica analysis of the finite 
volume correlations functions (and the results which can be stated in 
a simple and intuitive way by using the idea of decomposition into 
states of the Boltzmann-Gibbs measure) and the construction of pure 
states for an infinite system, give different information, which can 
be hardly compared one with the other.  In the replica method one 
obtains information only on those states whose weight $w$ does not 
vanish in the infinite volume limit\footnote{As it stands this 
sentence may be misleading because it could seem to describe the 
properties of a given state when we change the volume.  A more precise 
(but heavier) formulation is the following: for each system with large 
fixed volume $N$ the replica method gives information on the states 
(defined for that particular model) whose weight $w$ is not too 
small.}.  All local equilibrium states have the same free energy 
density: however the differences in the total free energy may grow as 
$L^{(D-1)}$.  From an infinite volume point of view all these states 
are equivalent, while from a finite volume point of view only the 
state with lower free energy and the states whose total free energy 
differ from the ground states by an amount of order one are relevant.

For example in the ferromagnetic case (in more than two dimensions at 
sufficient low temperature) there are equilibrium states which have 
positive magnetization in half of the infinite volume and negative 
magnetization in the other half.  These states are invisible in the 
replica method because their weight (when restricted to a finite 
volume system) goes to zero as $\exp(-A L^{D-1})$, $A$ being an 
appropriate constant (as we have already discussed special techniques, 
i.e.  coupling replicas, may be used to recover, at least partially, 
this information).  In the replica method the states are weighted with 
the corresponding Boltzmann-Gibbs weight and this weight can be hardly 
reconstructed from an analysis done directly at infinite volume.

Therefore one cannot expect a priori that formulae like the relation 
(\ref{E-WSUM}) are valid in the infinite volume limit (although that 
could happen): on the contrary, according to the facts we have 
discussed in the last paragraphs, one would better expect that the 
mathematical properties of an infinite system could be quite different 
from the ones of a finite large system.  The usual formalism of the 
replica theory does not concern the properties of the states of an actual
infinite system.

\subsection{A Comparison of the RSB Theory with Some Rigorous Results}

In a recent paper \cite{NS1} the authors have obtained new exact 
results about the behavior of finite dimensional spin glasses.  Here 
we discuss this interesting issue in some detail, and we observe that 
these results are in perfect agreement with the predictions of the RSB 
Ansatz that we have been discussing in the former sections.  We will 
stress how important it is to be careful to the real meaning of the 
objects that are defined.

\begin{itemize}

\item 
We consider a spin glass at low temperature.  The system is in a box 
of side $L$, volume $V=L^D$, with fixed boundary condition.  The 
Hamiltonian depends on a set of quenched variables $J$.  We consider 
the overlap $q$ and its probability distribution $P_J(q)$.  In the 
spin glass phase the function $P(q)$ (\ref{E-PQ-B}) is not a simple 
delta function, but it has a more complex structure.

\item
If the function $P(q_1)$ is not a delta function, one finds that the 
function $P_J(q_1)$ does depend on $J$ and it is not a self-averaging 
quantity.  In other words the quantity $P(q_1,q_2) \equiv 
\overline{P_J(q_1)P_J(q_2)}$ is not given by the product 
$P(q_1) P(q_2)$ (rigorous arguments in this direction have been 
recently given in \cite{GUERRA,AIZCON}).

\item 
In the same realization of the quenched random couplings we consider 
two systems which have a mutual overlap $\overline{q}$ in the $L \to 
\infty$ limit.  In this case the probability of finding a region of 
size $R$ where the overlap is $p\ne \overline{q}$ goes to zero as 
$\exp(-R^\alpha f(p,\overline{q}))$.  The authors of \cite{FRPAVB} 
have estimated the exponent $\alpha$ to be equal to $(D-\frac52)$.

\item 
In the infinite volume limit the two spin configurations $\sigma$ and 
$\tau$ defined with quenched disordered couplings $J$ which differ in 
a finite (arbitrarily small) portion of the lattice links 
turn out to have an overlap which is always equal to 
the minimum allowed overlap $q_{m}$ \cite{KONDOR}.

\end{itemize}

The work of \cite{NS1} describes an apparent contradiction with the 
results of the RSB Ansatz.  In \cite{NS1} the authors give two new 
definitions of a probability distribution of the overlaps $q$ (which 
we indicate, with abuse of language, again by $P_J(q)$).  Such 
$P_J(q)$ do not depend on $J$ in the large volume limit.  The point we 
want to stress is that the objects that the authors define are 
different from the ones we usually encounter in the literature.  Such 
$J$-independence turns out to be in perfect agreement with the RSB 
Ansatz (and a further support of its validity).

Let us list the predictions of the RSB approach for the quantities
that are defined in \cite{NS1}.  The authors present different
constructions: we will analyze them trying to translate them in a lay
language.

We consider a system of size $L$, and we focus our attention on what
happens in an internal box of size $R$.  We call $q_R$ the overlap of
two replicas in this box.  We call $I$ the couplings inside the box
and $E$ those outside the box.  The couplings $J$ are determined by
$E$ and $I$ ($J=I\oplus E$).

Following the first construction of \cite{NS1} we define

\begin{equation}
  P^{(1)}_I (q) \equiv \int d\mu(E) P_J^R(q)\ , 
\end{equation}
where $ P_J^R(q)$ is the probability distribution of the overlap 
$q_R$, i.e.  of the overlap restricted to the region $R$.

Let us first send $L \to \infty$, or if we prefer, let us consider the
case $L\gg R$.  When $R$ goes to infinity, still being much smaller
than $L$, replica theory implies that $q$ and $q_R$ are equal: the
equality of these two quantities has been discussed in detail in
section \ref{SS-CF} and it is a consequence of the clustering
properties of the correlations function in the ensemble at fixed $q$.
As a consequence $P^{(1)}_I (q)$ coincides with $\int d\mu(E) P_J(q)$.
This last integral does not depend on $I$, so for large $R$ replica
theory predicts that

\begin{equation}
  P^{(1)}_I=P(q)\ ,
\end{equation}
and it is independent from $I$, in perfect agreement with the general
results proven in \cite{NS1}.
        
Let us discuss a second definition, $P^{(2)}(q)$, which is inspired by 
the second construction of \cite{NS1}.  Here things become more 
interesting because we have a construction for the B-G measure in the 
infinite volume limit as the convergent limit of measures of finite 
volume.  If we skip technical details the main idea is the following.  
As in the previous case we consider a system of size $L$ and we 
concentrate our attention on what happens in a box of size $R$.  We 
call $I$ the couplings inside the box and $E$ those outside the box.  
If ${\cal C}$ is a configuration of the spins inside the smaller box 
and ${\cal D}$ that of the spin outside the box we can define a 
probability

\begin{equation}
  P_{I}({\cal C})\equiv \ba{P_{I\oplus E}({\cal C}\oplus{\cal D})}\ ,
\end{equation}
where the upper bar stands for an average over the external couplings 
$E$ and the external spins ${\cal D}$.  In other words we consider a 
system of size $R$ and we average its properties over the external 
world.  Compactness argument may be used to prove that the probability 
distribution $P_{I}({\cal C})$ has a limit (as usual at least by 
subsequences) when $L$ goes to infinity first and $R$ goes to infinity 
later.  Let us call the corresponding state $\rho^{B-G}_{J}$.

Armed with this infinite volume probability for an infinite system one 
can compute the probability distribution of the overlap and one can 
prove that it does not depend on the internal couplings $I$.  This 
would be hardly a surprise because we have obtained this B-G 
distribution by an average process over an infinitely large system, 
which naturally destroys non-self-averageness.  If we compute it in the 
framework of the RSB approach we find that the overlap is always zero.

The argument runs as follows: we consider two systems,
one with couplings $J_1=I\oplus E_1$ and the other with couplings
$J_2=I\oplus E_2$ (i.e.  the couplings are equal in the internal box
and different in the external box).  We consider the distribution
probability of the overlaps $q_R$ and $q$ among a configuration of the
first system and a configuration of the second system, the first
overlap ($q_R$) being restricted to the region of size $R$ (where the
couplings are $I$ for both systems).  We introduce the corresponding
probability distributions which obviously depend on the couplings $I$,
$E_1$ and $E_2$.  We define

\begin{equation}
  P^{(R2)}_I(q_R) \equiv \int d\mu(E_1)  d\mu(E_2)
  P_{I,E_1,E_2}^R(q_R)\ ,   \,\,\, \ 
  P^{(2)}_I(q) \equiv\int d\mu(E_1)  d\mu(E_2) P_{I,E_1,E_2}(q)\ .
\end{equation}
Also in this case $P^{(R2)}_I(q_{R})$ and $P^{(2)}_I (q)$ coincide in 
the large $R$ limit.  $P^{(2)}_I (q) =\delta(q)$, due to the chaotic 
nature of spin glasses.  $P^{(2)}_I$ is independent from $I$, as 
proven in \cite{NS1}, and it is different from $ P^{(1)}_I$.

It is clear that there are alternative definition of the function
$P(q)$, which are less interesting as far as they display a less rich
behavior.  The reader should notice that it is only a bad notation to
call these functions with the same name, as far as they describe
very different properties of the system\footnote{It would be not wise
to conclude that they must have the same properties because they have
been called with the same name.}. In studying the properties of all
these functions the RSB approach gives the correct answer, i.e. it
declares self-averaging objects which are self-averaging and
non-self-averaging quantities that are likely to be
non-self-averaging.

\subsection{On Infinite Volume States\protect\label{SS-META}}

We will discuss here about the metastate construction \cite{NS0,AIZWEH},
that has been introduced to allow a rigorous construction of the
infinite volume limit of the quenched state.  There are two main
possible starting points for such a task: the first is the
construction described in the previous section, where we average over
the couplings in an external region.  This approach generates the
states that we have called $\rho^{B-G}_{J}$.  In the second approach
(the metastate) we consider periodic boxes of size $L$, and we denote
by $\rho^{(L)}_{J}$ the expectation value with respect to the usual
B-G distribution.  Using equation (\ref{E-CESARO}) we define the
quenched state as the Ces\`aro average $\rho^{C}$.  Both procedures
construct the infinite volume quenched state.  One can prove that the two
constructions lead to the same infinite volume equilibrium state.  It
can be conjectured that if one changes the boundary conditions,
e.g.  if one uses antiperiodic boundary conditions, the result will
not change.  In this way we have a natural definition of an infinite
B-G state, which however involves in one way or another some kind of
average.

It is crucial to note that the limits mentioned before should be read in
a weak sense.  In other words for any given quantity $A$ which is a
function of only a {\em finite} number of spins we have that

\begin{equation}
  \rho^{C}_{J}(A)=\lim_{M\to \infty}\frac{1}{M} 
  \sum_{L=1}^M \rho^{(L)}_{J}(A) \ . 
\end{equation}
This equality does not have to hold for observable quantities which 
contain an explicit $L$-dependence.

This equilibrium state can be decomposed into extremal DLR states
(which we label by the index $\lambda$). We have
\begin{equation}
  \rho^{B-G}_{J} = \int d\mu(\lambda)\  \rho^{\lambda}_{J} \ .
  \label{eq99}
\end{equation}
In this way we find a natural measure $d\mu(\lambda)$ over the
extremal states. This measure is called the {\em metastate}.
In order to obtain the metastate we have to consider not only the expectation values themselves, but the products of expectation values. At this end we can define
\begin{equation}
\rho_J^2(A,B) = \lim_{M\to\infty} \frac1M \sum_{L=1}^M \rho_J^{(L)}(A) \rho_J^{(L)}(B)\ .
\end{equation}
One finds that \cite{NS0,AIZWEH}
\begin{equation}
\rho_J^2(A,B) = \int dv(s) \rho_J^s(A) \rho_J^s(B)\ ,
\end{equation}
where $s$ denotes a generic state (pure or a mixture) and $\nu(s)$ is a measure on the states. This measure is called the \emph{metastate} and it is the generalization of the measure $\mu(\lambda)$ defined in Eq.~(\ref{eq99}). This construction may be generalized took the products of more expectations values.

The results of the RSB Ansatz imply that the set of states on which
this measure $\mu(\lambda)$ is concentrated cannot be countable.  Indeed if that was 
the case the previous formula could be written as
\begin{equation}
  \rho^{B-G}_{J} = \sum_{\lambda=1}^{\infty} w(\lambda) \ 
  \rho^{\lambda}_{J}\ ,
\end{equation} 
and the probability of finding two configurations in the same pure
state would be given by
\begin{equation}
\sum_{\lambda=1}^{\infty} w(\lambda)^{2}>0 \ .
\end{equation}
But we have shown in the previous discussion that the RSB theory 
predicts that the overlap distribution in the state $\rho^{B-G}_{J}$ 
(the quenched state) is a delta function at zero overlap, so that the 
probability of finding two configurations in the same pure state must 
be zero, in contradiction with the previous formula.

We are ready now to illustrate the main flaw of a series of papers 
(see \cite{NS4,NS1} and references therein), where the metastate 
approach was used to claim the presence of an internal inconsistency 
in the predictions of the RSB theory.  The starting point of these 
papers was the assumption that the finite volume state can be 
approximately decomposed as the sum of infinite volume states:

\begin{equation}
  \rho^{L}_{J}\approx \sum_{\lambda=1}^{\infty} w^{L}(\lambda) 
  \rho^{\lambda}_{J} \ .
  \protect\label{WIT}
\end{equation}
This formula looks innocent  and similar to the one that it is used as a 
starting point of the RSB approach, (\ref{E-WSUM}), but it 
is indeed rather different: in the RSB theory the sum in the r.h.s.  
runs over the {\em finite volume states}, while in (\ref{WIT}) the sum 
runs over the {\em infinite volume states}.

Before further discussing if the heuristic results of \cite{NS4,NS1}
are correct we should investigate if their starting hypothesis
(\ref{WIT}) makes sense.  We will show here that (\ref{WIT}) can be
sometimes (not always) correct for ferromagnets, but it is unnatural
when dealing with spin glasses.  In the case of spin glasses
equilibrium configurations of a system of size $L$ with periodic
boundary condition when embedded in a larger system will have a much
higher energy than the equilibrium configurations.  Naively one would
expect an energy increase on the surface of the order of $L^{D-1}$,
but the actual increase may be smaller due to possible adjustments of
the interface.  The spins at $i=1$ and $i=L$ would be chosen in such a
way to minimize the free energy with periodic boundary conditions, but
would be out of place when merging the spin configuration in a larger
system.

The previous formula also fails in some ferromagnetic cases.  Let us
consider the case which we have discussed before in section
(\ref{SS-NODISG}) of a ferromagnet with antiperiodic boundary
conditions.  We consider the three dimensional case, with temperature
not too far from the critical one (or equivalently the two dimensional
case).  In these conditions the interface is rough and it has a width
increasing like $L^{\frac{1}{2}}$ in three dimensions (in two
dimensions the width is always proportional to $\log(L)$).

The divergence of the width of the interface has absolutely no
consequences on the analysis of states in finite volume (we classify
the configurations in states by the average position of the interface;
the correlation functions are clustering but for a region close to the
interface, which can be neglected since it is a fraction of the volume
which asymptotically goes to zero), but it implies that in the
infinite volume limit there are only two pure equilibrium states: the
ones with uniform positive and uniform negative magnetization.  The
state with positive magnetization in half space and negative
magnetization in the other half space exists only when the interface
is not rough.  That shows that the relation (\ref{WIT}) is not
satisfied even in this very simple case\footnote{A similar analysis of
the $3D$ ferromagnet implies that (\ref{WIT}) is never valid if we
take antiperiodic boundary conditions in two directions, e.g.  $x$ and
$y$.}. The opposite conclusion would be valid in three dimensions (but
not in two dimensions) at positive temperatures smaller than the
roughening temperature, because in this case the interface would have
a finite width also in the infinite volume limit.

Of course it is true that the probability distribution $P_{J}^{R,L}$
of the variables in a region of fixed size $R$ inside a box of size
$L$ with periodic boundary conditions can be written as

\begin{equation}
  \rho^{R,L}_{J}\approx \int d\mu ^{R,L}_{J}\ 
  \rho^{\lambda}_{J}\label{TRUE} \ ,
\end{equation}
where $L\gg R$, $\rho^{R,L}_{J}$ is the state associated to
$P_{J}^{R,L}$, and the difference among the r.h.s and the l.h.s.  goes
to zero when $L$ goes to infinite \footnote{To claim that the r.h.s
goes to the l.h.s.  when $L$ goes to infinity would be incorrect,
because both sides of (\ref{TRUE}) do not have a limit when $L\to
\infty$: only their difference (which goes to zero) has a limit.}.  As
we have discussed it is crucial that (\ref{TRUE}) fails in many cases
for $R=L$.  In the replica approach $\rho^{L,L}_{J}$ plays a central
theoretical role, and the other probability distributions (e.g.
$\rho^{R,L}_{J}$ for $R<L$) are derived quantities.

The existence of rigorous arguments which prove that the relation
(\ref{WIT}) cannot hold even in some simple cases, and of heuristic
arguments which imply that it is not valid for spin glasses strongly
suggests that it would be unwise to use it as the starting point of
any argument.  It is not a surprise that this formula (which sometimes
has been called the non-standard SK picture) leads to contradictory
conclusions.  The true RSB Ansatz (the so-called SK picture or mean
field picture) that we have discussed in the previous sections (and
that, following the wording of \cite{NS1,NS4}, is different both from
what \cite{NS1,NS4} call the standard SK picture and from what they
call the non-standard SK picture), does not lead to obvious
contradictions: as far as we know there are no generic arguments which
question its validity.

The metastate approach is telling us something about the behavior of
the properties of a fixed region of space in the infinite volume limit
(essentially that there is a natural way to define such a limit if we
average over the appropriate regions). This allows us to introduce a
natural measure over pure DLR states.  However this infinite volume
states are states in presence of a quenched random environment: they
are different from the finite volume states of the replica theory and
they are not directly related to the global behavior of systems in a
finite volume.

As stressed in \cite{GUEPRI} and in \cite{PARSIX} the crucial problem
in order to compare the rigorous approach to the RSB Ansatz
predictions consists in studying the behavior of the probability
distribution of two (or more) identical replicas, $P(\si,\tau)$ and to
verify the existence of correlations among the replicas.  In
particular if we study a system with the $n$-replicated Hamiltonian
(\ref{REPHAM}), with positive integer $n$, we can construct the
infinite volume probability distribution for this $n$-replicated
system.  The predictions of the replica approach can be translated in
predictions about the values of the overlaps among these replicas.
Replica symmetry breaking corresponds to the existence of correlations
among replicas which can be exposed by introducing additional replicas.

More precisely we can consider a system done of $n$ real replicas of 
size $L$.  We concentrate our attention on what happens in a box of 
size $R<L$.  We consider the $n\times n$ matrix $q_{R}^{a,b}$ of the 
overlaps in the box of size $R$.  Using the same notation as before 
we define:
 
\begin{equation} 
P_I ({\bf q}) \equiv \int d\mu(E) P_J^R({\bf q})\ ,
\end{equation} 
where $ P_J^R({\bf q})$ is the probability distribution of the overlap
matrix $q_R^{a,b}$.

The probability distribution should go to a finite limit when $L \to 
\infty$ first and $R \to \infty$ later.  We can also consider the 
probability distribution of only some elements of this matrix.  In 
particular we have already remarked that $P(q_{1,2})$ coincides with 
the probability distribution $P(q)$.  Moreover (and this is a crucial 
issue) the fact that $P^{L}_{J}(q)$ depends on $J$ translate in this 
language as a lack of factorization of this probability:

\begin{equation}
  P(q_{1,2},q_{3,4})\ne P(q_{1,2}) P(q_{3,4})\ .
\end{equation}
Spontaneous replica symmetry breaking corresponds to a 
non-independence of the replicas and to the presence of correlations 
among them.  It is possible to translate the predictions of the RSB 
Ansatz into this formalism, which can be also used to prove under 
general assumptions the validity of relations like (\ref{GUEREL}) 
(\cite{GUERRA,AIZCON}).

One can obtain similar results also using the metastate, defining

\begin{equation} 
  P^{R} ({\bf q}) \equiv \lim_{M\to \infty}
  \frac{1}{M-R} \sum_{L=R}^{M} P^{R,L}({\bf q})\ ,
\end{equation}
and sending $R\to\infty$ at the end.  

This approach is promising.  The framework can also be used for
studying systems without quenched disorder \cite{GLASSY}.  Using this
formalism it is possible to investigate the issue of replica symmetry
breaking also for systems without disorder like structural glasses.
For reasons of space we will not discuss here this important point in
more details.

\section{Numerical Results\protect\label{S-NUMRES}}

In this last part of the paper we will review briefly the large mass
of numerical results that support the fact that replica symmetry is
spontaneously broken in finite dimensional spin glasses. In particular
we will discuss numerical results for the three and four dimensional
Ising spin glass with Gaussian couplings and with binary couplings
(that can take the two values $\pm 1$).

We will extract from published papers only those evidences which are
relevant to show that the real finite dimensional systems behave as
implied from the predictions of the RSB Ansatz, and we will refer to
the original references for more details.  We will also discuss some
new numerical results (about a precise determination of the finite
size effects that affect the infinite volume sum rules of the theory
and about simulations with coupled replicas) that we will describe
here in detail.

We notice that the biggest systems which are fully thermalized in
numerical simulations contain at most $10^{4}$ spins.  This number is
not very different from what can be realized experimentally: there are
some indications that in a typical experiment it is possible to
thermalize only regions containing $10^{5}$ spins at most
\cite{JOORHA}.  If replica predictions would fail for systems of size
bigger than $10^{6}$, this fact would be unobservable both in computer
simulations done with the present computer technology and (more
important) in real experiments.

\subsection{The Phase Transition\protect\label{SS-PHATRA}}

\begin{figure}
\begin{center}
\includegraphics[width=0.5\textwidth]{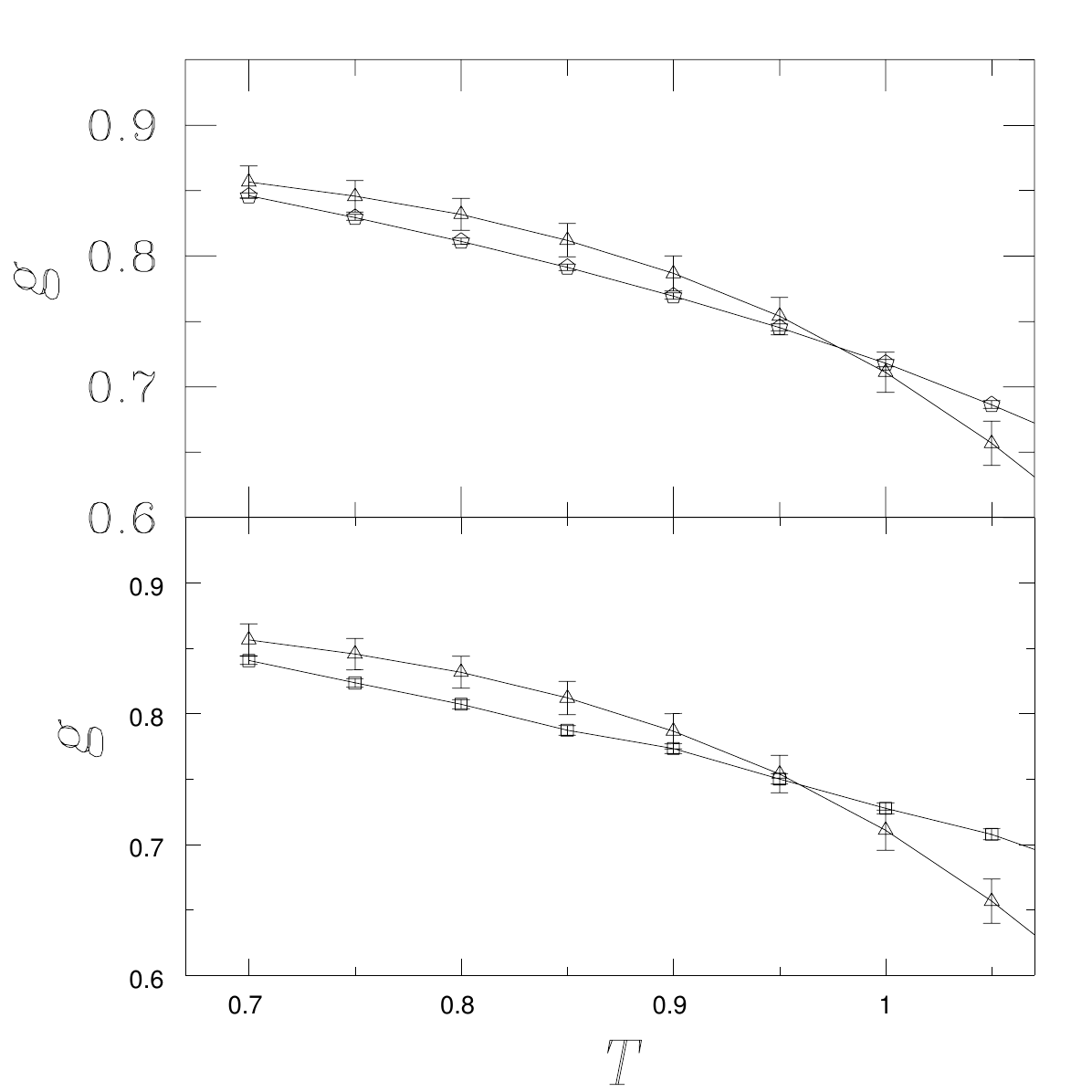}
\end{center}
\protect\caption[0]{The Binder cumulant versus $T$ for the $3D$ Ising
spin glass with Gaussian couplings. In the upper figure we show the
crossing of the $L=8$ and $L=16$ curves. In the lower figure are the
crossing of the curves for $L=4$ and $L=16$.}
\protect\label{fig:binder}
\end{figure}

\begin{figure}
\begin{center}
\includegraphics[width=0.5\textwidth]{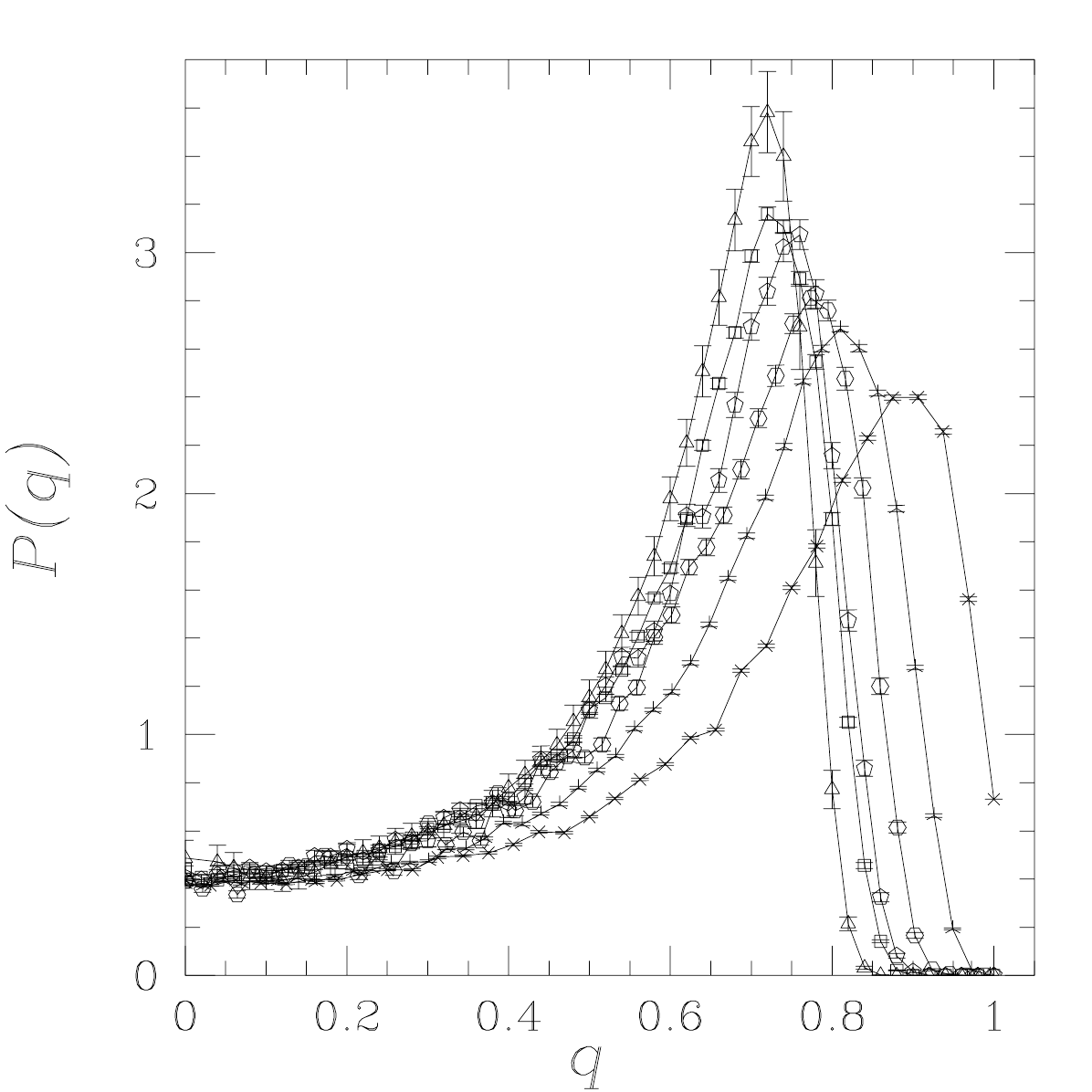}
\end{center}
\caption[0]{The probability distribution of the overlap for the $3D$
Ising spin glass with Gaussian couplings below the critical
temperature ($T=0.7\ T_c$).  The data are for $L=4$, $6$, $8$, $10$,
$12$ and $16$ and increasing the size the peak becomes sharper and higher.}
\label{fig:pq}
\end{figure}

\begin{figure}
\begin{center}
\includegraphics[angle=-90,width=0.5\textwidth]{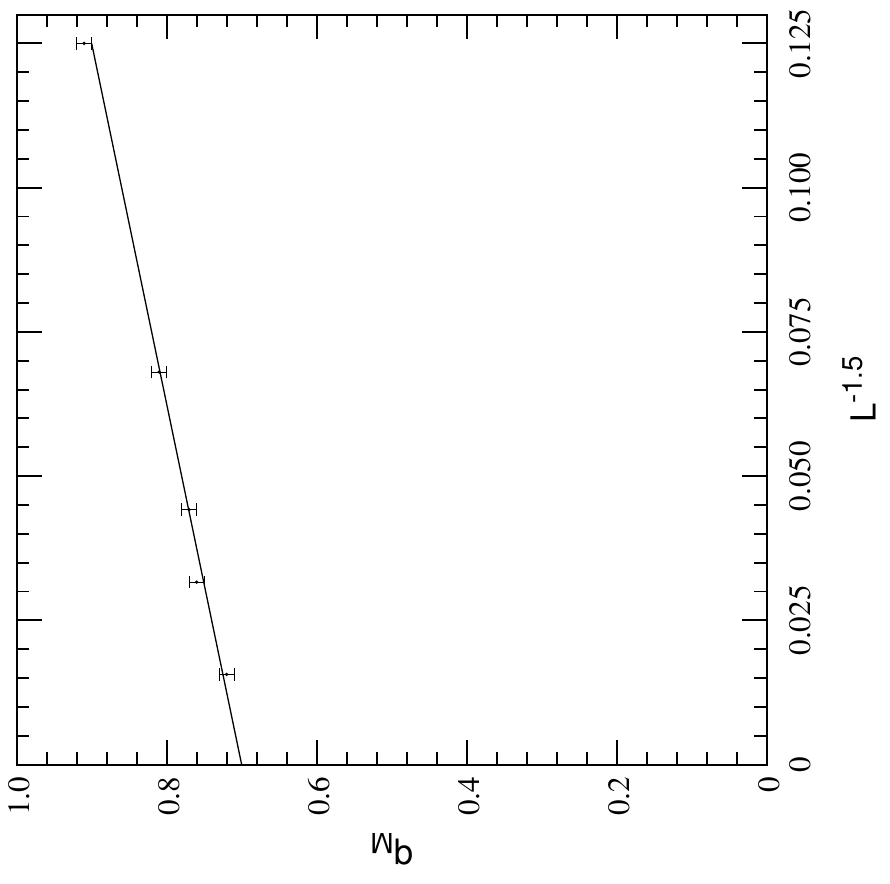}
\end{center}
\caption[0]{The value $q_M$ where $P(q)$ is maximum versus $L^{-1.5}$
in the $3D$ Ising spin glass with Gaussian couplings below the
critical temperature ($T=0.7\ T_c$).}
\label{fig:qea}
\end{figure}

\begin{figure}
\begin{center}
\includegraphics[width=0.5\textwidth]{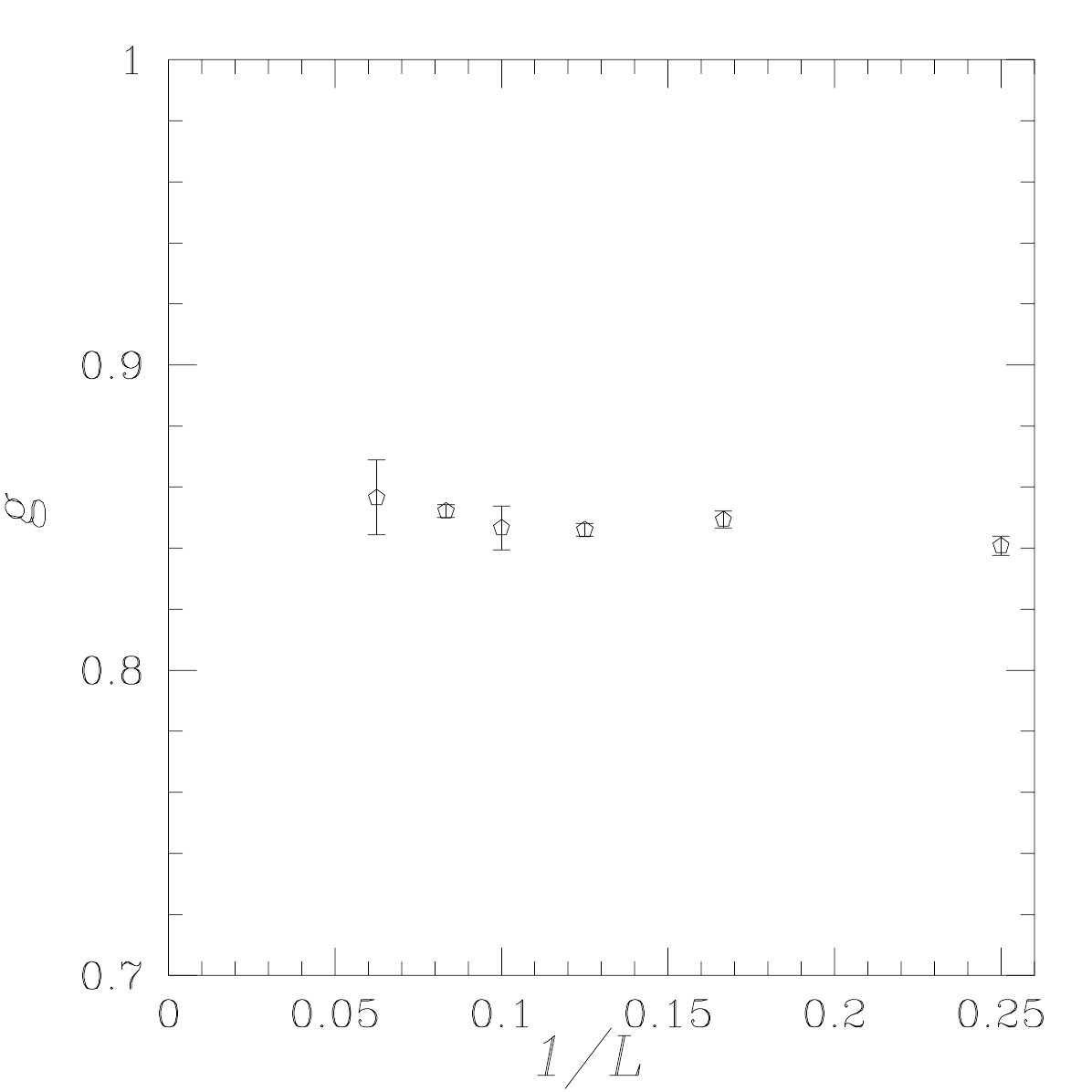}
\end{center}
\protect\caption[0]{The Binder cumulant as a function of $L$
for the $3D$ Ising spin glass with Gaussian couplings below the critical
temperature ($T=0.7\  T_c$).}
\protect\label{fig:binder07}
\end{figure}

In this section we have chosen four figures in order to make clear
some basic facts, i.e.:

\begin{enumerate}

\item  
The $3D$ EA spin glass undergoes a true phase transition (figure
\ref{fig:binder}).

\item The Edward-Anderson order parameter does not vanish in the
thermodynamical limit (figures \ref{fig:pq} and \ref{fig:qea}).

\item The low temperature phase is mean-field-like (figures
\ref{fig:pq} and \ref{fig:binder07}).

\end{enumerate}

In the next paragraphs we will discuss these three issues in better
detail.  For a complete discussion we address the reader to references
\cite{BOOKYO,MPRUNP,INMAPARU}.

In figure \ref{fig:binder} we show the best available evidence for the
existence of a phase transition. We plot the Binder cumulant $g$
defined as

\begin{equation}
  g\equiv \frac12\left(3 - {\overline{\lan q^{4} \ran}
  \over \overline{\lan q^{2} \ran}^{2}}
  \right)
  \ .
\end{equation}
The crossing of the curves of the Binder cumulant for different
lattice sizes is considered as the standard signature of the existence
of a phase transition.  This evidence has been first given by
Kawashima and Young \cite{KAWYOU}, and is also supported by
calculations of Berg and Janke \cite{BERJAN}.

As discussed in detail in the literature (see for example
\cite{BOOKYO}) the $3D$ case is atypical, in which we sit very close
to the lower critical dimension $D_c^L$. The signature for a
transition we have shown in \ref{fig:binder} is indeed atypical, since
more than a crossing of the different Binder parameter curves we
observe a merging. Again, this is a signature of the fact that $D_c^L$
is very close to $3$. For example in the $4D$ spin glass
\cite{4DIM,MARZUL} the crossing is a very typical, clear cut crossing
of the type one finds for the usual Ising model.

After establishing the existence of a phase transition we want to
characterize the low $T$ phase. A replica symmetry broken phase has to
be characterized by establishing in a clear way its peculiar
properties in the infinite volume limit.

For the (as we already said atypical) case of the $3D$ spin glass, we
start by discarding the possibility, suggested by the little
separation between different Binder cumulant curves for $T<T_c$, of
the presence of a Kosterlitz-Thouless phase transition without a
non-zero order parameter. In this case the order parameter would
become zero, in the infinite volume limit, even in the low $T$ region.
We have computed the probability distribution of the overlap, that we
show in figure \ref{fig:pq}. Using these probability distributions we
have computed the behavior of the positions of the maximum as a
function of the lattice size for a temperature well below $T_c$, as
shown in figure \ref{fig:qea}.  From figure \ref{fig:qea} it is clear
that the possibility that $\lim_{L\to\infty}q_M=0$ is unnatural (the
power $-1.5$ that we use in the horizontal scale comes from a best fit
to the data).  Our best fit to the form $q_{\infty}+a/L^b$ (that we
have also plotted in figure \ref{fig:qea}) gives us an infinite volume
order parameter, $q_{\rm EA} = 0.70\pm 0.04$ very close to the
estimate one finds using dynamical methods (given by the $q$ value
where the equilibrium regime ends, see figure \ref{fig:fdt_1} and later in
the text). In spite of this accurate estimate, and of the strong
qualitative evidence suggested by figure \ref{fig:qea}, a fit with
$q_{\infty}=0$ and a convergence with a very small power exponent
(i.e.  $0.2$) would have a greater $\chi^{2}$ but
cannot be excluded from our data.

Once we have characterized the transition point and we have ruled out
the scenario of a transition with no order parameter we show our
evidence of a broken replica symmetry in $3D$ spin glasses.

We plot in figure \ref{fig:binder07} the value of the Binder cumulant
at $T=0.7\ T_c$ as a function of the lattice size.  In an usual
ferromagnetic phase this points extrapolate to $1$ in all the broken
phase, for $T<T_c$. It is clear from the figure that in our data we do
not see any evidence of such a limit value. In a broken replica
symmetry phase one predicts, on the contrary, a non trivial shape of
$P(q)$ in the whole broken phase, and a non one limit of the Binder
parameter, that is, according to figure \ref{fig:binder07}, far more
plausible.

For sake of completeness we report here our best estimates of the
critical exponents and for the critical temperature:

\begin{equation}
  T_c=0.95\pm 0.04            \ ,\ \ 
  \nu=2.00\pm 0.15            \ , \ \ 
  \frac{\gamma}{\nu} = 2.36\pm 0.06 \ .
\end{equation}
These estimates of $\nu$ and $\gamma$ agree very well with those of
\cite{KAWYOU,BERJAN} for the $3D$ model with binary couplings.

\subsection{Sum Rules\protect\label{SS-SUMRULES}}

We describe here some original work dealing with numerical
verification (in the finite dimensional case) 
of some typical sum rules first derived for mean field
spin glasses \cite{MPSTV,GUERRA,AIZCON,PARSIX}.

These equalities are related to the stochastic stability of the
system, and give relations among the joint overlap averages of {\em
real replicas} (i.e. a finite number of copies of the system in the
same realization of the quenched disorder).  One can show
\cite{GUERRA} that under very general assumptions of continuity (that
turn out to be well verified \cite{MARINA}) they are also valid in
finite dimensional models.  In the following we will show numerically,
through Monte Carlo simulations, that they are very accurately
verified in the $3D$ Ising spin glass with binary couplings.

We define by

\begin{equation}
  E(\ldots)\equiv  \overline{\langle \ldots \rangle}
\end{equation}
the global average, taken both over the thermal noise and over the
quenched disorder.  In the infinite volume limit the following
equalities, among others, hold:

\begin{equation}
               E(q_{1,2}^{2} q_{3,4}^{2})    =
          \frac23  E(q_{1,2}^{2})^{2} +
          \frac13  E(q_{1,2}^{4})\ ,
\protect\label{G-A}
\end{equation}        
\begin{equation}
               E(q_{1,2}^{2} q_{2,3}^{2})    =
          \frac12  E(q_{1,2}^{2})^{2} +
          \frac12  E(q_{1,2}^{4})\ .
\protect\label{G-B}
\end{equation}        
In order to study this problem we have run simulations of the $3D$
Ising spin glass with binary couplings (that allows $32$ or $64$
systems to be updated at the same time, with a large amount of
computer time saving).  We have used the {\em parallel tempering}
technique \cite{MAPA93,PARTEM,MARI96} that allows to thermalize large
systems deep in the broken phase.  Thanks to Cray T3e runs of a
multi-spin-coded program we have been able to obtain a very high
statistics. For each value of the lattice size $L=6$, $8$, $10$, $12$,
we have considered $2048$ different realization of the quenched
disorder to define the sample averages.  For each sample we have
simulated $4$ replicas with separate and independent evolutions.  At
every sweep we have measured the energy and the mutual overlaps of the
four replicas.  Monte Carlo simulation parameters such as the number
of sweeps used for thermalization, the number of sweeps during which
observables were measured, the number of samples, the number of
$\beta$ values allowed in the parallel tempering
procedure\cite{PARTEM,MARI96}, the minimum and maximum $T$ value and
the temperature increment are summarized in the following table:

\begin{center}
\begin{tabular}{|c||c|c|c|c|c|c|c|} \hline
$L$ & Thermalization & Equilibrium & Samples & $N_{\beta}$ & $\delta T$ &  
$T_{min}$& $T_{max}$\\ \hline \hline
6    & 50000  &  50000    & 2048   &  19  &  0.1  & 0.5 & 2.3\\ \hline
8    & 50000  &  50000    & 2048   &  19  &  0.1  & 0.5 & 2.3\\ \hline
10   & 70000  &  70000    & 2048   &  37  &  0.05 & 0.5 & 2.3\\ \hline
12   & 70000  &  70000    & 2048   &  37  &  0.05 & 0.5 & 2.3\\ \hline
\end{tabular}
\end{center}

Parameters were tuned in order to have high ($\ge 0.5$) acceptance
rate for the temperature swapping process (the parallel tempering
needs this condition to be satisfied in order to have an acceptable
efficiency).  The allowed temperature range (assuming $T_c\simeq 1.1$
as in \cite{KAWYOU,BHATTYOUNG}) is approximately $0.5 T_c < T < 2
T_c$.

\begin{figure}
\begin{center}
\includegraphics[width=0.7\textwidth]{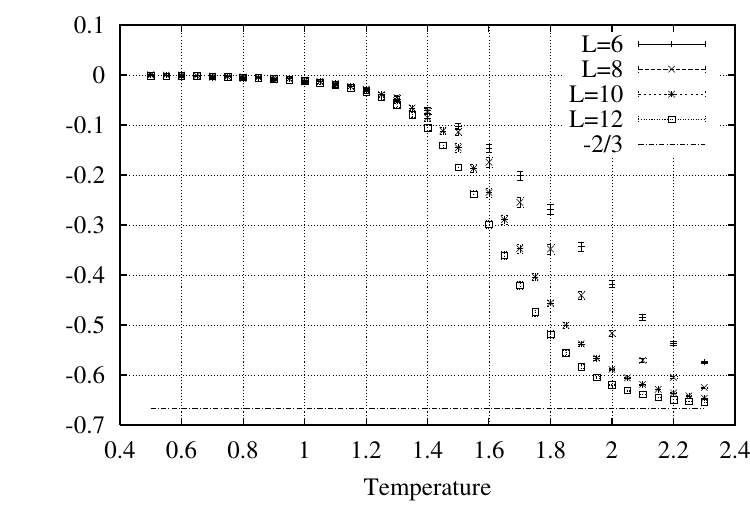}
\end{center}
\protect\caption[0]{The quantity (\ref{sumrule_ratio_rulea}) versus $T$
for different values of $L$.}  
\protect\label{final_ratio_rulea}
\end{figure}

\begin{figure}
\begin{center}
\includegraphics[width=0.7\textwidth]{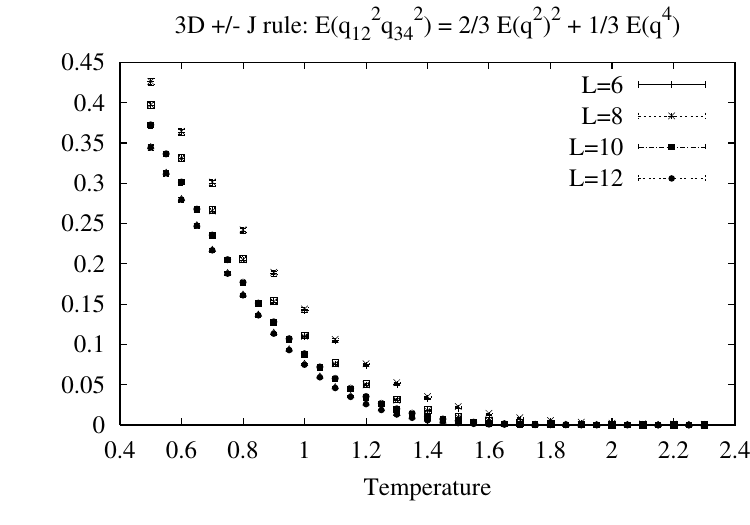}
\end{center}
\protect\caption[0]{The (indistinguishable on this scale) left hand
side and right hand side of the relation  (\ref{G-A}) for different values
of $L$. For all values of $L$ on this scale the curves are perfectly
superimposed within the error bars.}
\protect\label{final_sumrule_tutti_i_lati}
\end{figure}

\begin{figure}
\begin{center}
\includegraphics[width=0.7\textwidth]{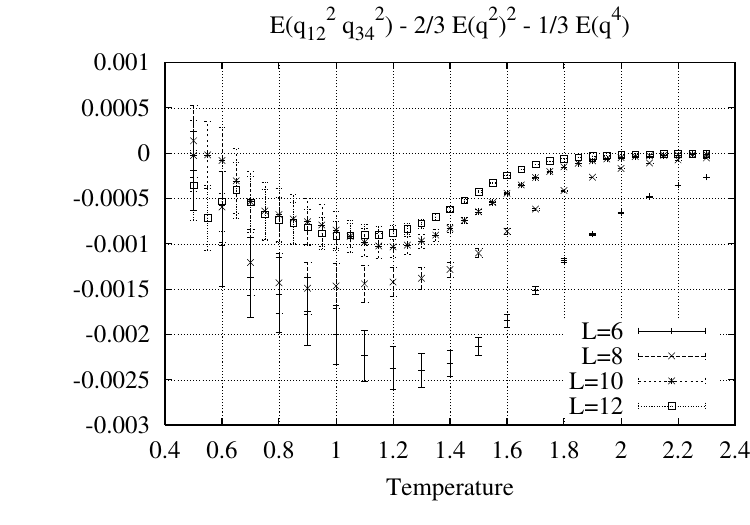}
\end{center}
\protect\caption[0]{Difference between the left
hand side and the right hand side of the relation (\ref{G-A}) versus $T$
for different values of $L$.}  
\protect\label{final_sumrule1234}
\end{figure}

Let us now discuss the results.
In figure \ref{final_ratio_rulea} we plot the quantity 

\begin{equation}
1.0 - \frac { \frac23  E(q_{12}^{2})^{2} +\frac13  E(q_{12}^{4}) }
            { E(q_{12}^{2} q_{34}^{2}) }
\label{sumrule_ratio_rulea}
\end{equation}
(i.e. $1$ minus the ratio of the left hand side and the right hand
side of the equality (\ref{G-A})) versus $T$ for each value of $L$. As
it should be at high temperature the limiting value is $-2/3$, since
in the high $T$ region $P(q)$ is a Gaussian centered around
$q=0$. Conversely at low $T$ the content of equation (\ref{G-A}) is
highly non-trivial.  Here the function $P(q)$ is not a simple
$\delta$-function: it is non-trivial, and for example $E(q^4)=O(1)$
and it differs from $E(q^2)^2$ of a quantity of order $1$ (typically
in the low $T$ phase their difference is of order $30\%$).  It is very
appealing that in this regime (\ref{G-A}) is verified up two
significant digits, as can be deduced from figures
\ref{final_sumrule_tutti_i_lati} and \ref{final_sumrule1234}.

Figure \ref{final_sumrule_tutti_i_lati} shows that in the broken phase
the relation (\ref{G-A}) is satisfied in a non-trivial way (i.e. not
through $0=0$). For high $T$ values (\ref{G-A}) just tells us that
zero equal zero, but in the broken phase the left hand side and the
right one are non-zero (we have already shown that by finding that the
value of $q_M$ such that $P(q_M)$ is maximum does not go to zero when
$L\to\infty$, and figure \ref{final_sumrule_tutti_i_lati} confirms it.

Figure \ref{final_sumrule1234} shows us how relation  (\ref{G-A}) is
violated on a finite lattice. Clearly (\ref{G-A}) is exact in the
infinite volume limit, and one get finite size corrections on finite
lattices. It is remarkable that these corrections are already small on
small lattices: they are maximum close to the critical point, decrease
when going far from $T_c$, and go smoothly to zero with increasing
lattice size. 

\begin{figure}
\begin{center}
\includegraphics[width=0.7\textwidth]{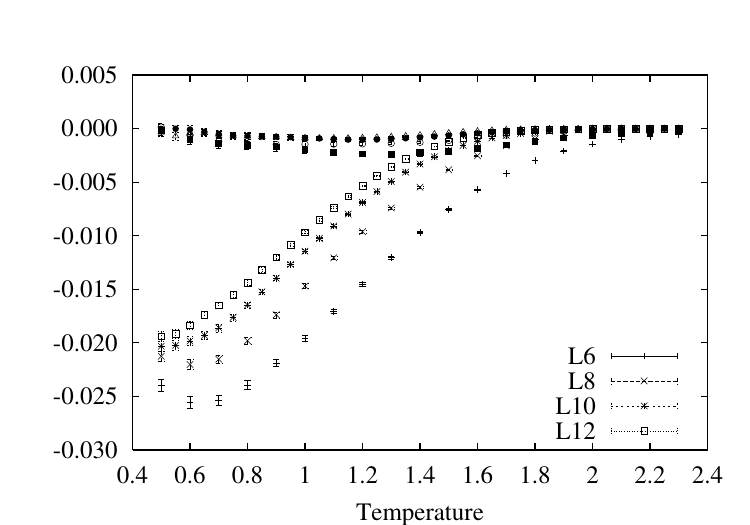}
\end{center}
\protect\caption[0]{
As in figure \ref{final_sumrule1234}, but we also plot the quantities
(\ref{THE-WRONG}) for different values of $L$.}
\protect\label{final_sumrule1234_gs}
\end{figure}

The cultivated reader can notice that relation (\ref{G-A}), for
example, is also satisfied in a droplet like situation, where all the
relevant expectation values are constant. In order to stress the
difference among this situation and the one observed in numerical
simulations we replot in figure \ref{final_sumrule1234_gs} the same
curves of figure \ref{final_sumrule1234}, adding the quantities

\begin{equation}
               E(q_{1,2}^{2} q_{3,4}^{2}) -
          \frac13  E(q_{1,2}^{2})^{2} -
          \frac23  E(q_{1,2}^{4})\ ,
\protect\label{THE-WRONG}
\end{equation}        
where we have interchanged the factors $\frac13$ and $\frac23$, by
putting them in the wrong places. In a droplet picture this relation
would work as well as relation (\ref{G-A}): figure
\ref{final_sumrule1234_gs} shows clearly that this is not the case, as
is implied by a RSB picture. Only in the warm phase both realizations
are satisfied (trivially, since all the relevant expectation values
are zero in the infinite volume limit).

The results related to the equality (\ref{G-B}) are even more
appealing. Here at least $3$ real independent replicas are needed to
implement (\ref{G-B}) ((\ref{G-A}) can also be implemented with only
two real replicas, since $E(q^2_{1,2}q^2_{3,4}) = \overline{\langle
q^2\rangle^2}$). These three replicas are constrained by the
ultrametricity constraints: the fact that the relation (\ref{G-B}) is
verified in $3D$ with an accuracy of two significant digits is a
strong hint toward the existence of a non-trivial (ultrametric)
structure of {\em pure states} in realistic short-range models.

\begin{figure}
\begin{center}
\includegraphics[width=0.7\textwidth]{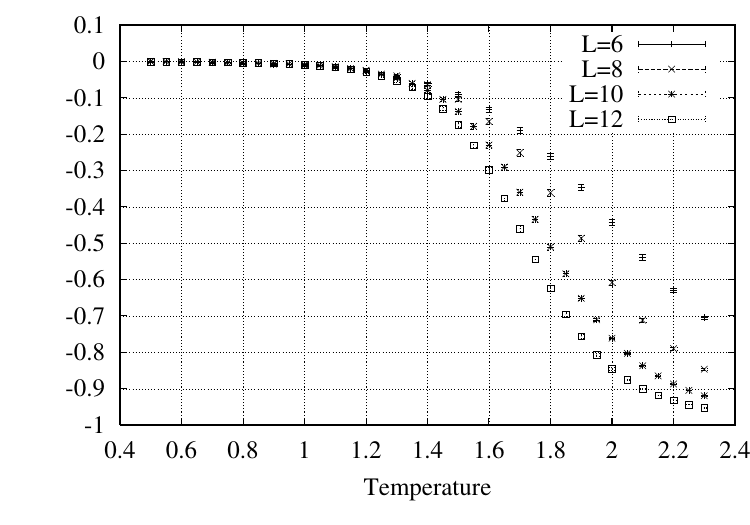}
\end{center}
\protect\caption[0]{\
As in figure \ref{final_ratio_rulea} but for the relation  (\ref{G-B}).
}
\protect\label{final_ratio_ruleb}
\end{figure}

As before the high $T$ region of figure \ref{final_ratio_ruleb} tells
us that in the high $T$ phase $P(q)$ is Gaussian, while the low $T$
is not. 

\begin{figure}
\begin{center}
\includegraphics[width=0.7\textwidth]{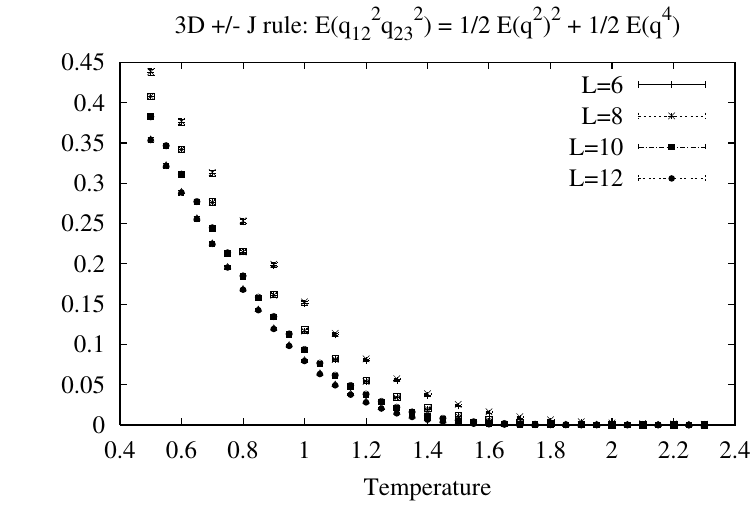}
\end{center}
\protect\caption[0]{
  As in figure \ref{final_sumrule_tutti_i_lati} but 
  for the relation  (\ref{G-B}).
}
\protect\label{final_sumruleb_tutti_i_lati}
\end{figure}

Figure \ref{final_sumruleb_tutti_i_lati} shows (like figure
\ref{final_sumrule_tutti_i_lati} does for (\ref{G-A})) the values of
the two sides of the relation (\ref{G-B}). The numerical values are
again perfectly superimposed within the limit of our statistical
errors.

\begin{figure}
\begin{center}
\includegraphics[width=0.7\textwidth]{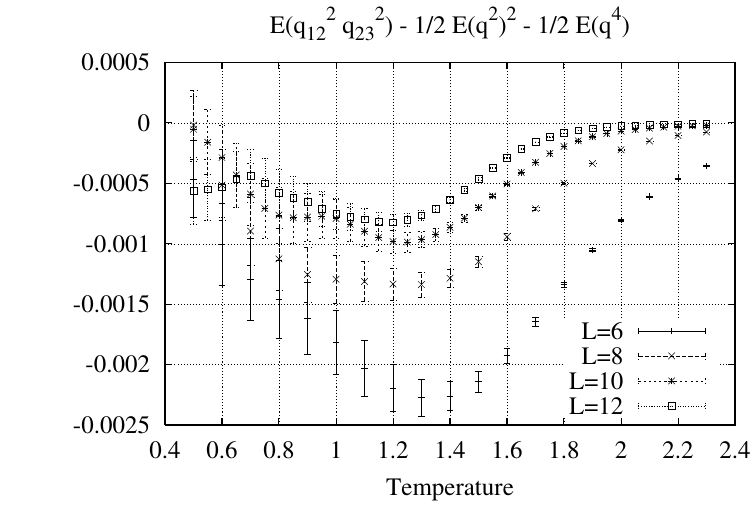}
\end{center}
\protect\caption[0]{
  As in figure \ref{final_sumrule1234} but 
  for the relation (\ref{G-B}).
}
\protect\label{final_sumrule1223}
\end{figure}

Finite size effect related to (\ref{G-B}) are shown in figure 
\ref{final_sumrule1223}. Also here the same discussion
done for figure \ref{final_sumrule1234} is valid.

We have always estimated the statistical errors that we have used in
our plots by using a direct jack-knife analysis (see for example
\cite{FLYVBJ}).  The amount of the violation of the sum rule that one
can see in figures \ref{final_sumrule1234} and \ref{final_sumrule1223}
close to $T_c$ decreases in a statistically significant way with $L$:
scaling fits are possible, and a power law explains it all, but there
are not enough points (and the statistical error over such a very
small number is not small enough) to make the fit reliable.

\subsection{Correlation Functions\protect\label{SS-CORFUN}}

In this subsection we study the equilibrium and the quasi-equilibrium
overlap-overlap correlation functions. These results have been
published in \cite{MPRR,MPRUNP}.  We have discussed the issue of
correlation function in the RSB formalism in section \ref{SS-CF}: we
will discuss here a few numerical results supporting our theoretical
framework.

Let us discuss first the definitions of quasi-equilibrium correlation
functions.  We have run quasi-equilibrium simulations following two
different procedures.  The first method is based on a sudden quench of
the system from a $T=\infty$ configuration to $T=T_f<T_c$.  In the
second approach (an annealing procedure) one starts the simulation at
a large value of $T$ and slowly cools the system down to $T_f$,
systematically decreasing the temperature of a small $dT$.  We have
always used $T_f=0.7\ T_c$.  In both cases we run a simulation for a
very large system of a size $L$ which can be considered infinite in a
first approximation.  At a given time $t$ we measure the correlations
function of the local overlap among two replicas of the systems
($C_{L}(x,t)$).

We can define the quasi-equilibrium correlation function as

\begin{equation}
  C(x)=\lim_{t\to \infty} \left(\lim_{L\to\infty}C_{L}(x,t)\right)\ .
\end{equation}
The order of the two limits matters. The overlap density among the two
replicas remains zero. It can be argued that $C(x)$ is equal to the
equilibrium correlation function of a system where two replicas are
constrained to have zero overlap.

\begin{figure}
\begin{center}
\includegraphics[width=0.5\textwidth]{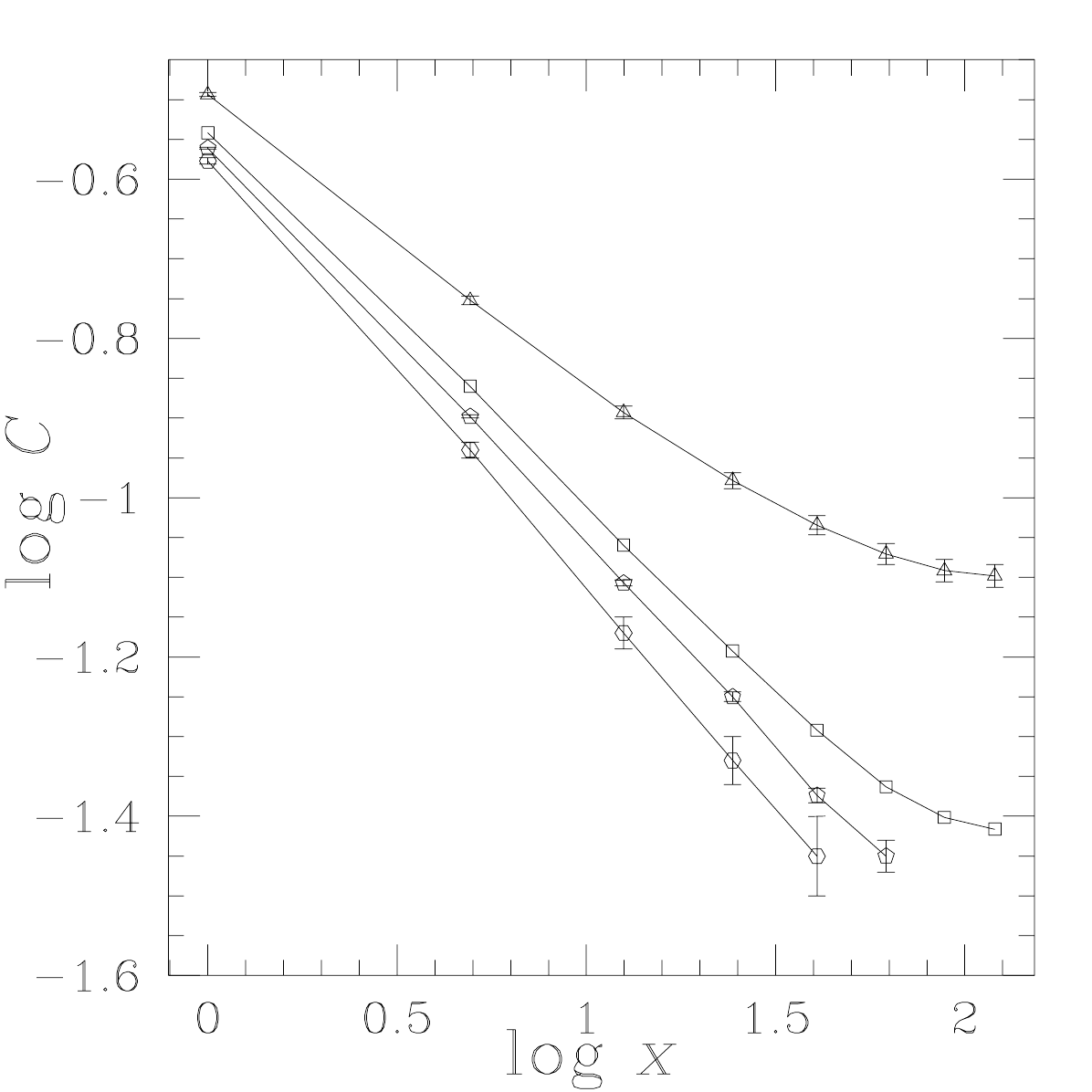}
\end{center}
\protect\caption[0]{
  Correlation functions at equilibrium and quasi-equilibrium 
  for the $3D$ Ising spin glass with Gaussian couplings
  in a double logarithmic scale.  The two lower curves correspond to an
  quasi-equilibrium simulation with an infinite time extrapolation. The
  third curve (from the bottom) corresponds to an equilibrium
  computation selecting couples of configurations with $q\simeq 0$.  The
  upper curve is for the equilibrium correlation function. $T=0.7\ T_c$. 
} 
\protect\label{fig:cor_1}
\end{figure}

We show in figure \ref{fig:cor_1} four different curves
that represent the $q-q$ correlation function for a $3D$ spin glass
with Gaussian couplings, computed using different approaches.

\begin{enumerate}
\item
The lower curve is the infinite time extrapolation of the correlation
function ($C(x,t)$) computed by using the first quasi-equilibrium method
described in the previous paragraph.

\item
The next curve (bottom to top in the figure) has been obtained with
the second quasi-equilibrium method after an extrapolation to infinite
time.

\item
The third curve is the equilibrium $q-q$ correlation function at
equilibrium computed only on configurations that have a mutual overlap
$q<0.01$.

\item
Finally the uppermost curve is the total $q-q$ correlation
function at equilibrium without any constraint. 
\end{enumerate}

The lowest two curves (quasi-equilibrium curves) have been
obtained simulating a very large system ($L=64$, which does not
approaches equilibrium on the time scale of our numerical
simulations), while the two upper curves have been obtained at
equilibrium (using parallel tempering\cite{MAPA93,PARTEM,MARI96}) on a
$L=16$ lattice (so that they can undergo sizable
finite size effects, and they surely do when the distance becomes
close to $\frac{L}{2}$.).

The quasi-equilibrium correlation function does not depend much on the
method one uses to estimate it.  During the quasi-equilibrium runs,
since the lattice size is very large and the initial conditions are
random ($T=\infty$, i.e. $q=0$), the overlap stays zero (in our
statistical accuracy) during the whole simulation.  As we have
remarked it looks safe to assume that in these conditions the infinite
time extrapolation gives us the true equilibrium value restricted to
the $q=0$ subspace.

In the infinite time limit $C(x) \propto x^{-\alpha}$ with
$\alpha \sim 0.5$.  If replica symmetry is broken it is useful to write
the correlation function in the form

\begin{equation}
  C(x) = \int dq\  P(q)\  C_q(x) \ , 
\end{equation}
where $C_q(x)$ is the correlation function restricted to the set of
equilibrium configurations with fixed mutual overlap $q$.

In the RSB approach $C_q(x)$ depends on $q$, which can assume any
value between $0$ and $q_{\rm EA}$.  The two curves obtained via the
infinite time extrapolation of the quasi-equilibrium data give an
estimate of $C_0(x)$.  In the droplet approach the system at
equilibrium is supposed to be always at $q_{\rm EA}$.  If the picture
of a droplet phase was valid the $q=0$ three lower curves of
figure\ref{fig:cor_1} would coincide with the full equilibrium upper
curve. On the contrary figure \ref{fig:cor_1} shows the clear
difference between $C(x)$ and $C_0(x)$.

The $q=0$ correlation functions shown in figure \ref{fig:cor_1} are
{\em non connected} (i.e.  $\lim_{x \to \infty} C_q(x) = q^2$): the
asymptotic value of $C_0$ is zero in the RSB approach and $q_{\rm
EA}^2$ in the simple Migdal-Kadanoff picture.  Computations by de
Dominicis et al.  (see for example \cite{DKTYOU} and references
therein) show that in the RSB approach one expects a pure power law
decrease for the $q=0$ ergodic component of the correlation function.
The pure power law decrease observed in figure \ref{fig:cor_1} is very
clear, calling for an asymptotic approach to zero, as in the RSB
theory.

The infinite time extrapolation of the quasi-equilibrium correlation
functions ($C(x,t)$) coincides indeed with the true equilibrium
result.  This is clear from figure \ref{fig:cor_1} where the three
lower curves follow a power law behavior with the same exponent.

An expected discrepancy from a pure power law appears in the large $x$
region for the $L=16$ equilibrium runs.  This is the point where the
correlation function starts to feel the effect of the periodic
boundary conditions (on the large, $L=64$ lattice where we have
computed the quasi-equilibrium curves, we expect to observe the effect
of the periodic boundary conditions for $\log x \simeq 3.5$).

\begin{figure}
\begin{center}
\includegraphics[width=0.5\textwidth]{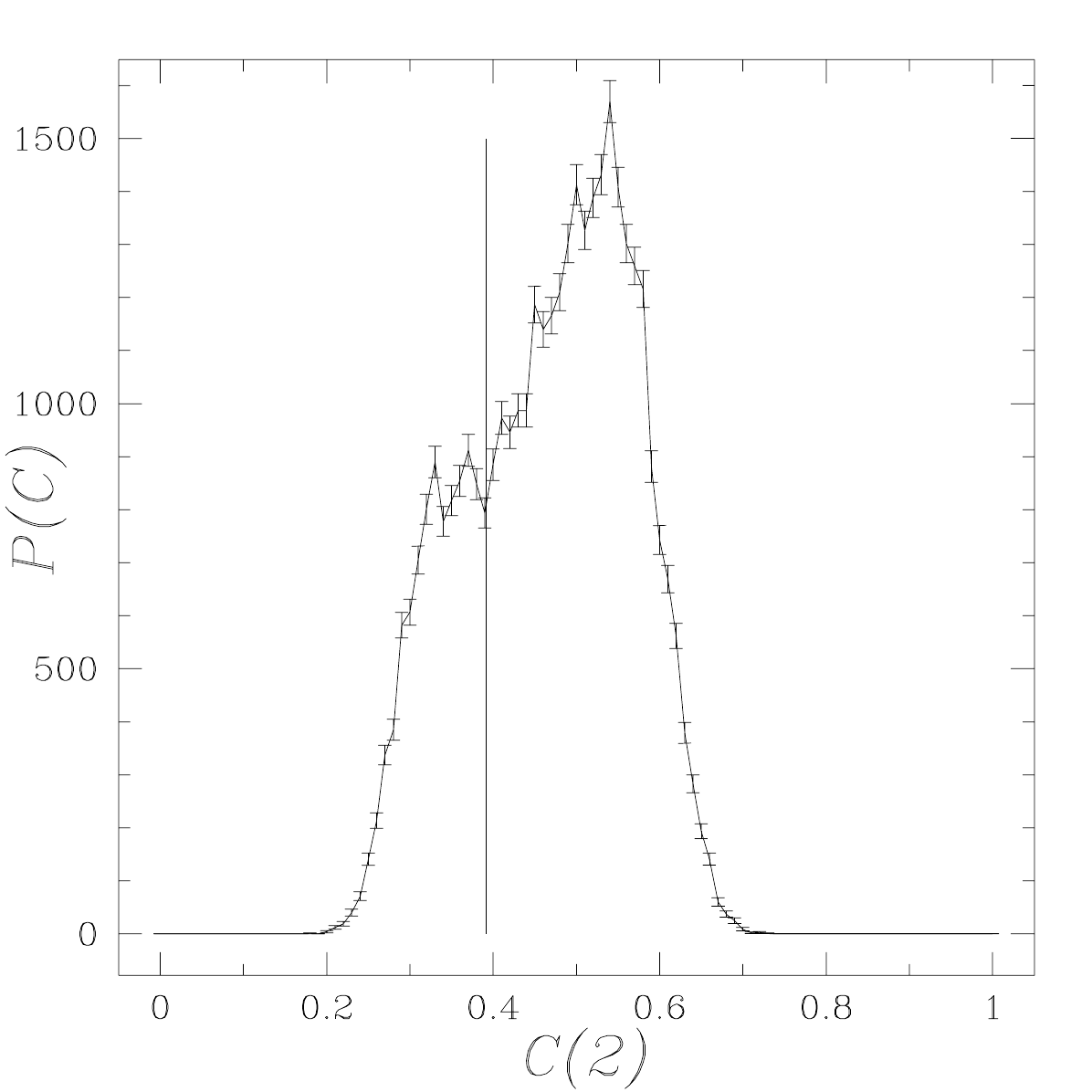}
\end{center}
\protect\caption[0]{Probability distribution of the $q-q$ equilibrium
correlation function $C(x)$ at distance $x=2$.}
\protect\label{fig:cor_2}
\end{figure}

We have also analyzed the probability distribution of $C(x)$ at
thermal equilibrium.  We have shown in the upper curve of figure
\ref{fig:cor_1} the expectation value of $C(x)$, but there is more
information in the full distribution probability (for fixed $x$).  We
show in figure \ref{fig:cor_2} the histogram of the $x=2$ component of
the equilibrium correlation function. We have also marked in this
figure the extrapolated value of the quasi-equilibrium correlation
function (the one of the lowest curve in figure \ref{fig:cor_1}). One
can distinguish a clear peak (in the region of large $C(2)$ values)
and an incipient maxima (or flex point) located close to the
quasi-equilibrium value.

From figure \ref{fig:cor_2} we can extract another interesting
conclusion.  If we assume, naively, a dependence of the form
$C(x)=a(x) + b(x)\ q ^2$ and a mean-field-like shape for the $P(q)$ we
expect the probability distribution of $C(x)$, to have the form of a
double peak shape, with the two peaks located at non-zero value of
$q$. This is what we see in figure \ref{fig:cor_2}: the equilibrium
histogram and the quasi-equilibrium values of the $q-q$ correlation
function agree well with all predictions of the broken replica
symmetry theory.

\subsection{Quasi-Equilibrium Window Overlap\protect\label{SS-OEWO}}

\begin{figure}
\begin{center}
\includegraphics[width=0.5\textwidth]{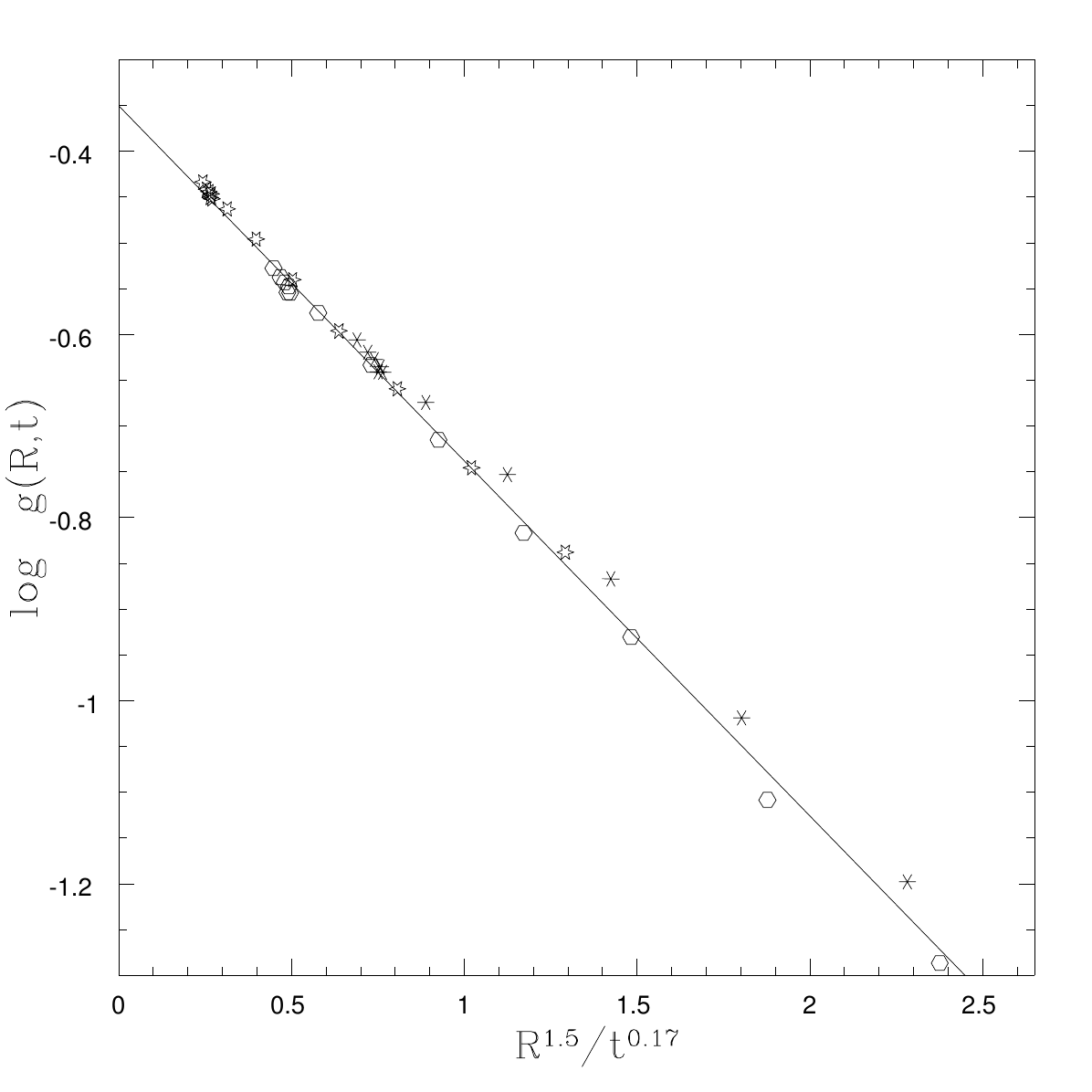}
\end{center}
\protect\caption[0]{Logarithm of the Binder cumulant $g(R,t)$ of
the window overlap $q_R$ (measured in a cubic box of linear size $R$)
versus the rescaled ratio of the window size $R$ and the Monte Carlo
time $t$. Stars are for $R=2$, hexagons are for $R=3$ and asterisks
for $R=4$.}  
\protect\label{fig:block_dyn}
\end{figure}

The so called {\em window overlap} has an important role in qualifying
the behavior of the system independently from the boundary conditions.
We have already introduced and discussed the window overlaps in
sections \ref{SS-CF} and \ref{SS-NODISG}. We will show here some
numerical results that help in clarifying the picture.

Let us repeat that the window overlap is computed using only the spins
belonging to part of the $3D$ lattice:

\begin{equation}
  q_B=\frac{1}{B^3} \sum_{x=0}^{B-1}\sum_{y=0}^{B-1}\sum_{z=0}^{B-1}
  \sigma(x,y,z)\  \tau(x,y,z),
\end{equation}
where $\sigma$ and $\tau$ are defined under the same realization of
the quenched disorder. We will denote the probability distribution of
$q_B$ by $P_B(q)$.

Here we consider the behavior of the window overlap during a
quasi-equilibrium simulation (see \cite{MPRR}).  As usual we start from
two random configurations, which we quench suddenly well below the
critical temperature. We monitor the value of the window overlap (for
different sizes of the small region, $B$, where we have computed the
overlap). By computing its probability distribution, $P_B(q,t)$, we
derive the Binder cumulant of $P_B(q,t)$: $g(B,t)$.

We find that the values of $g(B,t)$ collapse on a single curve when
using for the dynamical exponent the value found at this temperature
by analyzing the correlation functions (see the former section and
references therein). In figure \ref{fig:block_dyn} we show this
scaling plot.  It is clear that the data collapse very well for all
different values of the size of the small region where the overlap was
computed.  Moreover it is possible to obtain a safe extrapolation for
the infinite volume value of this Binder cumulant.  We estimate a
value of $0.64$, that is clearly different from the droplet model
prediction, $g=1$, and is not too far from the value of full volume
Binder cumulant $g$ at the critical point.

Recent experimental data \cite{JOWHV} find that real spin glasses are
well fitted by the same scaling law we have shown in figure
\ref{fig:block_dyn}.

\subsection{Equilibrium Window Overlap\protect\label{SS-EWO}}

\begin{figure}
\begin{center}
\includegraphics[width=0.5\textwidth]{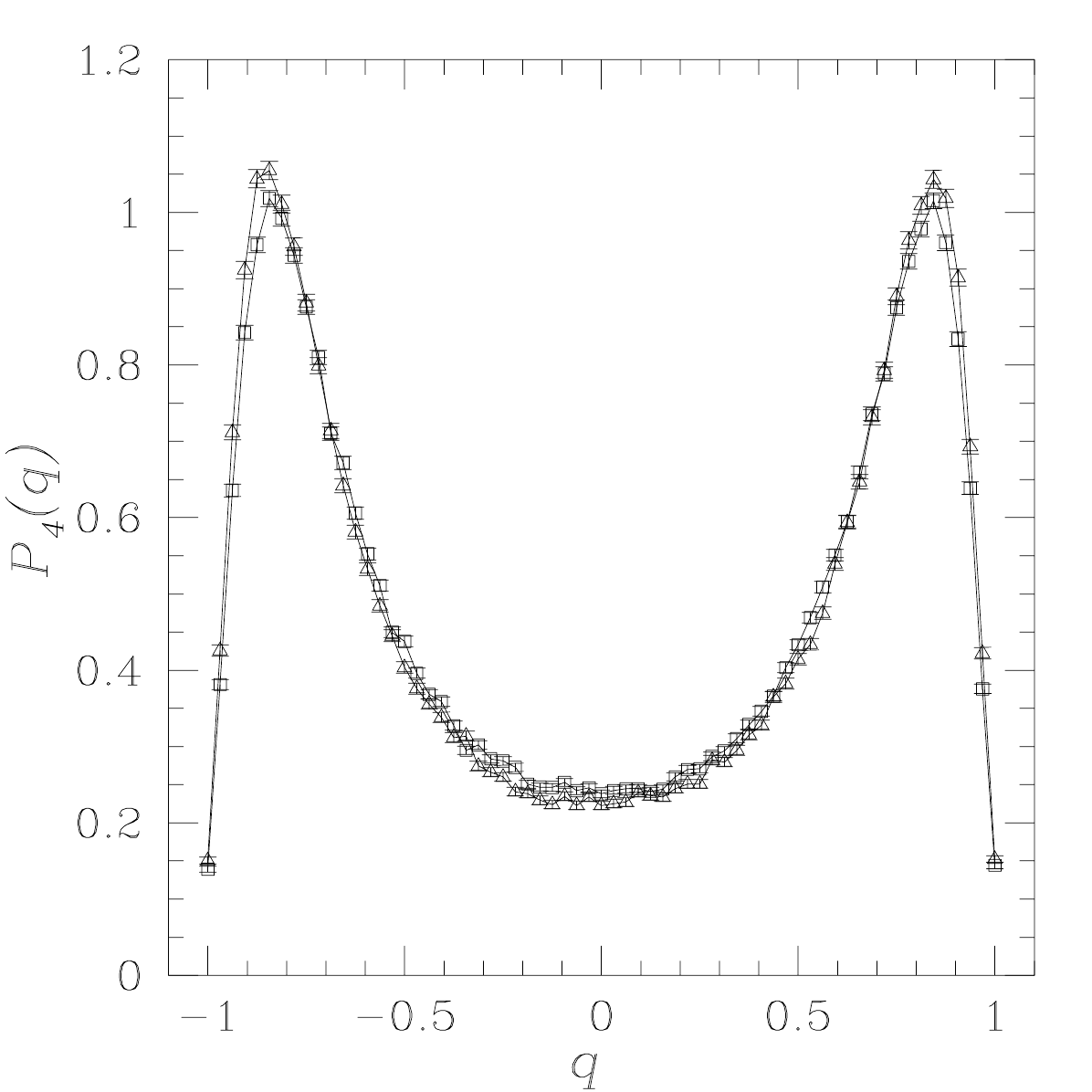}
\end{center}
\protect\caption[0]{
$P_4(q)$ for the $3D$ Ising spin glass with Gaussian 
couplings below the critical temperature ($T=0.7\  T_c$): 
triangles for $L=8$, squares for $L=12$.}
\protect\label{fig:block_ns0}
\end{figure}

Window overlaps have also been analyzed at equilibrium \cite{MPRRu2}.
The main goal of such a measurement is to check the theoretical ideas
discussed in sections \ref{SS-CF} and \ref{SS-NODISG}: it turns out to
give crucial hints about the behavior of the system.

\begin{figure}
\begin{center}
\includegraphics[width=0.5\textwidth]{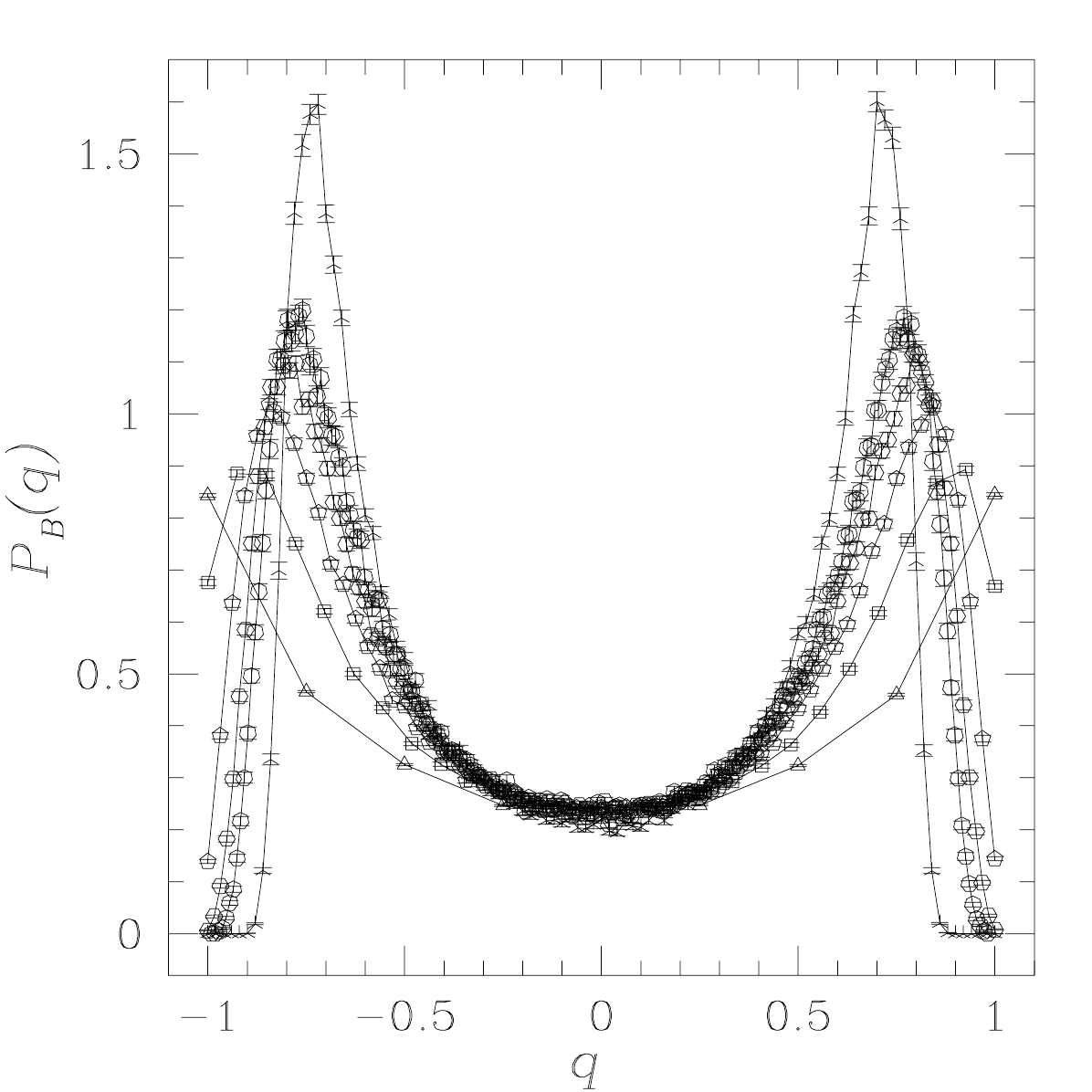}
\end{center}
\protect\caption[0]{ 
  $P_B(q)$ for the $3D$ Ising spin glass with
  Gaussian couplings below the critical temperature ($T=0.7\ T_c$) and
  $L=12$: $B=2$ (triangles), $B=3$ (squares), $B=4$ (pentagons), $B=5$
  (hexagons), $B=6$ (heptagons) and $B=12$ (three line stars).  $B$ is
  increasing with the height of the two maxima.  
}
  \protect\label{fig:block_ns}
\end{figure}

In figure \ref{fig:block_ns0} we show the probability distribution of
the window overlap (measured in a box of volume $4^3$) for two
different lattices: $L=8$ and $L=12$.  If interfaces were playing a
crucial role (see the discussion in section \ref{SS-NODISG}) one would
expect this distribution to tend toward a pair of $\delta$ function as
$L\to\infty$ for fixed window size $B$.  From figure
\ref{fig:block_ns0} it is clear that this is not the case. The shape
of the two probability functions is basically the same independently
of the lattice size.

In figure \ref{fig:block_ns} we show the probability distribution
of the window overlaps as a function of the window size for a fixed
lattice size, $L=12$.  The shape of the probability distribution does
not change much with $B$ (the largest change is when $B$ reaches $L$,
and we recover the usual overlap that is more sensitive to the
presence of a finite lattice than smaller observables).  Moreover the
behavior of the value of the probability distribution near the origin
(in the region of small overlaps) is opposite to the one we would get
if the model was building a simple droplet structure (i.e. this value
does not go to zero for small $B$ values , and it is on the contrary
slightly enhanced due to the finiteness of the window size).

It is clear that the distinctive features characterizing the broken
phase do not depend on the exact definition of the overlap: the full
and the window overlaps describe the same picture of the low
temperature phase.

\subsection{Zero Temperature}

An interesting and open problem is what happens in the zero temperature limit. In this limit it is clear the minimum energy configuration $\sigma^*$ dominates the partition function. It is less clear if the limit $T\to 0$ and the limit $N\to\infty$ can be freely interchanged (always {\it cum grand salis}).

A naive picture is the following: let us call $\sigma^r$ the local minima of the Hamiltonian and let us order them according to increasing value of the energy. Let us restrict the discussion to the case of models where the coupl- ing constant change continuously in such a way that with probability one no degeneracy of the states is allowed. At small finite temperature we would like to write
\begin{equation}
Z(\beta) \approx \sum_r e^{-\beta E^r}\ .
\end{equation}
This formula looks very similar to the relation
\begin{equation}
Z(\beta) = \sum_\alpha e^{-\beta F^\alpha}\ ,
\end{equation}
where the sum runs now over the finite volume states, and $F^\alpha$ is the corresponding free energy. One is therefore tempted at low temperature to identify the finite temperature finite volume states with the local minima of the Hamiltonian, and to use the properties of the firsts to deduce the properties of the latters \footnote{We neglect the fact that some of the minima correspond to two spins excitation of the ground states, and at finite temperature they are mapped into the same finite volume states where the ground state is mapped. A more detailed analysis is needed to cope with this problem, and we will not discuss it here.}.

However there may be some problems in this proposals: indeed in a little more refined approach one would write
\begin{equation}
Z(\beta) \approx \sum_r e^{-\beta E^r+S^r}\ ,
\end{equation}
where the entropy $S^r$ could be approximatively computed taking care of one spin excitations, i.e.,
\begin{equation}
S^r = \sum_i s(m_i^r)\ ,
\end{equation}
where
\begin{equation}
s(m) = \frac{1+m}{2} \log\left(\frac{1+m}{2}\right) + \frac{1-m}{2} \log\left(\frac{1-m}{2}\right)\ ,
\end{equation}
and the local magnetization can be computed at low temperature using mean field like equations
\begin{equation}
m_i^r = \tanh\left(\beta \sum_k J_{ik} m_k^r \right)\ .
\end{equation}
The point is not if this particular choice of the entropy is adequate, but that some entropy corrections are present. It may happen (this effect should be carefully investigated) that these entropy corrections, that are proportional to N, have a variation from state to state proportional to $N^{1/2}$, and these terms completely upset the mapping among finite temperature states and the zero temperature minima. As far as we expect that $S \propto N T$ in short range models, we could expect a crossover region at $T \propto N^{-1/2}$.

This could not happen at least for two different reasons:
\begin{itemize}
\item For some reasons, the fluctuations in the entropy are smaller than
$N^{1/2}$. If they are finite, they give only an irrelevant reweighting of the states.
\item Although it is not possible to have a one to one correspondence of the finite temperature states with the local minima, the statistical properties of the two sets are similar.
\end{itemize}

If any of the last two scenario is valid, one has that the function $P(q,T)$ and the corresponding function $x(q,T)$ are smooth functions of the temperature close to zero temperature also for large $N$, and no crossover region exists.

No evidence does exist in either directions, but, just for the sake of the discussion, let us assume that this zero temperature smoothness conjecture is correct, as it is often done in the literature. Let us define $E(q)$ as the energy gap between the ground state and all the states with overlap in modulus less than $q$ with the ground state. If we call $P_q(E)$ its distribution, the previous approach can be used to derive that at small temperature
\begin{equation}
\int dE P_q(E) e^{-\beta E} = x(q,T)\ .
\end{equation}
In this way we find that \footnote{In the usual mean field approximation, the probability distribution $P_q(E)$ can be computed analytically: one finds that $\langle E(q)\rangle \approx y(q)^{-1}$.}
\begin{equation}
P_q(0) = \lim_{T \to 0} \frac{x(q,T)}{T} \equiv y(q)\ .
\end{equation}
If $y(q)$ is different from zero we have that $P_q(E)$ is different from zero at finite $E$ when the volume goes to infinity, and local minima with different $q$ have a finite energy gap and difference in energy: if we close our eyes on energy differences which are of order 1 the ground state turns out to be non degenerate. Obviously the opposite conclusion holds if $y(q)=0$.

What do we know about the function $y(q)$? In the SK model one finds that $y(q)\neq 0$, and it is qualitatively of the form $y(q) \simeq q(1-q)^{-1/2}$. We do not have analytic informations on other models (like diluted ones) and numerical simulations in the low temperature region are not frequent \cite{BHC_94}. Quite recently new optimization techniques have been used which allow the computation of the ground states on relative large lattices (i.e., up to $10^3$) \cite{Hart_99,HM_99,MPZ_98,PY_99,PY_99bis}. For example the results by Palassini and Young \cite{PY_99,PY_99bis} seem to favor the hypothesis that in the infinite volume limit the first excited state is similar to the ground state (even if fitting ambiguities are in this case relevant, and the $3d$ result is largely compatible with the survival of a non-degenerate scenario). If replica symmetry breaking is present at non zero temperature, such a result could be interpreted by saying that $y(q)=0$: however a more carefully analysis is needed, especially on the extrapolation to the infinite volume limit. Moreover the possible existence of a crossover region in $T$ should be investigated before reaching definite conclusions.

\subsection{Coupled Replicas\protect\label{SS-COUREP}}

\begin{figure}
\includegraphics[width=0.6\textwidth]{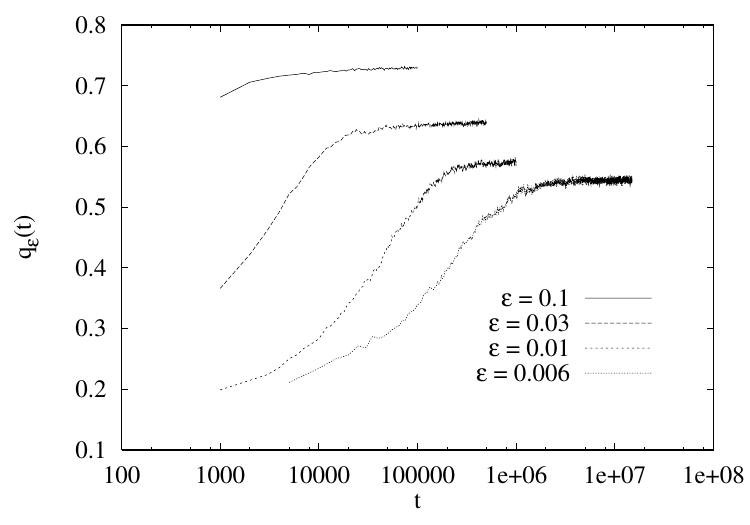}
\caption[0]{Time evolution of the overlap between two coupled
replicas for the $4D$ Ising spin glass with Gaussian couplings. 
The coupling strength, $\epsilon$, is smaller for the lower
curve.}
\label{F-conver}
\end{figure}

A very useful approach is based on coupling two lattice spin glass 
systems (of the same size and with the same realization of the 
quenched random couplings) by their mutual overlap. We define an 
Hamiltonian:

\begin{equation}
  {\cal H}_{J,\epsilon}[\sigma,\tau] \equiv 
  {\cal H}_{J}[\sigma] + {\cal H}_{J}[\tau] -
  \frac{\epsilon}{V} \sum_i \sigma_i \tau_i \ ,
\end{equation}
where ${\cal H}_{J}$ is the usual disordered Hamiltonian defined in 
equation (\ref{ham}).  We present here original results for the $4D$ 
Ising spin glass with Gaussian couplings: we will define in detail the 
procedure we have followed.  Older results of the same kind (with 
smaller lattices and a smaller statistical sample) can be found in 
references \cite{PA_RI} and \cite{CIPARIRU}.

We have studied (on a parallel computer of the APE series 
\cite{APE}) $4D$ systems of very large size: we have set $L=24$ 
($V=L^{4}$), and averaged over $6$ different coupled systems.  We have 
worked at zero magnetic field and
set $T = 1.35 \simeq 0.75\ T_c$.  We have investigate a set 
of $\epsilon$ values going from $0.1$ down to $0.006$.  We have used 
$10^{5}$ Monte Carlo sweeps of the system for $\epsilon=0.1$, $5\cdot 
10^{5}$ for $\epsilon=0.03$, $10^{6}$ for $\epsilon=0.01$ and 
$1.5\cdot 10^{7}$ for $\epsilon=0.006$.  For each run we have checked 
that the measured overlap had reached a stable plateau (see figure 
\ref{F-conver}).

\begin{figure}
\begin{center}
\includegraphics[width=0.6\textwidth]{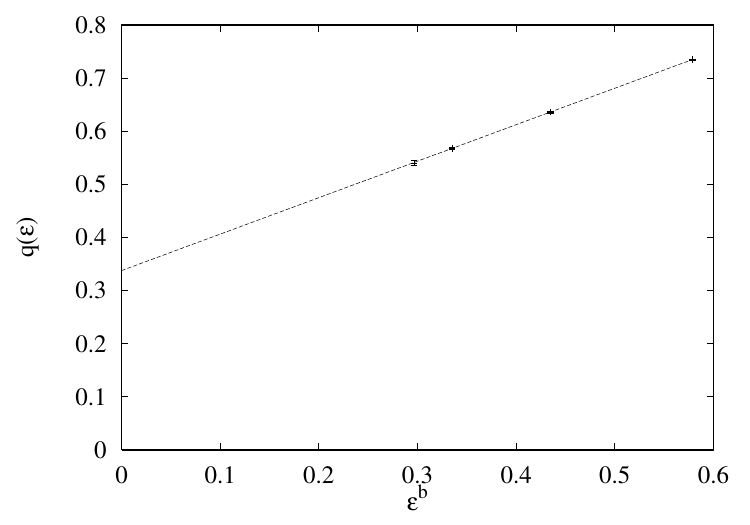} 
\end{center}
\caption[0]{The value $q_\epsilon(t=\infty)$ extrapolated to 
infinite time as a function of $\epsilon^{b}$.  We used the exponent 
found in our best fit: $b = 0.24\pm 0.03$.}
\label{F-q_eps}
\end{figure}

The presence of the interaction term $\frac{\epsilon}{V} \sum_i 
\sigma_i \tau_i$ with a positive value of $\epsilon$ forces the two 
copies of the system to stay closer than for $\epsilon=0$: 
$q_\epsilon(t) \equiv \frac{\sum_i \sigma_i(t) \tau_i(t)}{V}$ will 
tend as $t\to\infty$ to a value greater than the equilibrium value 
with $\epsilon=0$.

For $\epsilon$ not too small taking the infinite time limit of 
$q_\epsilon(t)$ is easy.  The forcing makes the correlation length 
(and time) small: the overlap tends to its asymptotic value in a 
reasonable time and we can obtain $q_\epsilon(t=\infty)$ directly from 
the value of the plateau, without any risky extrapolation procedure.  
We plot in figure \ref{F-q_eps} the value $q_\epsilon(t=\infty) \equiv 
q_\epsilon$ extrapolated to infinite time.

Next we have to estimate the limit for $\epsilon\to 0$ of $q(\epsilon)$.
In the simplified picture of a rugged free energy landscape,
made of many valleys, the coupling between two copies forces them into
the same valley, and so we expect to obtain in that limit an
information about the mean size of a valley.  Such  information, in
the theory of spin glasses, is encoded in the Edward-Anderson order
parameter ($q_{\rm EA}$) defined as the average overlap between two
replicas within the same pure state.

We can reach the same conclusion following a different way of
reasoning: the presence of the $\epsilon > 0$ term is equivalent to a 
potential favoring bigger overlaps, so that at equilibrium ($t 
\to \infty$) $q(\epsilon)$ will be the maximum overlap allowed 
for $\epsilon = 0$ plus a term due to the interaction,

\begin{equation}
  q(\epsilon) = q_{\rm max} + a \epsilon^b \ .
  \label{E-q_eps}
\end{equation}
For $\epsilon = 0$ the probability distribution function of the
absolute value of the overlap has support $[0,q_{\rm EA}]$:
the order parameter of the theory, $q_{\rm EA} = q_{\rm max}$, can be
obtained via the limit $\epsilon \rightarrow 0$.

We have fitted the behavior of (\ref{E-q_eps}) for the data shown in 
figure \ref{F-q_eps}.  On the best fit:

\begin{eqnarray}
  q_{\rm EA} = & 0.34 \pm 0.03 \ , \\
  a = & 0.69 \pm 0.02 \ , \\
  b = & 0.24 \pm 0.03 \ .
\end{eqnarray}
In the mean-field theory $b = \frac{1}{2}$, while in previous work 
\cite{PA_RI} in $4D$ the authors found $b \simeq \frac{1}{3}$.  The 
value found for the order parameter $q_{\rm EA}$ is in good agreement 
with the one found in \cite{4DIM}, where off-equilibrium measurement 
were suggesting $q_{\rm EA}(T=1.35) = 0.30 \pm 0.05$.

\subsection{Energy Overlap\protect\label{SS-ENEOVE}}

\begin{figure}
\begin{center}
\includegraphics[width=0.7\textwidth]{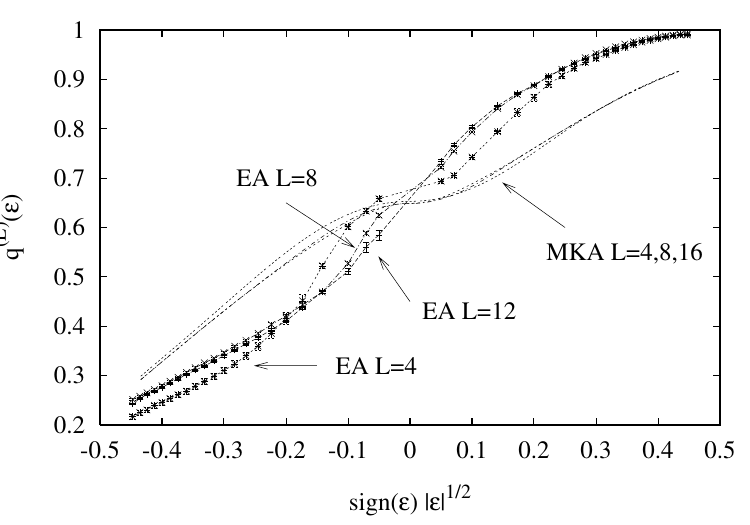}
\end{center}
\protect\caption[0]{ $q^{(L)}$ from numerical
simulations in the $3d$ EA spin
glass (lines with points) and in the Migdal Kadanoff approximation.}
\protect\label{fig:eo}
\end{figure}

We will show here results that implement the ideas discussed in
section \ref{S-ENEOVE} and \ref{COUREP}. We will show numerical
computations based on the Hamiltonian (\ref{E-HAMENRERE}). The data we
show here are from the work of some of us in reference
\cite{MOBODR}. They have been used in the debate about the hints that
one can hope to get from the Migdal-Kadanoff approximation when
studying spin glass systems \cite{MOBODR,MNPPRZ}.

We have analyzed the $3d$ EA model by numerical simulations (using a
tempering algorithm and an annealing scheme, checking convergence and
averaging over $64$ or more samples), with binary couplings and the
Hamiltonian (\ref{E-HAMENRERE}). We compare these data to the results
obtained in the Migdal Kadanoff approximation of the same model.

It turns out that one can see a clear difference, already at $T=0.7$ on
medium-size lattices, among the MKA and the EA model.  

In figure \ref{fig:eo} we show our results for $q^{(L)}(\epsilon)$
versus $\epsilon^{\frac12}$.  The MKA gives a smooth behavior: for
small $\epsilon$, $q^{(L)}(\epsilon)$ behaves like
$\epsilon^{\lambda}$, with $\lambda\simeq 1$.  Finite size effects
look very small for these sizes (from $4$ to $16$).  The EA model
behaves in a completely different way.  Here finite size effects are
large, and the behavior for small $\epsilon$ becomes more singular for
larger sizes.  The $L=4$ lattice is reminiscent of the MKA behavior,
but already at $L=8$ the difference is clear.  From our data we are
not able to definitely establish the existence of a discontinuity, but
the numerical evidence is strongly suggestive of that.  The data are
suggestive of the building up of a discontinuity as $L\to\infty$, i.e.
$q=q_{+}+A_{+}\epsilon^{\lambda}$ for $\epsilon>0$ and
$q=q_{-}+A_{-}|\epsilon|^{\lambda}$ for $\epsilon<0$, with $q_{+}\ne
q_{-}$ and an exponent $\lambda$ close to $\frac12$: a continuous
behavior (i.e.  $q_{+}= q_{-}$) cannot be excluded from these data,
but in this case we find an upper limit $\lambda < 0.25$, totally
different from the behavior of MKA, $\lambda \simeq 1$.  

\subsection{Ultrametricity\protect\label{SS-ULTRAM}}

The ultrametric organization of the equilibrium states is one of most
distinctive and striking features of the RSB solution of the spin
glass mean field theory. Checking if an organization of the same kind
still rules the state distribution in finite dimensional models is a
task of large interest \cite{CAMAPA}.
 
\begin{figure}
\begin{center}
\includegraphics[angle=90,width=0.7\textwidth]{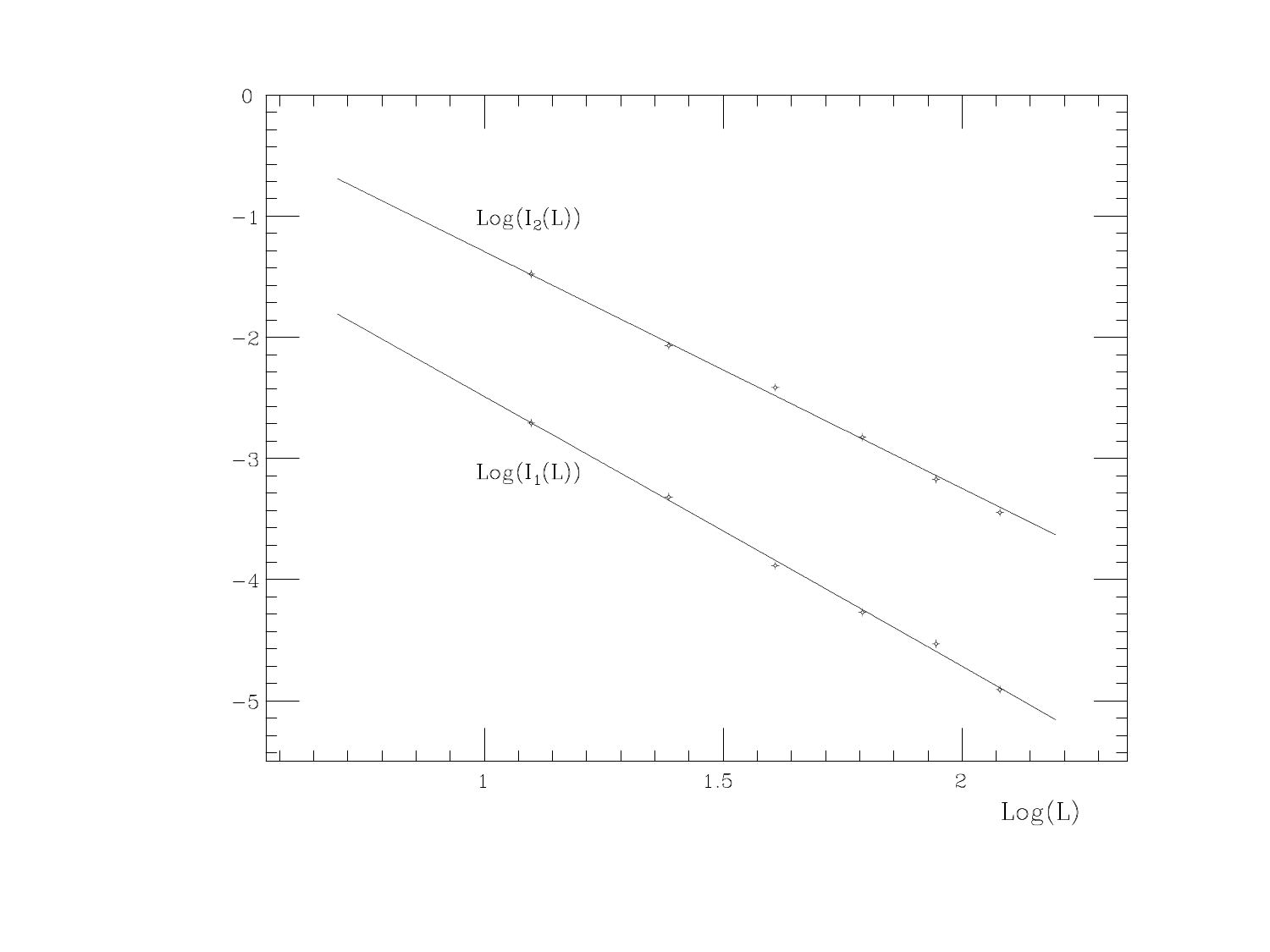}
\end{center}
\protect\caption[0]{The probability of a forbidden overlap
$I^L$ as a function of $L$. 
For the lower points we have
fixed $q_{1,2}=q_{2,3}$, for the upper points $q_{1,2} \neq q_{2,3}$.}
\protect\label{fig:ultra}
\end{figure}

The issue of the ultrametricity in a finite number of dimensions has
been checked numerically in \cite{CAMAPA}.  The most convincing
numerical tests have been done in four dimensions, since there the
phase transition is more clear. In figure \ref{fig:ultra} we
plot the probability ($I^L$) of the appearance of an overlap which is
forbidden by ultrametricity as a function of the lattice size.  This
probability goes to zero as $L\to\infty$ with a power law in $L$, and
the power law (i.e an exponent of $2.21$) is consistent with
expectations coming from mean field analytic computations (for more
details see \cite{CAMAPA}).  The support to the existence of a low
temperature phase with an ultrametric structure of states is very
strong.

Other numerical simulations done with a different technique in three
dimensions on systems with side from $4$ to $12$ are compatible with
ultrametricity, but the approach to zero seems to be much slower that
in four dimension and at the present moment it is difficult to reach a
definite conclusion \cite{HARTMA,ULTRA3}.  This problem should be better
investigated in the future.

We recall that is also possible to show \cite{INPARU96}, by using the
set of sum rules we have discussed before, that if the states are
organized ultrametrically in a finite dimensional spin glass the
detailed structure of the ultrametric organization has to be exactly
the same of the mean field solution.  This fact does not prove
ultrametricity, but once ultrametricity has been detected (for
example numerically) it allows to establish many other important
results about the structure of the system.

\subsection{Dynamics\protect\label{SS-DYNAMI}}

\begin{figure}
\begin{center}
\includegraphics[width=0.75\textwidth]{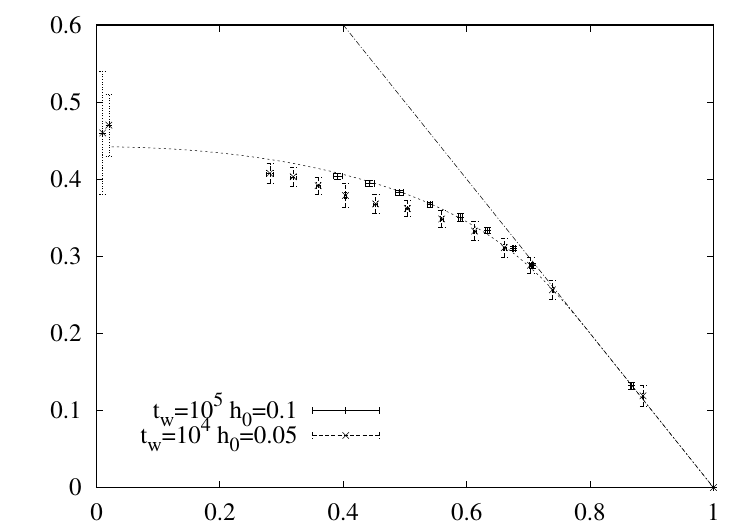}
\end{center}
\protect\caption[0]{The function $X(C)\equiv
\frac{T\,m}{h_0}$ versus $C$ (see text) for a $3D$ 
Ising spin glass with Gaussian couplings.  $L=64$ and $T\simeq 0.7\ 
T_c$.  The curve is the function $x(C)$ obtained integrating the 
equilibrium probability distribution of the overlap.  The straight 
line is the FDT prediction (i.e.  the straight line $1-C$).  Data are 
from two different numerical simulations, one with $t_w=10^5$ and 
$h_0=0.1$ and the other one with $t_w=10^4$ and $h_0=0.05$}
\protect\label{fig:fdt_1}
\end{figure}

The analysis of the dynamical behavior of spin glasses can give a 
large amount of information (see section \ref{SS-DYNAPP} earlier in 
the text for a first discussion and references).  Here we summarize 
some nice numerical results concerning the dynamical determination of 
the overlap probability distribution, from reference \cite{MPRRu1}.

As we have already discussed, the basic ingredients of the dynamical 
approach are the autocorrelation function $C(t,t')$ and the integrated 
response function.  We use a magnetic field that jumps at time 
$t_{w}$:

\begin{equation}
	h(t)=h_0 \theta(t-t_w)\ .
	\label{E-HHT}
\end{equation}
In this way by computing the function $X[u]$ one can test numerically
the validity of the relation (\ref{mag_fdt}), that relates the
magnetization and the correlation function.  In other words we try to
check that the function $X$ depends only on the value of the
correlation function $C(t,t')$ (and not on the times $t$ and $t'$)
even for large but finite times.  A second interesting goal is to
check that $X$ is given by the integrated probability distribution of
the overlap.

We have simulated four couples of spin samples, with Gaussian 
couplings, on a $L=64$ lattice in three dimensions.  We have used two 
different values of the magnetic fields in order to control the 
validity of the linear response approximation.

We show in figure \ref{fig:fdt_1} the results for the function $X$.  
We have also plotted the equilibrium behavior, where the FDT theorem 
holds (i.e.  the straight line $1-C$).  This method also gives us a 
way to estimate the order parameter of the system as the point where 
the function $X$ departs from the FDT regime.  From figure 
\ref{fig:fdt_1} we can estimate the value of the order parameter as 
$q_{\rm EA}\simeq 0.68$, in very good agreement with the equilibrium 
value quoted before ($q_{\rm EA}=0.70\pm 0.02$) (see figure 
\ref{fig:qea}). We remind the reader that

\begin{equation}
  x(q)=\int_0^q \d q^\prime\  P(q^\prime) \ ,
\end{equation}
where $P(q)$ is the equilibrium probability distribution of the
overlap.

>From \ref{fig:fdt_1} it is clear that for not too large waiting time
the function $X$ depends only on the value of the correlation
function. Moreover the equilibrium $x(q)$ matches extraordinarily well
the $X$ function.

It is clear that $x(q)$ does not have the form suggested by a droplet 
like picture.  The droplet picture predicts a linear region with slope 
$-1$ (the equilibrium regime, $q \in [q_{\rm EA},1]$) and another 
regime where we have a horizontal line ($q\in [0, q_{\rm EA}]$).  
Figure \ref{fig:fdt_1} rules out this possibility.

\section{Conclusions}

In this note we hope we have succeeded in clarifying better the
precise nature of the predictions of the Replica Symmetry Breaking
approach, and the numerical and analytical evidences available to
support its validity in finite dimensional, realistic spin glasses. It
is clear that there are many issues that need clarification: for
example working out a field theory describing finite dimensional
systems would be crucial. Also numerical simulations need to be
improved, and pushed to lower $T$ values (where the effect of the
critical point at $T_c$ is not obscuring too much the real low $T$
structure of the system).

Let us stress again the points of bigger relevance. Correlation
functions have been studied in detail in finite dimensional systems,
and they confirm a RSB picture. The proposal that a nontrivial
function $P(q)$ is an artifact due to the presence of interfaces among
two equilibrium states has been analyzed and falsified: the study of
window overlaps confirms that the non-trivial shape of $P(q)$ is not
due to the presence of interfaces. Coupling replicas helps again in
observing clear signatures of replica symmetry breaking. We have
discussed in detail about states, by stressing the importance of
finite volume states. We believe this is a crucial notion for
understanding a disordered system. We have analyzed the implications
of RSB on the finite volume states and discussed the hierarchical
organization of states. We have discussed in detail the difficulties
connected to the study of infinite volume states, and why we believe
that the recent rigorous results by Newman and Stein \cite{NS0,NS4}
strongly support RSB.  We have discussed numerical results that
confirm this point of view.

\section*{Acknowledgments}
The numerical computations that show the validity of the sum rules
have been run at the Konrad-Zuse-Zentrum f\"ur Informationstechnik
Berlin (ZIB) and at the HLRZ-J\"ulich. We thank H. St\"uben for
crucial help, and W. Janke and D. Johnston for discussions.


\end{document}